%% file: main.tex
\documentclass[twocolumn,prd, aps, notitlepage, nofootinbib, showpacs, superscriptaddress,longbibliography]{revtex4-1}
\usepackage[utf8]{inputenc}
\usepackage{graphicx,mathtools,xcolor,latexsym,rotating,soul}
\usepackage[T1]{fontenc}
\usepackage{tikz}
\usepackage{comment}
\usepackage{amsmath, amssymb, amsthm, amsfonts,breqn}
\usepackage{multirow}
\usetikzlibrary{decorations.pathmorphing}
\usepackage[colorlinks=true,citecolor=green,urlcolor=blue,linkcolor=blue]{hyperref}
\usepackage{pifont}
\usepackage{gensymb}
\usepackage{enumitem}
\usepackage{mathrsfs}
\newtheoremstyle{named}{}{}{\itshape}{}{\bfseries}{.}{.5em}{\thmnote{#3 }#1}
\theoremstyle{named}

\newcommand*{\rom}[1]{\expandafter\@slowromancap\romannumeral #1@}

\newcommand{\stf}[1]{\langle #1 \rangle}
\newcommand{\llp}{\ell (\ell + 1)}
\makeatletter
\let\cat@comma@active\@empty
\makeatother
\begin{document}
\title{Dynamical tidal response of non-rotating relativistic stars}

\author{Abhishek Hegade K. R.}
\email{ah30@illinois.edu}
\affiliation{Illinois Center for Advanced Studies of the Universe, Department of Physics, University of Illinois Urbana-Champaign, Urbana, IL 61801, USA}

\author{Justin L. Ripley}
\email{ripley@illinois.edu}
\affiliation{Illinois Center for Advanced Studies of the Universe, Department of Physics, University of Illinois Urbana-Champaign, Urbana, IL 61801, USA}

\author{Nicol\'as Yunes}
\email{nyunes@illinois.edu}
\affiliation{Illinois Center for Advanced Studies of the Universe, Department of Physics, University of Illinois Urbana-Champaign, Urbana, IL 61801, USA}
\begin{abstract}
Accurately modeling the tidal response of neutron stars is crucial to connecting gravitational wave observations of binaries to ultra-dense nuclear physics.
Most current models of the tidal response of relativistic stars either assume a static response model, or use phenomenological models inspired by Newtonian gravity.
In this work, we present a general formalism for computing the linear \textit{dynamical} tidal response function of relativistic, spherically symmetric stars.
Our formalism incorporates stratification due to thermal and chemical imbalances, allowing one to study the effects of $g$ modes on the tidal response function.
We also describe how to incorporate sources of dissipation due to shear and bulk viscosity.
To showcase the utility of our approach, we present several applications for polytropic stars in general relativity.
We show how our formalism can capture the dynamical tidal resonance due to the $f$ and $g$ modes of inviscid stars and explore the sensitivity of the dynamical tidal response to the compactness of the star.
We also compute the dissipative tidal deformability due to bulk and shear viscous dissipation assuming a simple viscous profile for the bulk and shear viscosity.
\end{abstract}
\maketitle
%-----------------------------------------------------
\section{Introduction}
Determining the properties of neutron star matter remains an outstanding goal of nuclear physics and astrophysics~\cite{Ozel:2016oaf,Lattimer:2021emm,Burgio:2021vgk}.
The physics of these objects can be indirectly probed through the gravitational waves (GWs) emitted from neutron star binaries.
A neutron star in a binary system is tidally deformed by the gravitational field of its companion, and its tidal response is dictated by the internal properties of the star. 
Thus, accurately modeling the tidal response of a neutron star is essential for connecting GW astronomical observations to nuclear physics \cite{Flanagan:2007ix}.

Several different approaches have been employed to model the impact of tides on the gravitational waveform of neutron star binaries.
Some waveform models are based on intuition from Newtonian theory~\cite{andersson2021phenomenology}.
Other models treat the stars' tidal response as a harmonic oscillator, with a fundamental oscillation mode set to the $f$-mode frequency of the star~\cite{Hinderer_2016,Steinhoff_2016,Gupta_2021}.
Both classes of waveform models are then calibrated to results from numerical relativity simulations of neutron star binaries~\cite{Hinderer_2016,Steinhoff_2016,abac2023nrtidalv3}.
These models have been successfully applied to model $f$-mode resonances in the tidal response and to place constraints on the $f$-mode frequencies of a neutron star, using current and future GW detectors~\cite{PhysRevD.100.021501,Pratten_2020}.

Though successful, the dynamical tidal models of~\cite{Hinderer_2016,Steinhoff_2016} use a number of simplifying assumptions.
These models of dynamical tidal response \textit{assume} that the dynamical tidal response in general relativity is similar to the Newtonian tidal response, i.e., that the tidal response can be spectrally decomposed in terms of a complete set of basis functions, consisting of the fundamental oscillation modes of a neutron star.
Additionally, these models cannot incorporate viscous effects or model effects due to thermal and chemical inhomogeneities, which might be important during the late inspiral due to Urca reactions~\cite{Alford:2017rxf,Alford:2019qtm,Most:2021zvc,Most:2022yhe,Chabanov:2023blf}, or other processes \cite{Jones:2001ya,Lindblom:2001hd,Gusakov:2008hv,Alford:2020pld}.
Moreover, these models use numerical relativity simulations for calibration, which can be computationally expensive.
To improve the models of~\cite{Hinderer_2016,Steinhoff_2016}, a prescription of dynamical tidal response that can go beyond intuition from Newtonian theory, that can incorporate viscous effects, and that is computationally less expensive to evaluate is necessary.

While there is a large body of work on the tidal response of Newtonian stars, (e.g.~\cite{Lai:1993di,Andersson_2020,andersson2021phenomenology,Ogilvie-review}), most calculations of the tidal deformability of relativistic stars assume that (i) the tidal response is static~\cite{Hinderer:2007mb,Binnington:2009bb,Damour:2009vw} and (ii) conservative, and,
(iii) that the neutron star is convectively unstable or neutrally stable, i.e., there are no low-frequency $g$ modes~\cite{Hinderer:2007mb,Binnington:2009bb,Damour:2009vw}.
The first and third assumptions require that the external tidal field experienced by a neutron star in a binary changes much more slowly than any relevant internal timescale of the star---that is, that the star is not resonantly excited during the inspiral. 
The second assumption requires that the neutron star tides only cause reversible changes to their internal energy.
These assumptions are sufficient to accurately model the tidal response of stars during the early inspiral stage, provided the viscous effects are small and there is no substantial stratification of different layers.

Due to the highly dynamical nature of spacetime during the late inspiral phase, both of these effects (i.e., viscous effects and stratification) may be large enough to be physically relevant for modeling the GWs emitted from the binary.
The consequences of stratification (through, e.g., temperature gradients or gradients in the chemical/nuclear composition at different densities) will become more pronounced should the orbital frequency approach the frequency of any modes of the star.
Viscous effects due to out-of-equilibrium nuclear reactions within the star (such as Urca reactions) may also be large enough to measurably affect the orbital motion of the stars \cite{Ripley:2023qxo}.

In this paper, we first describe how to compute the relativistic tidal response of neutron stars.
The formalism we develop is flexible enough to model the low-frequency behavior of the g-mode resonances due to stratification within the star, and the dissipative tidal response. 
Our approach relies on reducing the internal problem of the polar perturbations of a spherically-symmetric neutron star background into three master equations, involving the perturbation to the gravitational potential and the radial and angular components of the Lagrangian displacement vector~\cite{Cox-book,Lindblom-Mendell-Ipser-1997}.
These equations are essentially relativistic generalizations of the tidal equations for Newtonian stars.
To treat the external problem of imposing the tidal boundary conditions, we use the approach proposed recently by Poisson~\cite{Poisson:2020vap} and show how one can re-sum this expansion in the frequency domain to obtain the full dynamical tidal response.
We include viscous effects by using a perturbative expansion in Reynolds number~\cite{Terquem_1998} and provide general expressions for the viscous source functions.

We present several applications of our formalism for quadrupolar deformations of neutron stars obeying a polytropic EoS, which we use as a toy model for illustrative purposes.
We first show that our formalism reproduces the results of static tides in general relativity in the low-frequency regime for stars that are unstable/neutrally stable against convection---that is, for stars that \emph{do not} have any low-frequency resonant (g) modes~\cite{Hinderer:2007mb,Binnington:2009bb,Damour:2009vw}.
We next show how to obtain the $f$-mode and $g$-mode resonances of convectively stable stars using our formalism.
Finally, we calculate the dissipative tidal lag due to shear and bulk viscosity.

Finally, let us discuss the recent work of Pitre and Poisson~\cite{Pitre:2023xsr}, which also treats the problem of dynamical tidal response in full general relativity, and place it in the context of our work.
Our approach differs from theirs in a number of ways.
Pitre and Poisson treat both the internal and external perturbations of the spacetime in the small-time/small-frequency approximation.
Working with such an assumption inside the star only works if there are no low-frequency $g$ modes.
While we treat the external problem in the same way they do (in the time domain), we re-sum the external problem (i.e.~the master equations) in the frequency domain to allow for the possibility of resonances in the tidal response, both in the low-frequency limit due to $g$ modes and, in the high-frequency limit due to $f$ modes. 
For the case of an $f$-mode resonance, Pitre and Poisson had to re-sum the tidal response function   
\emph{after} solving for the tidal coefficients to obtain the complete tidal response near the resonance. 
We also incorporate effects, such as viscosity and stratification, which provides a more flexible and comprehensive account of stellar tides.
Finally, the normalization choice we make for the particular solutions of the gravitational master equation is different from that of Pitre and Poisson.
We discuss how this normalization choice affects the tidal response function in, and we also compare our work to~\cite{chakrabarti2013new} in Appendix~\ref{appendix:compare-Eric-to-us}.

The outline for the rest of the paper is as follows: In Sec.~\ref{sec:dissipation-review}, we review the major sources of dissipation inside a neutron star.
Next, in Sec.~\ref{sec:tidal-response-review}, we review how the tidal response of a neutron star is defined and review the properties probed by the tidal coefficients in the low-frequency regime.
In Sec.~\ref{sec:non-radial-perturbations}, we describe the master equations and the treatment of our boundary conditions for the tidal problem. In Sec.~\ref{sec:Polytropic-Star-Dissipative-Love}, we apply our formalism to compute the dynamical tidal response of polytropic stars and the low-frequency tidal lag coefficient of viscous polytropic stars.
Finally, in Sec.~\ref{sec:conclusions}, we present our conclusions and directions for future work.
Henceforth, we use the following conventions: the signature of our metric is $(-,+,+,+)$, and we use geometric units $G=1=c$.
%-----------------------------------------------------
\section{Physical sources of dissipation in neutron stars}\label{sec:dissipation-review}

Here, we review the various sources of dissipation that could be present in the interior of neutrons stars.
\emph{Microscopic} sources of dissipation arise from interactions between fundamental particles inside a neutron star.
Other \emph{anomalous} sources of dissipation arise from macroscopic sources, such as turbulence or crust dynamics, and they contribute as an \textit{effective} source of viscosity.
In giant stars and planets, microscopic sources of dissipation are too small to be relevant for the orbital dynamics of those objects. 
Effective sources from convective damping are considered to be major sources of dissipation in these systems~\cite{Ogilvie-review,Zahn:2008fk}.
In neutron stars, there are only rough order-of-magnitude estimates for sources of anomalous viscous processes.
For example, there could be dissipation due to the melting of the solid crust during the late stages of inspiral. There could also be dissipation due to turbulence near the surface of neutron stars~\cite{1992ApJ...398..234K}.
Such sources are speculative, but they ought to contribute to an averaged shear viscosity as large as $\stf{\eta}\sim 10^{29}\mathrm{g}\;\mathrm{cm}^{-1}\mathrm{s}^{-1}$ \cite{1992ApJ...398..234K} where brackets stand for the volume average over the entire star. 

By contrast, several microscopic sources of neutron star viscosity have been studied in great detail. 
Viscosity generated by microscopic processes in neutron stars depends sensitively on the local stellar temperature profile $T$.
The temperature profile affects not only the theoretically predicted values of viscosity, but also the mechanism that drives dissipation. 
Electron-muon scattering contributes to the shear viscosity, which scales as $T^{-2}$~\cite{Shternin:2008es}.
The peak value of this contribution is expected to be $\stf{\eta}\lesssim 10^{22} \mathrm{g}\;\mathrm{cm}^{-1}\mathrm{s}^{-1}$.
Bulk viscous contributions due to the presence of hyperons scale as $T^{-2}$ and may be the dominant source of dissipation in neutron stars at very low ($\sim$keV) temperatures.
The maximum value of hyperon bulk viscosity is estimated to be $\stf{\zeta}\sim 10^{30}\mathrm{g}\;\mathrm{cm}^{-1}\mathrm{s}^{-1}$ in some models \cite{Jones:2001ya,Lindblom:2001hd,Gusakov:2008hv,Alford:2020pld}. 

Bulk viscous contributions from direct and modified Urca processes are expected to dominate at higher temperatures, because those reactions scale as $T^{4}$ and $T^6$, respectively~\cite{physrevd.39.3804}\footnote{We note that beyond a resonant peak, these reactions become less strong at higher temperatures; current estimates place this peak at $T\sim 5\mathrm{MeV}$ \cite{Alford:2023gxq}, that is at higher temperatures than we expect the stars to reach during the inspiral.}.
Typical predictions for the bulk viscosity for Urca-process-driven viscosity range from $\stf{\zeta}\sim 10^{26}\;\mathrm{g}\;\mathrm{cm}^{-1}\mathrm{s}^{-1}$ when $T\sim0.1\mathrm{MeV}$, to $\stf{\zeta}\sim 10^{31}\;\mathrm{g}\;\mathrm{cm}^{-1}\mathrm{s}^{-1}$ when $T\sim 1 \, \mathrm{MeV}$, depending on the equation of state  (EoS)~\cite{Most:2021zvc,Yang:2023ogo}.
Given such high values of bulk viscosity, several groups have started to use numerical simulations to understand the impact of bulk viscosity during the late merger/post-merger phase.
Some groups have found that bulk viscous effects (from Urca-processes) are enhanced during the late inspiral~\cite{Most:2021zvc,Chabanov:2023abq,chabanov2023impact}. Meanwhile, other groups, working with moment-based treatments of neutrino transport, have not found evidence for large out-of-thermodynamic-equilibrium effects, which are necessary for generating an effective bulk viscosity during the late inspiral~\cite{Radice_2022}. Nevertheless, these  moment-based treatments of neutrino transport have revealed evidence of bulk viscous effects within a small window after the merger before matter returns to equilibrium~\cite{espino2023neutrino}.
GW observations from GW170817 have also been used to constrain the amount of bulk and shear viscosity during the inspiral~\cite{Ripley:2023lsq}.
Preliminary estimates indicate that the bulk viscosity during the inspiral is less than $\sim 10^{31} \mathrm{g}/{\mathrm{cm s} }$; similarly, shear viscosity during the inspiral was constrained to be less than $\sim 10^{28} \mathrm{g}/{\mathrm{{cm s}}}$. These constraints are expected to improve by 2-3 orders if GW170817-like events are observed using 3G detectors. 

Clearly, the strength of dissipative processes within a neutron star in a binary depends sensitively on the temperature profile of the star, and also on how the stellar temperature evolves over time.
As the binary enters the late inspiral, heating from tidal friction due to Urca reactions may increase the temperature of the two stars to tens of keV \cite{Arras:2018fxj}, with additional hyperonic bulk viscous contributions possibly heating the stars to higher temperatures \cite{Alford:2020pld}. 
Numerical relativity simulations of neutron star mergers additionally suggest tidal heating could increase the stellar temperature to a few MeV during the last few orbits \cite{Perego:2019adq}. 
This being said, there remains considerable uncertainty in the profile, value, and evolution of the temperature of neutron stars in binaries. 

%-----------------------------------------------------
%-----------------------------------------------------
\section{Tidal response of neutron stars and the properties probed by the tidal coefficients}\label{sec:tidal-response-review}

In this section, we define the tidal response function of compact objects, discuss the impact of the low-frequency tidal coefficients on the gravitational waveform, and review distinct physical processes probed by the low-frequency tidal coefficients.
Our discussion is based on~\cite{Poisson:2020vap,Ripley:2023qxo}, and we refer the reader to those references for a more detailed discussion.
In this section, we adopt the following notation.
We work with Cartesian coordinates $\left(t, x, y, z \right)$.
The indices $\left(i,j,\ldots\right)$ are used to denote spatial coordinates only.
We denote the $L^{th}$multipole moment of body $A$ by $I^{L}_{A}$ where $L=\ell_1,...\ell_{L}$ denotes a multi index of size $L$.
We denote the mass and radius of objects $A$ and $B$ by $m_{A/B}$ and $R_{A/B}$.
We occasionally denote time derivatives with an overhead dot.
We use $\stf{\cdots}$ in index lists to denote the symmetric trace-free (STF) combination of tensorial indices.
We transform from physical space to Fourier space by replacing $\partial_t \to -i \omega$.

To describe the tidal response of two compact objects in a binary, we first review their motion in the center-of-mass frame in Newtonian theory.
We model the binary as a system of two point particles, and introduce finite size corrections order by order in a multipolar expansion of each object (for a review, see e.g. \cite{Poisson-Will}).
To leading order in the multi-polar moments of objects $A$ and $B$, the center-of-mass acceleration of the binary is
\begin{align}
\label{eq:newtonian-eob-acceleration}
    a_i
    =
    &
    -
    \frac{M}{d^2}n_i
    +
    \sum_{\ell = 2}^{\infty}
    \frac{M}{2}\left(
        \frac{I_A^{\stf{L}}}{m_A}
        +
        \frac{I_B^{\stf{L}}}{m_B}
    \right)
    \partial_i\partial_{L}\frac{1}{d}
    ,
\end{align}
where $a^i\equiv \ddot{x}^i_A - \ddot{x}^i_B$ is the relative acceleration, $M$ is the total mass, $d$ is the distance between the two objects and $n_i$ is the normal vector.
We assume the binary is evolving in a circular orbit of frequency $\omega_{\mathrm{orb}}$.

To close the system of equations, we must relate the multipolar moments of each star to the gravitational field external to it.
We assume that the tidal response can be modeled via linear response theory.
We can then relate the STF multipole moments of object $A$ to the external field it is immersed in, $\mathcal{E}_A^{L}$, via the tidal response function $K_{\ell}(t-t')$ through 
\begin{align}
    I_A^{\stf{L}}(t) =  -\frac{2 }{(2\ell-1)!!} R_A^{2\ell + 1}\int_{-\infty}^{\infty} K_{\ell} (t-t') \mathcal{E}_A^{L} (t') dt'\,.
    \nonumber
\end{align}
In Fourier space this integral equation becomes
\begin{align}\label{eq:tidal-response-definition-v1}
    \hat{I}^{\stf{L}}_A(\omega)
    =
    -\frac{2}{(2\ell-1)!!} R_A^{2\ell + 1} \hat{K}_{\ell}(\omega) \hat{\mathcal{E}}_A^{L}
    \,.
\end{align}
We now switch from an STF-tensor description to a spherical-harmonic description, using the relation (see, e.g. Sec. 1.5.2 of~\cite{Poisson-Will})
\begin{subequations}\label{eq:multipole-moment-driving-potential-def}
\begin{align}
    \hat{I}^{\stf{L}}_A(\omega) \equiv \frac{4 \pi \ell!}{(2\ell + 1)!!} \sum_{m=-\ell}^{\ell} \mathscr{Y}^{* \stf{L}}_{\ell m} \hat{I}_{A,\ell m}(\omega) \,,\\
    \hat{\mathcal{E}}_A^{L} \equiv -\frac{4\pi \ell!}{2 \ell + 1} \sum_{m=-\ell}^{\ell} d_{A,\ell m} (\omega) \mathscr{Y}^{* \stf{L}}_{\ell m}
\end{align}
\end{subequations}
where $\mathscr{Y}^{* \stf{L}}_{\ell m}$ is the STF tensor that transforms Cartesian STF tensors into spherical harmonics, $\hat{I}_{A,\ell m}$ is the Fourier transform of the multipole moment of object $A$, and $d_{A,\ell m}$ is the tidal driving potential felt by object $A$ due to object $B$.
From here on, we drop the labels $A/B$ to simplify our notation, unless the context makes it unclear.

Using the above definitions, Eq.~\eqref{eq:tidal-response-definition-v1} can be simplified to 
\begin{align}\label{eq:tidal-response-definition}
    \hat{I}_{\ell m}(\omega)
    =
    2 R^{2\ell + 1} \hat{K}_{\ell}(\omega) d_{ \ell m}(\omega)
    \,.
\end{align}
The tidal response function $\hat{K}_{\ell}(\omega)$ is, in general, a complex function that depends on the internal structure of the compact object and the frequency of the orbit.
An elementary property of $\hat{K}_{\ell}(\omega)$ follows from the fact that $K_{\ell}(t)$ is a real function
\begin{align}
    \hat{K}_{\ell}^{*}(\omega) = \hat{K}_{\ell}(-\omega)\,,
\end{align}
and therefore, 
\begin{subequations}
\begin{align}
    \mathrm{Re}\left(\hat{K}_{\ell}(\omega)\right)
    =
    \frac{\hat{K}_{\ell}(\omega) + \hat{K}_{\ell}(-\omega)}{2} \,,\\
%--------------------------------------
    \mathrm{Im}\left(\hat{K}_{\ell}(\omega)\right)
    =
    \frac{\hat{K}_{\ell}(\omega) -\hat{K}_{\ell}(-\omega)}{2i} \,.
\end{align}
\end{subequations}
The real part of the tidal response is an even function of $\omega$ and quantifies interactions that conserve orbital energy, while the imaginary part probes dissipative interactions.
Henceforth, we will call $\mathrm{Re} \left(\hat{K}_{\ell}\right)$ the \textit{conservative tidal response function} and $\mathrm{Im} \left(\hat{K}_{\ell}\right)$ the \textit{dissipative tidal response function}.
We also introduce the following definitions
\begin{subequations}
\begin{align}
    k_{\ell}(\omega) &\equiv \mathrm{Re}\left(\hat{K}_{\ell}(\omega)\right)
    \,,\\
    \omega k_{\ell}(\omega) \tau_{d,\ell}(\omega)
    &\equiv 
    \mathrm{Im}\left(\hat{K}_{\ell}(\omega)\right)
    \,.
\end{align}
\end{subequations}
The function $k_{\ell}(\omega)$ is called the \textit{Love number}~\cite{Hinderer_2010,Binnington:2009bb,Damour:2009vw}, while we call $\tau_{d,\ell}(\omega)$ the \emph{tidal lag function}. 

For perfect fluids in Newtonian theory, one can show that the tidal response function is a real function, and that it can be written as ~\cite{Smeyers-Book,Andersson_2020} 
\begin{align}
    \label{eq:Newtonian-theory-tidal-response-perfect-fluid}
    \hat{K}_{\ell,\mathrm{Newt}}(\omega)
    &=
    \frac{1}{2(2 \ell + 1)R_A^{\ell + 1}}
    \sum_{j} \frac{Q_j^2}{\omega_j^2 - \omega^2} 
\end{align}
where, $Q_j$ is related to the ``overlap integral'' between the external tidal field and the fluid oscillation modes of the star~\cite{Press-Teukolsky}.
For polytropic stars (and when one ignores the self-gravity of linearized perturbations---the \emph{Cowling approximation}), the frequencies $\omega_j$ can be classified into the classical $f$-, $p$-, and $g$-mode frequencies~\cite{1941MNRAS.101..367C,Cox-book}.
For perturbations that include self-gravitational forces, and for more general EoSs, many stellar modes $\omega_j$ can be thought of as being close analogues to an $f$, $p$, or $g$ modes (in the sense that they approach an $f$, $p$, or $g$ modes in the Newtonian, polytropic limit).
Given this, we will often refer to $f$, $p$, or $g$ modes without further qualification. The frequencies of the $f$ mode and $p$ modes are mostly determined by pressure restoration forces within the star.
Generally, the family of $p$ modes includes modes of arbitrarily large frequencies.
The frequencies of the $g$ modes are mostly determined by gravitational forces, and are only present when there is stratification due to temperature or chemical gradients in the star.

Quantitatively, the criteria for the presence of $g$ modes is determined by the Schwarzschild-Ledoux criteria~\cite{1958ses..book.....S}, which determines if the star is stable against convection.
Suppose that the equilibrium sound speed is given by $c_e^2 = {dp}/{de}$ where $p$ is the pressure and $e$ is the energy density.
The adiabatic sound speed is given by $c_s^2 = \gamma p/(e+p)$ where $\gamma$ is the adiabatic index.
We say that a star is stable against convection if
\begin{align}\label{eq:Schwarzschild-criteria}
    c_s^2 > c_e^2 \implies
    \gamma > \frac{e + p }{p } \frac{dp}{de}\,.
\end{align}
This is the Schwarzschild-Ledoux criteria. 
If this is satisfied, then small vertical displacements of a stratified boundary will oscillate (at the Brunt-Va\"is\"al\"a frequency) and there are $g$ modes; if this criterion is not satisfied, stratified boundaries are unstable to convection and there are no $g$ modes. 
The $g$ modes can have extremely small frequencies; for polytropic stars within the Cowling approximation, the $g$ modes have an accumulation point at zero frequency \cite{1941MNRAS.101..367C}.
These low-frequency oscillations can greatly complicate the tidal response of a star, as in principle, very small frequency perturbations may resonantly excite the star.

\emph{Assuming} there are no $g$ modes, we can expand the conservative and dissipative tidal response functions in a Taylor series about $\omega=0$,
\begin{subequations}
\begin{align}
    k_{\ell}(\omega) &= k_{\ell}(0) + \omega^2 k_{\ell}^{(2)} + \mathcal{O}(\omega^4) \,,\\
    \tau_{d,\ell}(\omega) &= \tau_{d,\ell}(0) + \mathcal{O}(\omega^2)\,.
\end{align}
We can combine these two expansions to obtain
\end{subequations}
\begin{align}\label{eq:low-frequency-limit}
    \hat{K}_{\ell}(\omega) = k_{\ell}(0) + i\omega k_{\ell}(0) \tau_{d,\ell}(0) + k_{\ell}^{(2)} \omega^2 + \mathcal{O}(\omega^3) \,.
\end{align}
The constant $k_{\ell}(0)$ is the \textit{equilibrium} tidal Love number, $\tau_{d,\ell}(0)$ is the tidal lag number, and $k_{\ell}^{(2)}$ is a ``dynamical tidal Love number'' \cite{Lai_1997,Chakrabarti:2013lua,Steinhoff:2016rfi}.
The dynamical tidal Love number can be thought of as a correction to the equilibrium tidal Love number due to dynamical process, such as stellar oscillations and viscous relaxation.

Each of these low-frequency tidal coefficients probe distinct physical processes and affect the gravitational waveform at different post-Newtonian (PN)\footnote{In the PN framework, one solves the Einstein equations perturbatively in weak fields and small velocities. A term of NPN order scales as $v^{2N}$ relative to the controlling factor of the expansion.} orders.
The $\ell=2$ equilibrium tidal Love number first appears in the gravitational waveform at $5$ PN order and has been used to constrain the different equilibrium nuclear EoSs~\cite{Hinderer_2010,Chatziioannou_2020}.
The tidal lag number affects the gravitational waveform at $4$ PN order and has been used to inform the magnitude of dissipative effects, such as bulk and shear viscous dissipation~\cite{Ripley:2023lsq}.
The dynamical tidal Love number $k_{\ell}^{(2)}$ affects the gravitational waveform at $8$ PN order and is determined mainly by the $f$-mode oscillations, and the viscous relaxation times that damp the $f$-mode excitation.
While it appears at very high PN order in the waveform, leaving out the dynamical Love tidal number from a waveform model may bias the measurement of $k_{\ell}\left(0\right)$ over a population of GWs from binary neutron stars \cite{Pratten:2021pro}.
The value of the dynamical tidal number has been additionally constrained with GW data from the event GW170817~\cite{PhysRevD.100.021501,Pratten_2020}.

The low-frequency expansion of Eq.~\eqref{eq:low-frequency-limit} fails near resonances.
In the absence of $g$ modes, the largest resonance first occurs near the $f$-mode frequency of the star \cite{Smeyers-Book}. We can approximately represent the tidal response function near this resonance as 
\begin{align}\label{eq:K-Newt-v2}
    \hat{K}_{\ell} (\omega) \approx \frac{k_{\ell} }{\omega_f^2 - i \omega \left(\tau_{d,\ell} \omega_f^2\right) - \omega^2}
\end{align}
where $\omega_f$ is the $f$-mode frequency.
Most estimates of the $f$-mode frequency give values of $\omega_f/(2\pi) \sim 2 \mathrm{kHz}$, which lies very close to the peak (merger) frequency of binary neutron star coalescence \cite{Kokkotas:1999bd}. 
Observe that the above equation is equivalent to Eq.~\eqref{eq:Newtonian-theory-tidal-response-perfect-fluid} when we set $\tau_{d,\ell} =0$ and truncate Eq.~\eqref{eq:Newtonian-theory-tidal-response-perfect-fluid} to just the $f$-mode frequency.
The tidal lag function acts a damping time to the star oscillating at a driven frequency $\omega$.

There is a robust and thoroughly developed formalism to calculate the tidal response function $\hat{K}_{\ell}\left(\omega\right)$ within Newtonian theory~\cite{Lai:1993di,Andersson_2020,Pitre:2023xsr,Ogilvie-review}.
For fully relativistic stars, a robust formalism exists only for the calculation of the equilibrium tidal love number ($k_{\ell}(0)$)~\cite{Hinderer:2007mb,Binnington:2009bb,Damour:2009vw}.
Recently, Pitre and Poisson introduced a method to calculate $k_{\ell}^{(2)}$ in the absence of $g$ modes~\cite{Pitre:2023xsr}.
Some references have used effective models for calibrating the $k_{\ell}^{(2)}$ number, using numerical relativity data~\cite{Hinderer_2016}.
In the following sections, we remedy this situation by developing a formalism for calculating the frequency-dependent tidal response function in general relativity.
%--------------------------
%-----------------------------------------------------
\section{Non-radial polar perturbations of spherically symmetric stellar spacetimes}\label{sec:non-radial-perturbations}
In this section, we describe an approach to treat the time-dependent tidal excitation of initially spherically symmetric stars in full general relativity using the technique of matched asymptotic expansions~\cite{Poisson:2020vap}.
Extracting the tidal response of compact objects requires that one solve the Einstein equations in three distinct spacetime regions, and that one matches these solutions inside regions of common overlap.
\begin{figure}[h!]
    \centering
    \includegraphics[width = 1 \columnwidth ]{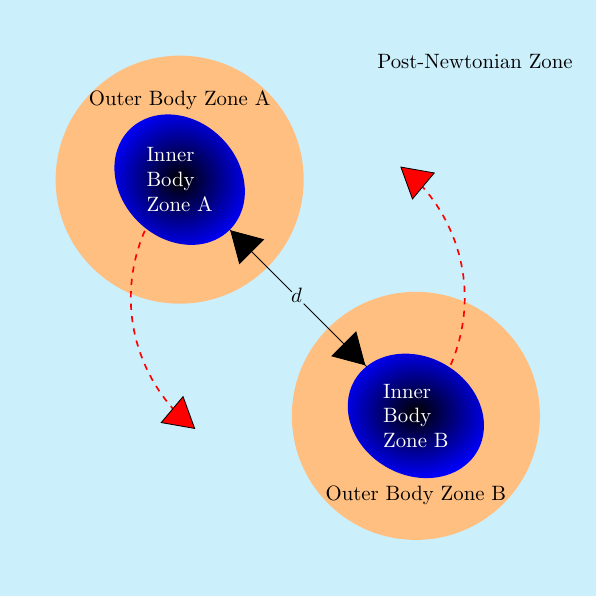}
    \caption{Cartoon (not to scale) depicting the motion of two tidally interacting neutron stars in a circular orbit of radius $d$ on a constant time slice.
    The spacetime is separated into three distinct zones: the inner body zone, the outer body zone and the post-Newtonian zone.
    The mutual gravitational attraction between the objects is weak in the post-Newtonian zone, and they are described as skeletonized compact objects with a multipolar structure in the post-Newtonian zone.
    The gravitational fields in the inner and outer body zone are strong, and the matching between the body zone metric and the post-Newtonian metric determines the dependence of the multipole on the internal structure of the objects.}
    \label{fig:zones-cartoon}
\end{figure}

To provide a concrete realization of these regions, we consider the motion of two neutron stars of radius $R_{A,B}$ and mass $M_{A,B}$, moving in a circular orbit at an orbital separation of $d$, as described by the relative acceleration in Eq.~\eqref{eq:newtonian-eob-acceleration}.
A cartoon depicting the objects on a fixed time slice with different regions of interest is shown in Fig.~\ref{fig:zones-cartoon}.
The objects are moving around each other in an ambient \textit{PN zone} (colored in light blue in the figure, and sometimes also called the ``PN near zone''), which we define as the region in which $\lambda \gg \bar{r}_{A,B} \gg R_{A,B}$, where $\bar{r}_{A,B}$ is the field point distance as measured from star $A$ or $B$ and $\lambda$ is the characteristic GW wavelength. In the PN zone, the neutron stars can be viewed as skeletonized compact objects with a multipolar structure that obeys the equation of motion of Eq.~\eqref{eq:newtonian-eob-acceleration}.
The value of the multipole moments of the object are unknown constants inside the PN zone and on a given time slice.

Close to each compact object, we can define body zones.
To describe these regions, let us focus on object $A$ (upper-left corner in the figure) with mass $m_A$ and radius $R_A$ and move into a frame where the compact object is at rest.
Let us further divide the body zone into an \textit{inner body zone}, $\bar{r}_A < R_A$, and an \textit{outer body zone}, $d \gg \bar{r}_A >R_{A}$. The outer body zone of body $A$ extends into the PN zone near body $A$, thus defining an overlap region that we denote the \textit{buffer zone} (and which will be important later on). In the inner/outer body zones,
the gravitational field of body $A$ is strong. 
We model the dynamics of this region by solving the full Einstein equations---that is, we do not make any weak-gravity assumption.

In the inner and outer body zones and for weak tidal interactions, one can solve the Einstein equations linearized around a background equilibrium solution, but different approaches in each zone.
In the inner body zone, we must solve the linearized Einstein equations in the presence of matter and in conjunction with the linearized relativistic fluid equations of the star. In the outer body zone, we only have to solve the linearized Einstein equations in vacuum.
Once we solve the linearized Einstein equations in the inner and outer zones, we glue them together by demanding that the metric (and, thus, in particular, the gravitational potential) be continuous and differentiable at the surface of the star.

The gluing procedure leaves a number of constants undetermined in the outer body zone, which must be found through asymptotic matching to relate them to the multipole and the tidal moments. That is, we demand that the solution obtained in the outer body zone, asympotically expanded in the PN zone, be asymptotic to the solution obtained in the PN zone, asymptotically expanded in the outer body zone, inside of the buffer zone region, $d \gg \bar{r} \gg R_{A}$. Technically, asymptotic matching also requires that we ensure the outer body zone solution and the PN zone solution are in the same gauge and coordinate system, which requires a coordinate transformation that has already been worked out in~\cite{Poisson:2020vap}. Asymptotic matching, then, determines the constants and the multipole moments, which are related to the tidal response function via e.g.~Eq.~\eqref{eq:tidal-response-definition-v1}. The asymptotic matching procedure summarized above is essentially identical to what has been carried out for perturbed black holes in~\cite{Yunes:2005nn,Johnson-McDaniel:2009tvj}. 
 
The rest of this section explains the details of the calculation required of the matching procedure.
We first describe how to solve the Einstein equations in the inner body zone by revisiting non-radial polar perturbations of spherically symmetric neutron stars in Sec.~\ref{sec:Field-equations}.
We reduce the coupled Einstein-fluid equations into three master equations: a second-order equation for the gravitational potential, and two first-order equations for the radial and the angular components of the Lagrangian displacement vector.
We then solve the linearized Einstein equations in the outer zone in Sec.~\ref{sec:tidal-bcs} and discuss the matching between the outer zone and the PN zone to obtain the tidal boundary conditions in Sec.~\ref{sec:Matching-PN}.
Next, we perturbatively add viscous fluid corrections to the equations of motion, and solve the equations in the inner zone of the star to obtain the dissipative tidal response function in  Sec.~\ref{sec:viscous-source}.
We then discuss the general nature of the solutions near the origin and the surface of the star in Sec.~\ref{sec:boundary-conditions}.
Finally, we provide another summary of our calculational approach in Sec.~\ref{sec:summary-body-zone}.
%-----------------------------------------
\subsection{Field equations in the body zone}\label{sec:Field-equations}
The equations of motion for the non-radial, linear, polar perturbations of the Einstein equations coupled to a perfect fluid (about a spherically symmetric background) were first published by Thorne and Campolattaro~\cite{1967ApJ...149..591T}.
In that work, the authors additionally reduced the linearized, polar equations into a system of five coupled equations.
Lindblom and Detweiler later showed that these equations can be further reduced into a system of four coupled equations~\cite{Lindblom-Detweiler-1983,Detweiler-Lindblom-1985}.
Finally, in~\cite{Lindblom-Mendell-Ipser-1997} Lindblom, Mendell and Ipser showed that these four equations could be further simplified into a system consisting of two sub-systems of two coupled equations each, one for the gravitational potential and one for the fluid perturbation.
The Newtonian limit of these equations reduces to their Newtonian analogues ~\cite{Cox-book,Smeyers-Book}.
Here, we present a different but essentially equivalent reduction of the linearized, polar Einstein equations, which is valid for stably stratified stars---that is, for stars that contain $g$-mode oscillations. 
We find that this reduction is particularly convenient when including viscous corrections to the perturbative corrections; see Sec.~\ref{sec:viscous-source}.

Before we describe the equations of motion, we set up our notation. 
We are interested in describing the dynamics in the inner body zone.
We consider linearized perturbations to a spherically symmetric Tolman-Oppenheimer-Volkoff (TOV) background in polar coordinates $\left(t,r,\theta,\phi\right)$.
These coordinates will be used to describe the dynamics in both the inner and outer body zone.
Capital Latin indices $A,B\ldots$ are used to denote the coordinates $\left(\theta,\phi\right)$.
Note that in this subsection we drop the use of $A,B$ to denote body $A/B$; we only consider the perturbation of one object in this section.
We Fourier transform the time dependence of all the perturbed quantities, i.e.~we replace $\partial_t \to -i \omega$.
Scalar, vector and tensor polar spherical harmonics are denoted by $Y_{\ell m}$, $E_A^{\ell m }$ and $Z_{AB}^{\ell m }$ respectively; see Appendix~\ref{appendix:scalar_vector_tensor_spherical_harmonics} for a brief review.
We only consider polar perturbations in this article.
We set $u_{\mu}$ to be the fluid four velocity, $p$ to be the pressure, $n$ to be the baryon number density,
$\rho = m n$ the rest mass energy density, also known as the baryon (mass) density
, $e$ to be the total energy density, and $\eta,\zeta$ to be the shear/bulk viscosity coefficients.
Finally, we use $\delta$ and $\Delta$ to denote the Eulerian and Lagrangian fluid perturbations, respectively; we review relativistic fluid perturbation theory in Appendix~\ref{appendix:relativistic-perturbation-theory}.
Since the field equations we discuss have been described in detail in the literature before (e.g. \cite{Lindblom-Detweiler-1983,Detweiler-Lindblom-1985,Ipser-Lindblom-Newtonian-1990}), our discussion will be brief, and we relegate most of the details to Appendix~\ref{appendix:relativistic-perturbation-theory}.

We analyze the Einstein equation and the fluid conservation equations coupled to a stress-energy tensor $T_{\mu\nu}$ and a source term $\mathbb{S}_{\mu\nu}$,
\begin{subequations}\label{eq:Field-equations}
\begin{align}
    \label{eq:Einstein-equations}
    G_{\mu\nu} &= 8 \pi \left(T_{\mu\nu} + \mathbb{S}_{\mu\nu} \right)
    \,, \\
    \label{eq:Fluid-conservation-equations}
    \nabla_{\mu} T^{\mu\nu} &= -\nabla_{\mu} \mathbb{S}^{\mu\nu} \,.
\end{align}
\end{subequations}
The source term serves as a proxy for modeling out-of-equilibrium contributions to the stress-energy tensor.
The stress-energy tensor is that of a perfect fluid obeying a polytropic EoS $p(e)$ 
\begin{align}
    \label{eq:perfect-fluid-tensor}
    T_{\mu\nu} &= \left(e+ p\right) u_{\mu} u_{\nu} 
    + p g_{\mu\nu}
    \,.
\end{align}
The background metric is given by 
\begin{align}\label{eq:background-metric}
    ds^2_0 
    = 
    -
    e^{\nu(r)} dt^2 
    + e^{\lambda(r)} dr^2 
    + r^2 d\Omega^2 
    \,.
\end{align}
We will assume that the source tensor $\mathbb{S}_{\mu\nu}$ vanishes on the background and it is perpendicular to the fluid four-velocity vector,
\begin{align}\label{eq:S-perp-condition}
    u^{\mu}\mathbb{S}_{\mu\nu} &=0\,.
\end{align}
The reason for using this additional restriction is that, for small deviations from equilibrium, we want energy conservation [Eq.~\eqref{eq:Deltaebyeplusp}] to be valid.
We could instead demand that $u^{\mu}\nabla^{\nu}\mathbb{S}_{\mu\nu} =0$, but since the viscous sources we consider satisfy Eq.~\eqref{eq:S-perp-condition}, we will restrict our attention to sources that obey the less general Eq.~\eqref{eq:S-perp-condition}.
Extending our reduction to more general sources that do not satisfy Eq.~\eqref{eq:S-perp-condition} should be straightforward.

\subsubsection{Background metric\label{sec:background-metric}}
The background metric of Eq.~\eqref{eq:background-metric} and fluid variables satisfy the TOV equations
\begin{subequations}\label{eq:TOV-equations}
\begin{align}
    \label{eq:lambda-equation-TOV}
    \lambda'(r) &= \frac{1- e^{\lambda(r)} + 8 \pi r^2 e^{\lambda(r)} e(r)}{r} \,,\\
    \label{eq:nu-equation-TOV}
    \nu'(r) &= \frac{-1+e^{\lambda }+8 e^{\lambda } \pi  r^2 p}{r} \,,\\
    \label{eq:p-equation-TOV}
    p'(r) &= -\frac{(e+p) \left(-1+e^{\lambda }+8 e^{\lambda } \pi  r^2 p\right)}{2 r }
    \,.
\end{align}
\end{subequations}
We introduce the mass aspect $m(r)$, which is related to $\lambda(r)$ by the equation
\begin{align}
    \lambda(r) &= -\log\left(1 - \frac{2 m(r)}{r} \right) .
\end{align}
From Eq.~\eqref{eq:lambda-equation-TOV} we have that
\begin{align}
    \label{eq:M-equation-TOV}
    m'(r) &= 4 \pi r^2 e(r) \,.
\end{align}
For numerical integration, it is beneficial to use Eqs.~\eqref{eq:nu-equation-TOV}, \eqref{eq:p-equation-TOV} and \eqref{eq:M-equation-TOV} and invert $e(p)$ to $p(e)$.

To integrate the TOV equations, we need to understand the local behavior of the variables at the origin ($r=0$) and at the surface of the star $(r=R)$, which is defined by $p(R)=0$, $p(r<R) > 0$.
Series expanding about the origin and imposing regularity of the spacetime, the TOV equations admit a closed form, perturbative solution
\begin{subequations}
\begin{align}
    p(r) &= p_c - \frac{2 \pi r^2}{3} \left(p_c + e_c\right) \left(3 p_c + e_c \right) + \mathcal{O}\left(r^4\right) \,,\\
    M(r) &= \frac{4\pi}{3} r^3 e_c + \mathcal{O}\left(r^5\right) \,,\\
    \nu(r) &= \nu_c + \frac{4 \pi r^2}{3} \left(3 p_c + e_c \right) + \mathcal{O}\left(r^4\right) \,,
\end{align}
\end{subequations}
where $e_c,p_c$ and $\nu_c$ are the values of the energy density, pressure and $\nu(r)$ at $r=0$.
The behavior near the surface of the star is more complicated and depends on how the pressure goes to zero.
In this paper, we assume that the EoS near the surface of the star is locally a polytrope with polytropic index $n >0$
\begin{align}
    p &= p_0 \left(\frac{e}{e_0}\right)^{1 + \frac{1}{n}} \,.
\end{align}
The procedure for obtaining the polytropic index for a numerically integrated TOV solution is described in~\cite{Lindblom-Detweiler-1983}.
Once we have solved for $n$, we obtain the following closed-form, perturbative solution near the stellar boundary
\begin{subequations}
\begin{align}
    &z\equiv R-r\,,\\
    &p(r) = 
    p_{\mathrm{R},0}z^{n+1}
    + \mathcal{O}\left[z^{n+2}\right] \,,\\
    &M(r) = M+ \mathcal{O}\left[z^{n+1}\right] \,,\\
    &\nu(r) = \log\left(1-\frac{2M}{R}\right)
    -
    \frac{2 M z}{R(R-2M)}  
    +
    \mathcal{O}\left[z^{n+2}\right] \,,\\
    &\lambda(r) = -\log\left(1-\frac{2M}{R}\right)
    +
    \frac{2 M z}{R(R-2M)}
    \nonumber\\
    &
    \hspace{2cm}
    +
    \mathcal{O}\left[z^{n+2}\right]\,,
\end{align}
\end{subequations}
where $M$ and $R$ are the mass and the radius of the star, and where
\begin{align}
    p_{\mathrm{R},0} &\equiv 
    \frac{p_0}{R^{n+1}}
    \,.
\end{align}
When integrating the TOV equations [Eq.~\eqref{eq:TOV-equations}], a detailed knowledge of behavior of the variables $p,M,\nu$ and $\lambda$ near the surface is not necessary.
We present them here because these expansions will be needed when solving the time dependent perturbation equations presented in the next section.

\subsubsection{Master equations for non-radial perturbations}
We now consider non-radial polar perturbations of the background metric in Eq.~\eqref{eq:background-metric} in Regge-Wheeler gauge.
We decompose the linearly perturbed metric components into spherical harmonics $Y_{\ell m}$, so that the line element reads 
\begin{align}\label{eq:metric-polar-pert}
    ds^2 &= -e^{\nu(r)} \left(1 - 2 H(r) e^{-i\omega t}  r^{\ell} Y_{\ell m}\right) dt^2
    \nonumber\\
    &- 
    2 i H_1(r) e^{-i\omega t}  r^{\ell} Y_{\ell m} dt dr
    \nonumber\\
    &+
    e^{\lambda(r)}
    \left(1 + 2 H_2(r) e^{-i\omega t}  r^{\ell} Y_{\ell m }\right) dr^2
    \nonumber\\
    &+
    r^2 \left(
    1 - K(r) e^{-i\omega t} r^{\ell} Y_{\ell m }
    \right)
    d \Omega^2
    \,.
\end{align}
Equation~\eqref{eq:S-perp-condition} and the assumption that $\mathbb{S}_{\mu\nu}$ vanishes in equilibrium implies that $\mathbb{S}_{t\mu} = 0$.
We next decompose the non-zero components of source tensor in terms of scalar $Y_{\ell m}$, vector $E_A^{\ell m }$, and tensor spherical harmonics $Z_{AB}^{\ell m }$ 
\begin{subequations}\label{eq:spherical-decomposition-source-term}
\begin{align}
    \mathbb{S}_{rr} &= S_{0}(r) e^{-i\omega t} r^{\ell-2} Y_{\ell m } \,,\\
    \mathbb{S}_{rA} &= S_{1}(r) e^{-i\omega t} r^{\ell-1}  E_A^{\ell m } \,,\\
    \mathbb{S}_{AB} &= S_{Z}(r) e^{-i\omega t} r^{\ell} Z_{AB}^{\ell m } \nonumber\\
    &+ S_{\Omega}(r) e^{-i\omega t} \Omega_{AB} r^{\ell} Y_{\ell m } \,,
\end{align}
\end{subequations}
where $(S_0(r),S_1(r),S_Z(r),S_\Omega(r))$ are functions to be determined.
The non-zero components of the Lagrangian displacement vector are parameterized as 
\begin{subequations}
\begin{align}\label{eq:lagrangian-displacement-vectors}
    \xi^r &= W(r) e^{-i\omega t} r^{\ell - 1} e^{-\lambda/2} Y_{\ell m} \,,\\
    \xi_A &= - V(r) e^{-i\omega t} r^{\ell } E_{A}^{\ell m} \,,
\end{align}
\end{subequations}
where $(W(r),V(r))$ are functions to be determined.
We can now determine the Eulerian perturbation of the four-velocity vector by using Eq.~\eqref{eq:lagrangian_variation_fluid_velocity}.
\begin{subequations}
\begin{align}\label{eq:pertubed-four-veolocity-vectors}
    \delta u^t 
    &=  
    e^{-\frac{\nu }{2}} r^{\ell } H e^{-i \omega t} Y_{\ell m }
    \,,\\
    \delta u^r 
    &= 
    -i \omega e^{-(\lambda+\nu)/2} r^{\ell-1 } W e^{-i \omega t} Y_{\ell m } 
    \,,\\
    \delta u_{A} 
    &= 
    i \omega e^{-\nu/2} r^{\ell} V e^{-i \omega t} E^{\ell m}_A 
    \,.
\end{align}
\end{subequations}
We use Eq.~\eqref{eq:lagrangian-displacement-vectors} to obtain $\Delta n/n$ using Eq.~\eqref{eq:Deltanbyn},
\begin{align}\label{eq:n0-definition}
    &\frac{1}{r^{\ell} e^{-i\omega t} Y_{\ell m}}
    \frac{\Delta n}{n } \equiv  n_0(r) =-H_2 + K
    -\frac{\ell (1+\ell) V}{r^2}
    \nonumber\\
    &
    \hspace{2cm}
    -
    \frac{e^{-\frac{\lambda }{2}} (1+\ell) W}{r^2}
    -
    \frac{e^{-\frac{\lambda }{2}} W'}{r}
    \,.
\end{align}
The Lagrangian perturbations of the energy density and the pressure are parameterized by using Eqs.~\eqref{eq:Deltaebyeplusp} and \eqref{eq:Deltapbyp}.
We note that the assumption in Eq.~\eqref{eq:S-perp-condition} is crucial in deriving Eq.~\eqref{eq:n0-definition}, as that expression follows from the particle current conservation equation $\nabla_{\mu}\left(n u^{\mu}\right)=0$.

At this point, the unknown functions that we need to determine from the field equations are $\left(H, W, V, H_1,H_2,K, n_0\right)$.
We will now use the field equations to obtain three master equations only involving the variables $\left(H, W, V \right)$.
We first find a relation for $H_2$ by using the $(\theta,\phi)$ component of the Einstein equation [see Eq.~\eqref{eq:Einstein-equations}], 
\begin{align}\label{eq:H2-elim}
    H_2 &= H - 8 \pi S_Z\,.
\end{align}
Next, we use Eq.~\eqref{eq:H2-elim} to get rid of $H_2$ in the $(r,\theta)$ and $(t,\theta)$ components of Eq.~\eqref{eq:Einstein-equations} to obtain expressions for $H_1$ and $H_1'$
\begin{subequations}\label{eq:H1-elim}
\begin{align}
    H_1 &= 
    -\frac{e^{\nu } \ell K}{r \omega }
    +\frac{16 e^{\nu } \pi  S_1}{r \omega }
    -\frac{2 e^{\nu } H'}{\omega }
    -\frac{e^{\nu } K'}{\omega }
    \nonumber\\
    &+\frac{8 e^{\nu } \pi  S_Z \left(2+r \nu '\right)}{r \omega }-\frac{2 e^{\nu } H \left(\ell+r \nu '\right)}{r \omega } 
    \,,
    \\
%---------------------------------------------------
    H_1' &=
    16 e^{\lambda } \pi  \omega  (e+p) V
    -\frac{8 e^{\nu } \pi  S_1 \left(2 \ell-r \lambda '+r \nu '\right)}{r^2 \omega }
    \nonumber\\
    &+\frac{e^{\nu } H' \left(2 \ell-r \lambda '+r \nu '\right)}{r \omega }
    +\frac{e^{\nu } K' \left(2 \ell-r \lambda '+r \nu '\right)}{2 r \omega }
    \nonumber \\
    &+
    K \left(-e^{\lambda } \omega +\frac{e^{\nu } \ell \left(2 \ell-r \lambda '+r \nu '\right)}{2 r^2 \omega }\right)
    \nonumber\\
    &-
    S_Z \left(16 e^{\lambda } \pi  \omega +\frac{4 e^{\nu } \pi  \left(2+r \nu '\right) \left(2 \ell-r \lambda '+r \nu '\right)}{r^2 \omega }\right)
    \nonumber\\
    &
    +H \left(2 e^{\lambda } \omega +\frac{e^{\nu } \left(\ell+r \nu '\right) \left(2 \ell-r \lambda '+r \nu '\right)}{r^2 \omega }\right)
    \,.
\end{align}
\end{subequations}
From the $\theta$ component of the fluid conservation equation [see Eq.~\eqref{eq:Fluid-conservation-equations}], we obtain an expression for the Eulerian perturbation of the pressure $\delta p$ (and hence the Lagrangian perturbation $\Delta p$ as well).
Using the relation Eq.~\eqref{eq:Deltapbyp}, can then rewrite $n_0$ as
\begin{align}\label{eq:n0-elim}
    n_0(r) &= 
    \frac{H (e+p)}{p \gamma}
    +\frac{\left(-2+\ell+\ell^2\right) S_Z}{2 r^2 p \gamma}
    -\frac{S_{\Omega}}{r^2 p \gamma}
    \nonumber\\
    &-\frac{e^{-\nu } \omega ^2 (e+p) V}{p \gamma}+\frac{e^{-\frac{\lambda }{2}} W p'}{r p \gamma}
    -\frac{e^{-\lambda } S_1'}{r p \gamma}
    \nonumber\\
    &
    -\frac{e^{-\lambda } S_1 \left(2+2 \ell-r \lambda '+r \nu '\right)}{2 r^2 p \gamma}
    \,.
\end{align}
We can eliminate $K$ and $K'$ by using the $(r,r)$ component and the $(t,r)$ component of the field equations [Eq.~\eqref{eq:Einstein-equations}] to obtain
\begin{subequations}\label{eq:K0-elim}
\begin{align}
    K &=
    \alpha_0
    H 
    +
    \alpha_1 W  
    +
    \alpha_2 V
    +
    \alpha_{3} H'
    \nonumber\\
    &
    +
    \alpha_4 S_0  
    +
    \alpha_5 S_1  
    +
    \alpha_6 S_Z  
    +
    \alpha_7 S_{\Omega}  
    +
    \alpha_{9} S_1'
    \,,\\
%---------------------------------------
    K' &=
    \beta_0 H  
    + 
    \beta_1 W  
    +
    \beta_2 V  
    +
    \beta_3 H'  
    \nonumber\\
    &
    + 
    \beta_4 S_0 
    +
    \beta_5 S_1 
    +
    \beta_6 S_Z  
    +
    \beta_7 S_{\Omega}  
    +
    \beta_{9} S_1'
    \,.
\end{align}
\end{subequations}
In the above, we used Eqs.~\eqref{eq:H2-elim},~\eqref{eq:H1-elim}, and ~\eqref{eq:Deltapbyp} to remove $H_1$, $H_2$, and $\delta p$. 
The functions $\alpha_i$ and $\beta_i$ are completely determined by the background metric of Eq.~\eqref{eq:background-metric} and are provided in the  supplementary \texttt{Mathematica} file available at~\cite{github-code}.

Equations~\eqref{eq:H2-elim}, \eqref{eq:H1-elim}, \eqref{eq:n0-elim} and \eqref{eq:K0-elim} allow us to eliminate the variables $\left(H_1,H_2,K, n_0\right)$ and their derivatives from the other field equations.
The master equations are obtained by simplifying the $r$ component of the fluid conservation equation and the $(t,t)$ component of the Einstein field equation.
Schematically, they take the following form
\begin{subequations}\label{eq:master-equation-fluid-perturbations}
\begin{align}
%----------------------------------------------------
W' &=  H \alpha_{W,0}
+W \alpha_{W,1}
+V \alpha_{W,2}
+\alpha_{W,3} H'
\nonumber\\
&
+ \alpha_{W,4} S_0
+ \alpha_{W,5} S_1
+ \alpha_{W,6} S_Z
\nonumber\\
&+ \alpha_{W,7} S_{\Omega}
+ \alpha_{W,9} S_1'
\,,\\
%----------------------------------------------------
%----------------------------------------------------
V' &=H \alpha_{V,0}
+W \alpha_{V,1}
+V \alpha_{V,2}
+\alpha_{V,3} H'
\nonumber \\
&
+ \alpha_{V,4} S_0
+ \alpha_{V,5} S_1
+ \alpha_{V,6} S_Z
+ \alpha_{V,7} S_{\Omega}
+ \alpha_{V,8} S_0'
\nonumber\\
&+ \alpha_{V,9} S_1'
+ \alpha_{V,10} S_Z'
+ \alpha_{V,11} S_{\Omega}'
+ \alpha_{V,12} S_{1}''
\,,\\
%----------------------------------------------------
H'' &= H \alpha_{H,0}
+W \alpha_{H,1}
+V \alpha_{H,2}
+\alpha_{H,3} H'
\nonumber\\
&
+ \alpha_{H,4} S_0
+ \alpha_{H,5} S_1
+ \alpha_{H,6} S_Z
+ \alpha_{H,7} S_{\Omega}
+ \alpha_{H,8} S_0'
\nonumber\\
&+ \alpha_{H,9} S_1'
+ \alpha_{H,10} S_Z'
+ \alpha_{H,11} S_{\Omega}'
+ \alpha_{H,12} S_1''
\,.
\end{align}
\end{subequations}
The functions $\alpha_{H,i},\alpha_{V,i}$ and $\alpha_{W,i}$ are functions $\alpha_i,\beta_i,\beta_i'$, and the other background variables, and they are provided in the supplementary \texttt{Mathematica} file~\cite{github-code}.
For ease of notation, we write the master equation in vector form by defining 
\begin{subequations}
\begin{align}
    \Vec{Y} &\equiv \left(H,W,V,H'\right) \,,\\
    \Vec{S} &\equiv \left(S_0, S_1, S_Z, S_{\Omega}, S_0', S_1', S_{Z}', S_{\Omega}', S_1'' \right) \,,
\end{align}
\end{subequations}
and the matrices
\begin{widetext}
\begin{align}\label{eq:matrix-A-and-B}
    \boldsymbol{A} &\equiv \begin{pmatrix}
    0 & 0 & 0 & 1 \\
    \alpha_{W,0} & \alpha_{W,1} & \alpha_{W,2} & \alpha_{W,3} \\
    \alpha_{V,0} & \alpha_{V,1} & \alpha_{V,2} & \alpha_{V,3}\\
    \alpha_{H,0} & \alpha_{H,1} & \alpha_{H,2} & \alpha_{H,3}    
    \end{pmatrix} \,,
    \qquad
    \boldsymbol{B} \equiv \begin{pmatrix}
    0 & 0 & 0 & 0 &0 &0 &0 & 0 &0 \\
    \alpha_{W,4} & \alpha_{W,5} & \alpha_{W,6} & \alpha_{W,7}
    &0 & \alpha_{W,9} & 0 & 0
    & 0 \\
    \alpha_{V,4} & \alpha_{V,5} & \alpha_{V,6} & \alpha_{V,7}
    &\alpha_{V,8} & \alpha_{V,9} & \alpha_{V,10} & \alpha_{V,11}
    & \alpha_{V,12}\\
    \alpha_{H,4} & \alpha_{H,5} & \alpha_{H,6} & \alpha_{H,7}
    &\alpha_{H,8} & \alpha_{H,9} & \alpha_{H,10} & \alpha_{H,11} 
    & \alpha_{H,12}
    \end{pmatrix}
    \,.
\end{align}
\end{widetext}
The master equation can now be written as a system of four first order differential equations
\begin{align}\label{eq:master-equation-matrix}
    \Vec{Y}' &= \mathbf{A} \Vec{Y} + \mathbf{B} \Vec{S} \,,
\end{align}
where, the matrices $\mathbf{A}$ and $\mathbf{B}$ depend only on the background TOV solution.
When combined with suitable boundary conditions, these equations provide a complete description of the polar sector of the linearly perturbed field equations.

%-----------------------------------------------
\subsection{Solution in the outer body zone in a re-summed small-frequency expansion}\label{sec:tidal-bcs}
At the surface of the star, the gravitational field $H$ must be once differentiable. 
To match the interior and exterior solutions, we first find the general solution to the vacuum perturbed Einstein equations.
In vacuum the background metric reduces to
\begin{align}
    e^{\nu} = e^{-\lambda} = 
    \left(1-\frac{2M}{r}\right)
    ,
\end{align}
where $M$ is the mass of the star.
In the region exterior to the star, the master equations [Eq.~\eqref{eq:master-equation-matrix}] reduce to a single, second-order differential equation for $H$.
One can relate this equation to the Zerilli-Moncrief equation and the Regge-Wheeler equation as we show in Appendix~\ref{appendix:H-to-ZM-RW}.
To impose the tidal boundary conditions, we expand the potential $H$ as 
\begin{align}\label{eq:H-odd-even}
    H(r;\omega) = H_{+}(r;\omega) + H_{-} (r;\omega)
\end{align}
where the $\pm$ parts of the potential are even and odd functions of $\omega$,
\begin{align}
    H_{\pm}(r;\omega) = \pm H_{\pm}(r;-\omega) \,.
\end{align}
We expand the even and odd functions in a small frequency expansion, 
\begin{align}
    \label{eq:definition-epsilon}
    \varepsilon \equiv M \omega \ll 1
    ,
\end{align}
namely
\begin{subequations}\label{eq:H-odd-even-small-frequency}
\begin{align}
    H_{+} &= \sum_{k=0}^{\infty} \varepsilon^{2k} H_{+}^{(2k)}(r) \,,\\
    H_{-} &= \sum_{k=0}^{\infty} \varepsilon^{2k+1}H_{-}^{(2k)}(r)   \,.
\end{align}
\end{subequations}
We note that the vacuum master equations [Eq.~\eqref{eq:master-equation-matrix}] are even in $\omega$. 
Consequently, at each order in the frequency expansion, $H_{+}^{(2k)}$ and $H_{-}^{(2k)}$ take the same functional form [see Eq.~\eqref{eq:H-pm-sols}].

In our approach, we truncate the expansion at ${\cal{O}}(\varepsilon^2)$; we show how to extend the approach to higher orders in $\varepsilon$ in Appendix~\ref{appendix:resum}.
We note that to leading order ($\varepsilon^0$ and $\varepsilon^1$), the master equation for $H$ reduces to that for calculating the exterior, static tidal response of a star \cite{Hinderer:2007mb,Damour:2009vw,Binnington:2009bb}. 
The solutions at this order of truncation are given by 
\begin{subequations}\label{eq:H-pm-sols}
\begin{align}
   H_{\pm}^{(0)} &=  \frac{a_{P,\pm}^{(0)} M^{\ell}}{ r^{\ell}} \hat{P}^2_{\ell} \left(\frac{r}{M}-1\right) 
   +
   \frac{a_{Q,\pm}^{(0)} M^{\ell}} {r^{\ell}} \hat{Q}^2_{\ell} \left(\frac{r}{M}-1\right) 
   \,,\\
   H_{\pm}^{(2)} &= \frac{a_{P,\pm}^{(2)} M^{\ell}}{ r^{\ell}} \hat{P}^2_{\ell} \left(\frac{r}{M}-1\right) 
   +
   \frac{a_{Q,\pm}^{(2)} M^{\ell}} {r^{\ell}} \hat{Q}^2_{\ell} \left(\frac{r}{M}-1\right) 
   \nonumber\\
   &+
   \frac{a_{P,\pm}^{(0)} M^{\ell}}{r^{\ell}} \mathbb{H}_{P,\ell}^{(2)} + \frac{a_{Q,\pm}^{(0)} M^{\ell}}{r^{\ell}} \mathbb{H}_{Q,\ell}^{(2)} \,.
\end{align}
\end{subequations}
The functions $\hat{P}^2_{\ell}$ and $\hat{Q}^2_{\ell}$ are Legendre functions of the first and second kind\footnote{Note that the Legendre function of the second kind $\hat{Q}^m_{\ell}(x)$ is defined on the domain $x>1$ unlike the normal convention where it is defined on the interval $-1<x<1$. Practically, this amounts to replacing terms proportional to $\log(1-x)$ by $\log(x-1)$.}, with the normalization
\begin{subequations}
\begin{align}
    \hat{P}^2_{\ell}(x) &\xrightarrow{x\to \infty} x^{\ell} \,,\\
    \hat{Q}^2_{\ell}(x) &\xrightarrow{x\to \infty} x^{-\ell-1} \,.
\end{align}
\end{subequations}
The particular solutions $\mathbb{H}_{P,\ell}^{(2)}$ and $\mathbb{H}_{Q,\ell}^{(2)}$ are the same for both the even and odd parts in $\omega$ of the solution.
Although we cannot determine the general solution to $\mathbb{H}_{P/Q,\ell}^{(2)}$ for all $\ell$, finding the functional form for a given $\ell$ is straightforward~\cite{Poisson:2020vap}.
We list the functional form for $\ell = (2,3)$ in the supplemental \texttt{Mathematica} notebook~\cite{github-code}.

The particular solutions $\mathbb{H}_{P/Q,\ell}^{(2)}$ are not uniquely defined, as one can add a homogeneous solution and the solution would still be a solution to the differential equation.
To specify a unique solution, we need to impose additional constraints on the behavior of the solution.
To impose such a constraint, we change coordinates to a Lorentzian harmonic coordinate system, $\left(t,\bar{x}^{1},\bar{x}^{2},\bar{x}^{3}\right)$
\begin{align}\label{eq:harmonic-coord-body}
    \bar{x}^{a} &= \bar{r} \Omega^{a}   \,,\hspace{1cm} \bar{r} = r - M\ \,,\\
    \Omega^{a} &= \left[\sin(\theta)\cos(\phi), \sin(\theta) \sin(\phi),\cos(\theta) \right]\,.
\end{align}
This coordinate system is harmonic in the exterior background Schwarzschild metric.
The Schwarzschild metric admits a PN expansion compatible with the PN metric constructed in harmonic gauge in this coordinate system.
This fact will be used later when performing an asymptotic expansion of the metric given in Eq.~\eqref{eq:background-metric}.

To completely specify the functional form of the particular solutions, we choose the following normalization condition for $\bar{r} \gg M$:
\textit{we demand that the asymptotic expansion when $\bar{r}\gg M$ of the function $\mathbb{H}_{P/Q,\ell}^{(2)}(r)(1-2M/r)$ not contain any term proportional to $\bar{r}^{\ell}$ or $\bar{r}^{-\ell-1}$}. If such terms are present, then we add/subtract homogeneous solutions proportional to $\hat{P}_{\ell}^2(r)$ or $\hat{Q}_{\ell}^2(r)$ so that these terms are cancelled.
Additionally, the appearances of $\log(\bar{r})$ in the asymptotic expansion is normalized as  $\log(\bar{r}/M)$.
We denote the normalized solution by $\hat{N} \mathbb{H}_{P/Q,\ell}^{(2)}(r)$.

To see the normalization process more concretely, let us look at the specific case of $\ell=2$.
Consider the non-normalized particular solution used in~\cite{Poisson:2020vap},
\begin{widetext}
\begin{subequations}
\begin{align}
    \mathbb{H}_{P,2}^{(2)}(r) &= 
    \frac{107 u^6-860 u^5+1538 u^4-2074 u^3+3989 u^2-3430 u+70}{630 (u-1)^2 u}-4 u \text{Li}_2(u)-2 u \log ^2(u) \nonumber\\
    &+\left(\frac{u^3}{3}-\frac{8 u^2}{3}+\frac{214 u}{105}-\frac{1}{3 u}+\frac{8}{3}\right) \log (1-u)-\frac{\left(u^4-8 u^3+8 u-1\right) \log (u)}{3 u}
    \,,\\
    \mathbb{H}_{Q,2}^{(2)}(r)
    &=
    \frac{5 \left(u^4-8 u^3+8 u-1\right) \log ^2(u)}{32 u}+\frac{107 \left(u^4-8 u^3+8 u-1\right) \log (1-u)}{672 u} +\frac{15 u \zeta (3)}{2}-\frac{15 u \text{Li}_3(u)}{2}
    \nonumber\\
    &
    -\frac{857 u^5-7095 u^4-4 \left(642 \pi ^2-3017\right) u^3+8 \left(321 \pi ^2-1729\right) u^2+8655 u+3307}{4032 (u-1) u} +\frac{5}{8} u \log ^3(u)
    \nonumber\\
    &+
    \frac{1}{112} \left(-35 u^3+280 u^2-428 u+\frac{35}{u}-280\right) \text{Li}_2(u) + \frac{15}{4} u \text{Li}_2(u) \log (u)
    \nonumber\\
    &+\frac{1}{112} \left(-35 u^3+280 u^2-214 u+\frac{35}{u}-280\right) \log (1-u) \log (u)
    \nonumber\\
    &-\frac{\left(107 u^6-860 u^5+1538 u^4-2074 u^3+3989 u^2-3430 u+70\right) \log (u)}{672 (u-1)^2 u}
\end{align}
\end{subequations}
\end{widetext}
where $u=1-2M/r$ and ${\rm{Li}}_2(r)$ is the dilog function.
The asymptotic expansion of $\mathbb{H}_{P/Q,2}^{(2)}(r)(1-2M/r)$ for $\bar{r} \gg R > M$ are given by 
\allowdisplaybreaks[4]
\begin{subequations}\label{eq:asymptotic-expansion-Eric}
\begin{align}
\label{eq:Hp-Eric-asymptotic}
    &\mathbb{H}_{P,2}^{(2)}(r)(1-2M/r)
    =
    -\frac{11 \bar{r}^4}{42 M^4} -\frac{107 \bar{r}^3}{63 M^3}
    - \frac{214\bar{r}^2}{105} \log\left(\frac{\bar{r}}{M}\right)
    \nonumber\\
    &+\frac{a_{2,2}\bar{r}^2}{M^2}
    +
    \frac{\bar{r} \left(428 \log \left(\frac{\bar{r}}{M}\right)+140 \pi ^2+1105-428 \log (2)\right)}{105 M}
    \nonumber\\
    &+
    \frac{1}{630} \left(-1284 \log \left(\frac{\bar{r}}{2 M}\right)-420 \pi ^2-4789\right)
    \nonumber\\
    &+
    \frac{1507 M^2}{630 \bar{r}^2}+\frac{494 M}{315 \bar{r}}
    -\frac{64 M^3}{15 \bar{r}^3} \log\left(\frac{M}{\bar{r}}\right)
    \nonumber\\
    &
    +\frac{a_{1,2}M^3}{\bar{r}^3} + 
    \mathcal{O} \left( \bar{r}^{-4}\right)
    \,, \\
\label{eq:Hq-Eric-asymptotic}
    &\mathbb{H}_{Q,2}^{(2)}(r)(1-2M/r)
    =
    \frac{a_{4,2}\bar{r}^2}{M^2}
    -\frac{M}{2 \bar{r}}
    -\frac{M^2}{6 \bar{r}^2} 
    \nonumber\\
    &
    + \frac{214 M^3 \log \left(\frac{\bar{r}}{M}\right)}{105 \bar{r}^3}
    +
    \frac{a_{1,2}M^3}{\bar{r}^3}
    + \mathcal{O}\left( \bar{r}^{-4} \right)\,,
\end{align}
\end{subequations}
where
\begin{subequations}\label{eq:ai-vals-PP-2}
\begin{align}
    a_{1,2} &\equiv -\frac{717}{175}-\frac{1}{15} 64 \log (2)\,,\\
    a_{2,2} &\equiv \frac{214 \log (2)}{105}-\frac{2}{21} \left(37+7 \pi ^2\right)\,,\\
    a_{3,2} &\equiv \frac{2 \pi ^2}{3}+\frac{28076}{3675}-\frac{1}{105} 214 \log (2) \,,\\
    a_{4,2} &\equiv 0\,.
\end{align}
\end{subequations}
The asymptotic expansions provided in Eq.~\eqref{eq:asymptotic-expansion-Eric} contains terms proportional to $\bar{r}^{2}$ and $\bar{r}^{-3}$.
To cancel these terms, we add ``counter-terms'' proportional to the homogeneous solutions, namely $a_{1,2} (1-2M/r) \hat{Q}_{\ell}^{2} + a_{2,2} (1-2M/r)\hat{P}_{\ell}^{2} $. 
The subtraction of these counter-terms removes the terms proportional to $\bar{r}^{-3}$ and $\bar{r}^{2}$ in the asymptotic expansion of Eq.~\eqref{eq:Hp-Eric-asymptotic}. Similarly, we subtract the counter-term $a_{3,2} (1-2M/r) \hat{Q}_{\ell}^{2} + a_{4,2} (1-2M/r) \hat{P}_{\ell}^{2}$ to remove the terms proportional to $\bar{r}^{-3}$ and $\bar{r}^{2}$ in the asymptotic expansion of Eq.~\eqref{eq:Hq-Eric-asymptotic}. 
The normalized solutions are then
\begin{subequations}
\begin{align}
    \hat{N}\mathbb{H}_{P,2}^{(2)}(r)
    &\equiv 
    \mathbb{H}_{P,2}^{(2)}(r) - a_{1,2} \hat{Q}_{\ell}^{2} - a_{2,2} \hat{P}_{\ell}^{2} \,,\\
    \hat{N}\mathbb{H}_{Q,2}^{(2)}(r)
    &\equiv 
    \mathbb{H}_{Q,2}^{(2)}(r) - a_{1,2} \hat{Q}_{\ell}^{2} - a_{2,2} \hat{P}_{\ell}^{2}
    \,.
\end{align}
\end{subequations}
The asymptotic expansion of the normalized solutions for $\bar{r} \gg R > M$ are given by 
\begin{subequations}\label{eq:asympototic-expansion}
\begin{align}
%-------------------------------------
    &\hat{N}\mathbb{H}_{P,2}^{(2)}(r)\left(1- \frac{2M}{r}\right)
    =
    -\frac{11 \bar{r}^4}{42 M^4} -\frac{107 \bar{r}^3}{63 M^3}
    \nonumber\\
    &- \frac{214\bar{r}^2}{105} \log\left(\frac{\bar{r}}{M}\right)
    \nonumber\\
    &
    +
    \frac{\bar{r} \left(428 \log \left(\frac{\bar{r}}{M}\right)+140 \pi ^2+1105-428 \log (2)\right)}{105 M}
    \nonumber\\
    &+
    \frac{1}{630} \left(-1284 \log \left(\frac{\bar{r}}{2 M}\right)-420 \pi ^2-4789\right)
    \nonumber\\
    &+
    \frac{1507 M^2}{630 \bar{r}^2}+\frac{494 M}{315 \bar{r}}
    -\frac{64 M^3}{15 \bar{r}^3} \log\left(\frac{M}{\bar{r}}\right)
    +
    \mathcal{O} \left( \bar{r}^{-4}\right)
    \,, \\
%-------------------------------------
    &\hat{N}\mathbb{H}_{Q,2}^{(2)}(r)\left(1- \frac{2M}{r}\right)
    =
    -\frac{M}{2 \bar{r}}
    -\frac{M^2}{6 \bar{r}^2} 
    \nonumber\\
    &
    + \frac{214 M^3 \log \left(\frac{\bar{r}}{M}\right)}{105 \bar{r}^3}
    + \mathcal{O}\left( \bar{r}^{-4} \right)
    \,,
\end{align}
\end{subequations}
which clearly does not contain any terms proportional to $\bar{r}^2$ or $\bar{r}^{-3}$.
The normalization procedure exemplified above will be important when matching the metric obtained in the body zone to that in the PN zone inside the buffer zone.
A \texttt{Mathematica} notebook for generating the normalized solutions $\hat{N} \mathbb{H}_{P/Q,\ell}$ is available at~\cite{github-code}.

After the normalization is carried out, the solution for $H(r)$ is given by 
\begin{widetext}
\begin{align}\label{eq:H-normalized}
    H(r) &= \frac{M^{\ell}}{r^{\ell}}\left(a_{P,+}^{(0)} 
    + 
    a_{P,-}^{(0)} \varepsilon 
    +
    a_{P,+}^{(2)}\varepsilon^2 
    + 
    a_{P,-}^{(2)} \varepsilon^3 \right )
    \frac{}{}
    \hat{P}_{\ell}^{2}\left(\frac{r}{M}-1\right)
    +
    \frac{M^{\ell}}{r^{\ell}}\left(a_{Q,+}^{(0)} 
    + 
    a_{Q,-}^{(0)} \varepsilon 
    +
    a_{Q,+}^{(2)}\varepsilon^2 
    + 
    a_{Q,-}^{(2)} \varepsilon^3 \right )
    \hat{Q}_{\ell}^{2}\left(\frac{r}{M}-1\right)
    \nonumber\\
    &+
    \frac{M^{\ell} \varepsilon^2}{r^{\ell}} \left(a_{P,+}^{(0)} +a_{P,-}^{(0)} \varepsilon  \right) \hat{N} \mathbb{H}_{P,\ell}^{(2)}(r)
    +
    \frac{M^{\ell} \varepsilon^2}{r^{\ell}} \left(a_{Q,+}^{(0)} +a_{Q,-}^{(0)} \varepsilon \right) \hat{N} \mathbb{H}_{Q,\ell}^{(2)}(r)
    +
    \mathcal{O}\left(\varepsilon^4 \right)
    \,.
\end{align}
\end{widetext}
The above solution completes the description of the solution in the outer body zone.
The constants $\left(a_{P,\pm}^{(0)},a_{Q,\pm}^{(0)},a_{P,\pm}^{(2)},a_{Q,\pm}^{(2)}\right)$ are, as of yet, undetermined.
We relate these constants to the tidal potential and the multipole moment of the object by asymptotically matching the outer body zone solution to the solution in the PN zone. Before proceeding with this matching calculation, let us point out that the normalization described above was \textit{not} made in~\cite{Poisson:2020vap,Pitre:2023xsr}, as we have shown for the case of $\ell=2$ in Eq.~\eqref{eq:Hp-Eric-asymptotic} and \eqref{eq:Hq-Eric-asymptotic}.
We compare our normalization choice with that of~\cite{Poisson:2020vap,Pitre:2023xsr} in Appendix~\ref{appendix:compare-Eric-to-us}.

\subsection{Matching the outer body zone solution and the solution in the post-Newtonian zone}\label{sec:Matching-PN}
For the time-independent problem, the term proportional to $\hat{P}^2_{\ell}$ (the asymptotically ``growing'' potential) is usually interpreted as the imposed tidal moment, while the term proportional to $\hat{Q}^2_{\ell}$ (the asymptotically ``decaying'' potential) is the induced multipole moment.
In the time-dependent problem we are interested in, we can make the same separation if the object under consideration is embedded in an ambient PN environment~\cite{Poisson:2020vap}. 
As we described at the beginning of this section, the calculation of a tidally interacting binary involves solving the Einstein equations in three distinct zones on a fixed time slice: the inner body zone, the outer body zone and the PN zone.
In the body zone the object is in its rest frame and the dynamical description of the field equations are obtained by the master equation [Eq.~\eqref{eq:master-equation-matrix}] in the inner zone, while the solution in the outer body zone is given by Eq.~\eqref{eq:H-normalized}.
In the ambient PN zone, the object is viewed as a skeletonized compact object, which is part of a global spacetime where it is tidally interacting with an external environment.

To describe the solution in the PN zone, we set up harmonic coordinates, $\left(T,X^{a}\right)$, such that the metric to leading, Newtonian, order is given by
\begin{align}
    g_{TT,\mathrm{PN}} &= -1 + 2 U + {\cal{O}}(1/c^4) \,,\\
    g_{Ta,\mathrm{PN}} &= {\cal{O}}(1/c^3)  \,,\\
    g_{ab,\mathrm{PN}} &= \delta_{ab} + {\cal{O}}(1/c^2) \,.
\end{align}
Here, the function $U (T,X^{a})$ is the Newtonian gravitational potential, and the ${\cal{O}}(1/c^{2,3,4})$ are uncontrolled remainders at 1PN order. For two point-particles, the Newtonian potential to leading order takes the simple form $U_{\rm{pp}} = M_1/r_1 + M_2/r_2$, where $r_{1,2}$ are field point distances from the two particles. However, for extended bodies, this potential takes on a more complicated form that depends on the perturbing tidal interaction between the objects.  

The PN metric can be asymptotically expanded in the buffer zone. Doing so, leads to 
\begin{align}
   U(T,X^{a}) &= \frac{M}{s} -\frac{4 \pi \hat{I}_{\ell m} (\omega_{\mathrm{PN}})Y_{\ell m} e^{-i\omega_{\mathrm{PN}} T}}{(2\ell + 1)s^{\ell + 1}}
   \nonumber \\
   &
    -\frac{4 \pi d_{\ell m}(\omega_{\rm{PN}}) Y_{\ell m}e^{-i\omega_{\rm{PN}} t} }{2 \ell + 1} {s}^{\ell} + {\cal{O}}(1/s^{\ell+2},s^{\ell+1})\,, 
\end{align}
where $M$ is the mass of the object, $s = \left|X^{a}-Z^{a}(T) \right|$ is the field point distance from the compact object in the PN zone, with $Z^{a}(T)$ is the object's worldline,
$\hat{I}_{\ell m }$ is the multipole moment of the object, $d_{\ell m}(\omega)$ is the tidal moment of the object, and $\omega_{\mathrm{PN}}$ is the Fourier-domain frequency in the PN zone. As one can see, the asymptotically expanded PN metric in the buffer zone is a bivariate expansion in both $s \ll d$ and $s \gg M$.

To obtain a complete description of the Newtonian potential, we need to find the multipole moments $\hat{I}_{\ell m}(\omega_{\mathrm{PN}})$, which are obtained through asymptotic matching. The outer body zone metric, however, is not in the same coordinate system as the PN metric. The coordinate transformation from $(T,X^{a})$ to the harmonic coordinates $(t,\bar{x}^{a})$ in the body zone has already been worked out in Sec. VI C of~\cite{Poisson:2020vap} and Sec. 8.3 of~\cite{Poisson-Will}, and thus, we will not present it again here. 
The PN metric, asymptotically expanded in the buffer zone, and in the new coordinate system becomes
\begin{subequations}
\begin{align}
    \label{eq:gtt-PN-BZ}
    g_{tt,\mathrm{PN}} &= -1 + 2\bar{U} +  {\cal{O}}(1/c^4)\,,\\
    g_{t\bar{a},\mathrm{PN}} &=  {\cal{O}}(1/c^3) \,,\\
    g_{\bar{a}\bar{b},\mathrm{PN}} &= \delta_{\bar{a}\bar{b}} +  {\cal{O}}(1/c^2)\,.
\end{align}
\end{subequations}
where
\begin{align}\label{eq:Newt-potential}
   \bar{U}(t,\bar{x}^{a}) &= \frac{M}{\bar{r}}  - \frac{4 \pi \hat{I}_{\ell m}(\omega) Y_{\ell m}e^{-i\omega t}}{(2\ell + 1)\bar{r}^{\ell + 1}}  
   \nonumber\\
   &-\frac{4 \pi d_{\ell m}(\omega) Y_{\ell m}e^{-i\omega t} }{2 \ell + 1} \bar{r}^{\ell} + {\cal{O}}(1/\bar{r}^{\ell +2},\bar{r}^{\ell +1})\,,
\end{align}

With the discussion of the asymptotically-expanded PN metric complete, let us redirect our attention to the outer zone metric. This metric was presented in Eq.~\eqref{eq:metric-polar-pert} in Schwarzschild-like coordinates, so we must first transform it to harmonic coordinates through Eq.~\eqref{eq:harmonic-coord-body}. After doing so, and after asymptotically expanding this metric in $M/\bar{r} \ll 1$, one finds
\begin{align}
    &g_{tt}= -1 + \frac{2M}{\bar{r}} 
    \nonumber\\
    &+ 2
    \left(a_{P,+}^{(0)} 
    + 
    a_{P,-}^{(0)} \varepsilon 
    +
    a_{P,+}^{(2)}\varepsilon^2 
    + 
    a_{P,-}^{(2)} \varepsilon^3 \right) Y_{\ell m} \bar{r}^{\ell} e^{-i\omega t}
    \nonumber\\
    &+
    2
    \left(a_{Q,+}^{(0)} 
    + 
    a_{Q,-}^{(0)} \varepsilon 
    + 
    a_{Q,+}^{(2)}\varepsilon^2 
    + 
    a_{Q,-}^{(2)} \varepsilon^3  \right)\frac{M^{2\ell+1} Y_{\ell m} e^{-i\omega t}}{\bar{r}^{\ell+1}}
    \nonumber\\
    &+
    \frac{2(-1)^{\ell} M^{\ell+1}}{\bar{r}^{\ell+1}}
    \nonumber\\
    &+
    {\cal{O}}\left[1/c^4,1/\bar{r}^{\ell+2},\bar{r}^{\ell+1}\right]
    +
    \mathcal{O}\left( \log(\bar{r}/M)\right)\,,
\end{align}
while the other components are trivial. In the expansion above, we have \textit{only} retained those terms that scale either as $1/\bar{r}$, $\bar{r}^{\ell}$ or $1/\bar{r}^{\ell+1}$, since these are the only ones we will need for the asymptotic matching. Note that the coefficients of the $\bar{r}^{\ell}$ and $\bar{r}^{-\ell-1}$ terms are exact to $\mathcal{O}\left(\varepsilon^4 \right)$, because of the normalization condition, i.e.~no PN truncation has been made when writing these coefficients.
We note that there are also terms that scale with $\log(\bar{r}/{M})$.
These terms arise because of the phase shift from the tortoise coordinate and the appearance of logarithmic terms in the near zone expansion of gravitational potential~\cite{Sasaki_2003}. 
Though important at extremely high-frequencies, these terms will be ignored as they do not impact the tidal response in the regime we are interested in.

We can now match this asymptotically-expanded outer body zone metric to the asymptotically-expanded PN metric inside the buffer zone. Matching will therefore determine the values of the constants $a_{P,i}$ and $a_{Q,i}$ in the outer zone metric in terms of the $\hat{I}_{\ell m}$ and $d_{\ell m}$ of the PN metric. 
We first observe that the term ${2(-1)^{\ell} M^{\ell+1}}/{\bar{r}^{\ell+1}}$ is due to the gravitational potential of the body and is of a very high PN order, and we therefore discard it in the matching.
The matching procedure then shows that
\begin{subequations}\label{eq:apq-defs}
\begin{align}
    a_{P,+}^{(0)} 
    + 
    a_{P,-}^{(0)} \varepsilon 
    +
    a_{P,+}^{(2)}\varepsilon^2 
    + 
    a_{P,-}^{(2)} \varepsilon^3  
    &=
    \frac{4 \pi d_{\ell m}(\omega)}{2 \ell +1} 
    \,,\\
    a_{Q,+}^{(0)} 
    + 
    a_{Q,-}^{(0)} \varepsilon 
    + 
    a_{Q,+}^{(2)}\varepsilon^2 
    + 
    a_{Q,-}^{(2)} \varepsilon^3 
    &=
    % \nonumber\\
    \frac{4 \pi \hat{I}_{\ell m }(\omega)}{(2\ell +1)M^{2\ell+1}}  \,.
\end{align}
\end{subequations}
We have now asymptotically matched the metric of the isolated body to the PN metric. 

Let us now go back to the metric potential and carry out a resummation of coefficients of the particular solution, which is one of the new results of this paper. 
We rewrite Eq.~\eqref{eq:H-odd-even} using Eqs.~\eqref{eq:H-odd-even-small-frequency} and \eqref{eq:H-pm-sols} to find
\begin{align}
    \label{eq:H-intermediate-step}
    H  &= \frac{4 \pi d_{\ell m}(\omega) M^{\ell} }{2 \ell +1}
    \hat{P}^2_{\ell} \left(\frac{r}{M}-1\right) r^{-\ell}
    \nonumber\\
    &+
    \frac{4 \pi}{2\ell +1} \frac{\hat{I}_{\ell m }(\omega)}{M^{\ell+1}}
    \hat{Q}^2_{\ell}\left(\frac{r}{M}-1\right) r^{-\ell}
    \nonumber\\
    &+
    \varepsilon^2\left(a_{P,+}^{(0)} + a_{P,-}^{(0)} \varepsilon\right)\hat{N}\mathbb{H}_{P,\ell}^{(2)}\frac{M^{\ell}}{r^{\ell}}
    \nonumber\\
    &+
    \varepsilon^2\left(a_{Q,+}^{(0)} + a_{Q,-}^{(0)} \varepsilon\right)\hat{N}\mathbb{H}_{Q,\ell}^{(2)}\frac{M^{\ell}}{r^{\ell}}
    + \mathcal{O}\left(\varepsilon^4\right)
    \,.
\end{align}
Next, we note that using Eq.~\eqref{eq:apq-defs}, we can rewrite 
\begin{align}
    \varepsilon^2\left(a_{P,+}^{(0)} + a_{P,-}^{(0)} \varepsilon\right) 
    &= 
    \varepsilon^2 \frac{4\pi d_{\ell m}}{2\ell+1} 
    + 
    \mathcal{O}\left(\varepsilon^4\right)
    , \\
    \varepsilon^2\left(a_{Q,+}^{(0)} + a_{Q,-}^{(0)} \varepsilon\right)  
    &=
    \varepsilon^2 \frac{4\pi \hat{I}_{\ell m}}{(2\ell +1)M^{2\ell+1}}
    + 
    \mathcal{O}\left(\varepsilon^4\right)
    .
\end{align}
This is the critical step in which one effectively ``re-summs'' the $\varepsilon$ expansion for the coefficients to the particular solution. 
We can then write the external gravitational potential, Eq.~\eqref{eq:H-intermediate-step}, as
\begin{align}\label{eq:H0-ansatz-v1}
    H  &= \frac{4 \pi d_{\ell m}(\omega) M^{\ell} }{2 \ell +1}
    \hat{P}^2_{\ell} \left(\frac{r}{M}-1\right) r^{-\ell}
    \nonumber\\
    &+
    \frac{4 \pi}{2\ell +1} \frac{\hat{I}_{\ell m }(\omega)}{M^{\ell+1}}
    \hat{Q}^2_{\ell}\left(\frac{r}{M}-1\right) r^{-\ell}
    \nonumber\\
    &+
    \varepsilon^2\left(\frac{4 \pi d_{\ell m}(\omega) M^{\ell}}{2 \ell + 1}\right)\hat{N}\mathbb{H}_{P,\ell}^{(2)}r^{-\ell}
    \nonumber\\
    &+
    \varepsilon^2\left(\frac{4 \pi}{2\ell +1} \frac{\hat{I}_{\ell m }(\omega)}{M^{\ell+1}}\right)\hat{N}\mathbb{H}_{Q,\ell}^{(2)}r^{-\ell}
    + 
    \mathcal{O}\left(\varepsilon^4\right)
    \,.
\end{align}

With this re-summation, we have rewritten the external $H$ in terms of the tidal moment $d_{\ell m}$ and the induced multipole moment $\hat{I}_{\ell m}$.
In Appendix~\ref{appendix:resum} we show that we can perform this re-summation procedure at each order in the small frequency perturbation and the re-summation appears naturally in perturbation theory.
Let us now recall that the tidal response $\hat{K}_{\ell m}$ is defined by Eq.~\eqref{eq:tidal-response-definition}. 
Using this, we can rewrite Eq.~\eqref{eq:H0-ansatz-v1} in terms of the response and the tidal moment coefficient $d_{\ell m}$
\begin{align}\label{eq:H0-boundary-v1}
    H 
    &= 
    \frac{4 \pi d_{\ell m} (\varepsilon) }{2 \ell +1} \frac{M^{\ell}}{r^{\ell}}
    \bigg[ 
        \frac{2 \hat{K}_{\ell  } (\omega)}{C^{2 \ell +1}} 
        \left(\hat{Q}_{\ell}^2 + \varepsilon^2 \hat{N}\mathbb{H}_{Q,\ell}^{(2)}\right)  
        \nonumber\\
        &+ 
        \hat{P}^2_{\ell} 
        + 
        \varepsilon^2 \hat{N}\mathbb{H}_{P,\ell}^{(2)}  
    \bigg]
    + 
    \mathcal{O}\left(\varepsilon^4\right)
    \,.
\end{align}
We recall that the boundary condition at the stellar surface ($r=R$) for $H$ is that $H$ and $H'$ must be continuous across the boundary.
This means that we must use Eq.~\eqref{eq:H0-boundary-v1} as the exterior potential solution, which we match to the interior (stellar) potential. 
This matching of the interior solution for a given frequency $\omega$ provides the tidal response function $\hat{K}_{\ell m}(\omega)$.

%------------------------------------------------------------------------------
\subsection{Interior solution in a small-frequency and small-viscosity expansion}\label{sec:viscous-source}

To summarize our setup so far, our goal is to solve the master equations [Eq.~\eqref{eq:master-equation-matrix}], which consists of four, first-order differential equations for four unknowns ($V,W,H,H'$).
The exterior (vacuum) master equations were discussed and solved in Sec.~\ref{sec:tidal-bcs} using a small-frequency expansion.
This solution was then matched to the solution in the PN zone in Sec.~\ref{sec:Matching-PN}.
In this subsection, we will explain how to obtain a solution to the interior (non-vacuum) master equations in a small-frequency and small-viscosity expansion, which must be initialized at the origin by requiring the spacetime and fluid fields all remain regular.
This interior solution is important because it must be matched at the surface of the star with the external solution, i.e.~we must require that $H$ and $H'$ in the interior be the same as the exterior solution of Eq.~\eqref{eq:H0-boundary-v1} at the stellar surface. At the stellar surface, we also require that the Lagrangian displacement of the pressure vanishes.
This set of boundary conditions, for a given background and choice of $\varepsilon$, completely specifies the solution, and thus, the low-frequency limit of the general tidal response function $\hat{K}$ and the tidal response coefficients of Eq.~\eqref{eq:low-frequency-limit}, for non-stratified stars.

Let us begin by describing how to take the low-frequency limit and how to perturbatively incorporate viscous corrections to the stress-energy tensor; we will then use this expansion to compute the tidal lag parameter. For a viscous fluid, the stress-energy tensor can be decomposed as 
\begin{align}
    T_{\mu\nu}^{\mathrm{full}} &= T_{\mu\nu} + \mathbb{S}_{\mu\nu}
    \,,
\end{align}
where $T_{\mu\nu}$ is given by Eq.~\eqref{eq:perfect-fluid-tensor}, and $\mathbb{S}_{\mu\nu}$ captures the effects of out-of-equilibrium corrections to the perfect fluid stress-energy tensor.
For a fluid close to thermodynamic equilibrium, one can generally view these corrections as adding perturbative corrections to the perfect fluid equations of motion.

More specifically, let us denote the volume-averaged viscosity (shear/bulk), the volume-averaged pressure, and the volume-averaged energy density inside the neutron star by $\stf{\eta}$, $\stf{p}$, and $\stf{e}$, respectively, where volume average is here defined via 
\begin{align}
    \stf{A} = \frac{3}{R^3} \int_0^R A(r) \; r^2 dr \,.
\end{align}
for any function $A(r)$.
The spatial parts of the ideal fluid stress-energy tensor and the viscous stress-energy tensor then roughly scale as
\begin{subequations}
\begin{align}
    \left| T_{ij}^{\mathrm{ideal}} \right| &\sim \stf{p} + \stf{e} \omega^2 \delta L^2  \,,\\
    \left|T_{ij}^{\mathrm{visc}} \right| &\sim \stf{\eta} \omega \frac{\delta L}{L}  \,,
\end{align}
\end{subequations}
where $L$ is a typical length scale of oscillations inside the neutron star, and $\delta L$ is a typical value of the Lagrangian displacement vector.
In the regions of the star away from the surface, one can assume that $\delta L/L \ll 1$. When $\omega^2$ is small, one can also ignore the kinetic energy contribution to the ideal fluid stress-energy tensor.
Therefore, in the interior of the star (away from the surface), we assume that
\begin{align}\label{eq:epsilon-def}
    \epsilon &\equiv 
    \frac{\stf{\eta} \omega \; \delta L}{L \stf{p} }
    =
    \frac{\stf{\bar{\eta}} \stf{\Omega} \delta L}{L}
    \ll 1
\end{align}
where the dimensionless quantities $\stf{\bar{\eta}}$ and $\stf{\Omega}$ are defined by 
\begin{subequations}
\begin{align}
    \stf{\bar{\eta}} &\equiv \frac{\stf{\eta} \sqrt{4 \pi \stf{e}}}{\stf{p}} \ll 1\,,\\
    \label{eq:omega-small-condition}
    \stf{\Omega} &\equiv \frac{\omega}{\sqrt{4 \pi \stf{e}}} \ll 1\,.
\end{align}
\end{subequations}
Near the surface of the star, these assumptions can easily fail because $\delta L \sim L$ and $(p,e) \to 0$.
To avoid this potential issue, we \emph{assume} that the viscous coefficients go to zero more rapidly than the ideal part of the stress-energy tensor, e.g.
\begin{align}\label{eq:visc-to-zero-cond}
    \lim_{p\to 0}\frac{\eta}{p} = 0
    .
\end{align}

In the interior of the star, we wish to expand the master variables in a small $\epsilon$ expansion.
A practical way of doing so is to first expand the solutions in small viscosity $\stf{\Bar{\eta}}$
\begin{align}\label{eq:perturbative-solution-interior-to-the-star}
    \Vec{Y}(r;\omega) 
    &= \Vec{Y}_{(0)} + i \stf{\Bar{\eta}} \Vec{Y}_{(1)} + \mathcal{O} \left(\epsilon^2\right)\,,
\end{align}
where we treat $\stf{\Bar{\eta}}$ as a formal order counting parameter, i.e.~this parameter can be set to unity when numerically solving for the perturbations.
Inserting the expansion of Eq.~\eqref{eq:perturbative-solution-interior-to-the-star} into the master equation [Eq.~\eqref{eq:master-equation-matrix}], we obtain 
\begin{subequations}\label{eq:perturbative-scheme-master-equation}
\begin{align}
    \label{eq:master-equation-0-order}
    \Vec{Y}'_{(0)} &= \mathbf{A} \Vec{Y}_{(0)} \,,\\
    \label{eq:master-equation-1-order}
    \Vec{Y}'_{(1)} &= \mathbf{A} \Vec{Y}_{(1)} - \frac{i}{\stf{\Bar{\eta}}}\, \mathbf{B} \Vec{S} \,.
\end{align}
\end{subequations}
Note that $\Vec{S} \propto \stf{\Bar{\eta}}$, and therefore, the above expression is valid.
Finally, we solve the above system in the small frequency limit by choosing a range of value of $\omega$ that ensure Eq.~\eqref{eq:omega-small-condition} is valid.
The combined effect of these expansions (i.e.~of a small $\left<\bar{\eta}\right>$ and a small $\omega$ expansion) is a small $\epsilon$ expansion.

The source vector $\vec{S}$ depends on the decomposition of the viscous source tensor $\mathbb{S}^{\mathrm{bulk}}_{\mu \nu}$ through Eq.~\eqref{eq:spherical-decomposition-source-term}, so let us discuss this quantity in more detail. The viscous source tensor we use is 
\begin{subequations}\label{eq:source-tensors-visc}
\begin{align}
    \label{eq:bulk-viscous-source-tensor}
    \mathbb{S}^{\mathrm{bulk}}_{\mu \nu} &= 
    -
    \zeta \theta_{(0)} \Delta_{\mu\nu}^{(0)}
    \,,\\
    \label{eq:shear-viscous-source-tensor}
    \mathbb{S}^{\mathrm{shear}}_{\mu \nu} &= -2 \eta \sigma_{\mu\nu}^{(0)} \,,
\end{align}
\end{subequations}
where $\zeta(p)$ and $\eta(p)$ are the bulk and shear viscosities, and $\Delta_{\mu\nu}\equiv u_{\mu}u_{\nu} + g_{\mu\nu}$.
The fluid expansion and the shear tensor are given by
\begin{align}
    \theta_{(0)} &\equiv \left(\nabla_{\mu} u^{\mu}\right)_{(0)} \,,\\
    \sigma^{(0)}_{\mu\nu} &\equiv
        \left(\Delta^{\gamma}_{(\nu} \Delta^{\delta}_{\nu)} \nabla_{\gamma} u_{\delta}\right)^{(0)} 
        -
        \frac{1}{3} \left(\Delta_{\mu \nu}\Delta^{\gamma \delta} \nabla_{\gamma} u_{\delta}\right)^{(0)}
    \,.
\end{align}
We note that here the label $(0)$ indicates that the sources are evaluated using the zeroth-order (in $\epsilon$) solution to the master equation [Eq.~\eqref{eq:master-equation-0-order}].
That is, we substitute our solution to the perturbative perfect fluid equations (zeroth-order in the viscosity expansion) into the viscous source tensor Eq.~\eqref{eq:source-tensors-visc}.
We list the source functions $S_0,S_1,S_{Z}$ and $S_{\Omega}$ for the bulk and shear viscosities (Eq.~\eqref{eq:source-tensors-visc}) in Appendix~\ref{appendix:source-functions}. 

The expressions for the source functions $S_0,S_1,S_{Z}$ and $S_{\Omega}$ in terms of the viscosities present an important property:
\textit{The first-order master equation [Eq.~\eqref{eq:master-equation-1-order}] for a purely bulk-viscous perturbation ($\eta = 0$) depends only on $\zeta(p)$, while for a purely-shear viscous perturbation ($\zeta = 0$), the master equations depend only on $\eta, d \eta/d p$ and $d^2 \eta/d p^2$.}
This implies that shear perturbations are very sensitive to changes in the local values of the profile of the shear viscosity.
In particular, small changes to the shear viscosity profile near the surface of the star can lead to significant changes to the tidal response of the star.

Before we proceed further, we briefly discuss some facts about causal relativistic fluid theories.
The source functions in Eq.~\eqref{eq:source-tensors-visc} do not describe causal and stable fluid propagation if one considers non-perturbative solutions to the relativistic equations of motion~\cite{1983AnPhy.151..466H,1985PhRvD..31..725H}.
To remedy this situation, a number of relativistic viscous fluid theories have been proposed~\cite{1976AnPhy.100..310I,1979RSPSA.365...43I,Muller-book,Baier:2007ix,Denicol_2012,Bemfica:2020zjp,Kovtun_2019}.
The structure of the dissipative part of the stress-energy tensors of these fluid theories differ from the one given in Eq.~\eqref{eq:source-tensors-visc} to ensure the fluid theory remains causal (e.g. \cite{HegadeKR:2023glb}), but these differences matter only at high frequencies because they are of $\mathcal{O}\left(\omega^2\right)$.
Therefore, if one restricts attention to low frequencies and small deviations from equilibrium, only the ``hydrodynamic'' (low-frequency) modes should survive and the leading-order dissipative part of the stress-energy tensor should be equal to the one given in Eq.~\eqref{eq:source-tensors-visc}.
The perturbative scheme in Eq.~\eqref{eq:perturbative-scheme-master-equation} then is only theoretically consistent (as a relativistic description of the system) when $\omega/\sqrt{4 \pi \stf{e}} \ll 1$, which is the region in which we obtain the tidal lag parameter for highly relativistic stars.
Hence, we will ignore the contributions from the causal corrections to the viscous stress-energy tensor in this paper.

%---------------------------------------------
%-----------------------------------------------
\subsection{Boundary conditions for the internal problem near the origin and the surface}\label{sec:boundary-conditions}
To solve the master equations in the interior we need to understand the local behaviour of $\Vec{Y}$ near the origin and the surface.
The discussion in this section will be general and should highlight the general pattern of the calculation to obtain the tidal response function. 
For specific applications, see Sec.~\ref{sec:summary-body-zone}.
A general solution to the master equation [Eq.~\eqref{eq:master-equation-matrix}] is a sum of the homogeneous and particular solutions
\begin{align}
    \Vec{Y} &= \Vec{Y}_{\mathrm{hom}} + \Vec{Y}_{\mathrm{part}}
    \,.
\end{align}
The form of the particular solution depends on the source functions $\vec{S}$. 
To keep the discussion general, we focus on the boundary conditions for the homogeneous solution.
The local properties of the particular solution for bulk and shear viscous source are provided in the supplementary \texttt{Mathematica} file~\cite{github-code} and the source properties are provided in Appendix~\ref{appendix:source-functions}.

The master equations [Eq.~\eqref{eq:perturbative-scheme-master-equation}] possess two singular points, one at the origin and one at the surface of the star, $r=R$.
A physically acceptable solution must be regular (finite) at both of these singular points.
To solve the field equations, we divide the domain of the star into a left $r \in (0,r_f]$ and a right $r \in [r_f, R) $ domain, separated by a fiducial point $r_f$. This fiducial point is artificial, so one must ensure than any final physical solution that is obtained numerically is insensitive (up to numerical error) to the choice of $r_f$.

Let us now discuss the solution in the left domain. At the origin, we assume that the solution is analytic and so we expand it in a Taylor series
\begin{subequations}
\begin{align}
    H_{\mathrm{hom}}(r) 
    &=
    h_0 
    + 
    r^2 h_2(h_0, w_0) + \mathcal{O}\left(r^4\right) \,, \\
    W_{\mathrm{hom}}(r) &= w_0 + r^2 w_2(h_0,w_0) + \mathcal{O}\left(r^4\right) \,, \\
    V_{\mathrm{hom}}(r) &= -\frac{w_0}{\ell} + r^2 v_{2}(h_0,w_0) + \mathcal{O}\left(r^4\right) \,,
\end{align}
\end{subequations}
where $h_0$ and $w_0$ are constants.
The structure of the field equations show that all the Taylor series coefficients are completely determined by the constants $\left(h_0,w_0\right)$.
Therefore, near the origin, we have two linearly independent solutions corresponding to our choices for the values of these constants. 
To obtain the most general solution, we integrate the homogeneous equation with two linearly independent values of $(h_0,w_0)$ (for example $(1,0)$ and $(0,1)$).
Let us denote these linearly independent homogeneous solutions by $\Vec{Y}_{\mathrm{hom},\mathrm{L},1}$ and $\Vec{Y}_{\mathrm{hom},\mathrm{L},2}$.
Finally, we denote the particular solution on the left domain by $\Vec{Y}_{\mathrm{left}, \mathrm{part}}$.
The general solution in the left domain then is
\begin{align}
    \Vec{Y}_{\mathrm{left}}
    &=
    \Vec{Y}_{\mathrm{left}, \mathrm{part}}
    +
    a_1 \Vec{Y}_{\mathrm{hom},\mathrm{L},1}
    +
    a_2 \Vec{Y}_{\mathrm{hom},\mathrm{L},2} \,,
\end{align}
where $a_1$ and $a_2$ are arbitrary constants.

To determine the general solution in the right domain, we need to find local solutions near the surface of the star.
The solution may not be analytic near the stellar surface for an arbitrary EoS.
Despite this, we can show the near-surface solution is analytic if we \emph{assume} a polytropic EoS near the stellar surface $p \propto e^{1+1/n}$, with $n$ the polytropic index and $\gamma = 1+1/n$ the adiabatic index.
We parameterize the adiabatic index as 
\begin{align}\label{eq:chi_0_surface}
    \gamma(p) &= \frac{\chi_0 (e(p)+p)}{p e'(p)}\,,
\end{align}
where $\chi_0$ is a constant used to denote the degree of stratification [see Eq.~\eqref{eq:Schwarzschild-criteria}].
We can then expand the master variables near the stellar surface as follows:
\begin{subequations}\label{eq:surface-solution-homogenous}
\begin{align}
    H_{\mathrm{hom}}(r) &= h_{r,0} + z h_{r,1} + \mathcal{O} \left(z^2\right)\,,\\
    W_{\mathrm{hom}}(r) &= w_{r,0} + z w_{r,1}(h_{r,0},h_{r,1},w_{r,0})+ \mathcal{O}(z^2) \,,\\
    V_{\mathrm{hom}}(r) &= \frac{h_{r,0}(R-2M)}{R \omega^2}
    -
    \frac{M \sqrt{R-2M} w_{r,0}}{R^{5/2} \omega^2} \nonumber \\
    &+
    v_1(h_{r,0},h_{r,1},w_{r,0}) z + \mathcal{O}(z^2) \,,
\end{align}
\end{subequations}
where $z= R-r$.
The near-surface solution is completely determined if we specify the three independent constants $(h_{r,0},h_{r,1},w_{r,0})$\footnote{We provide the expressions for $h_2$, $w_2$, $v_2$, $w_{r,1}$ and $v_{r,1}$ in the supplementary \texttt{Mathematica} notebook~\cite{github-code}.}.
We pick three linearly independent values of $(h_{r,0},h_{r,1},w_{r,0})$ and integrate the homogeneous equations on the right domain; we denote the solutions by $\Vec{Y}_{\mathrm{hom},\mathrm{R},1}$, $\Vec{Y}_{\mathrm{hom},\mathrm{R},2}$ and $\Vec{Y}_{\mathrm{hom},\mathrm{R},3}$.
We write the particular solution as $\Vec{Y}_{\mathrm{right}, \mathrm{part}}$.
In principle, then, we have three linearly independent solutions in the right domain,
\begin{align}
    \Vec{Y}_{\mathrm{right}}
    &=
    \Vec{Y}_{\mathrm{right}, \mathrm{part}}
    +
    b_1 \Vec{Y}_{\mathrm{hom},\mathrm{R},1}
    \nonumber\\
    &+
    b_2 \Vec{Y}_{\mathrm{hom},\mathrm{R},2}
    +
    b_3 \Vec{Y}_{\mathrm{hom},\mathrm{R},3}\,.
\end{align}

To determine the perturbative solution for the entire star, we need to find the constants $a_i$ and $b_i$.
Demanding continuity of the solution at the fiducial point $r_f$ provides us with four equations (for the four components of $\vec{Y}$)
\begin{align}\label{eq:matching-fiducial}
    &\Vec{Y}_{\mathrm{left}, \mathrm{part}}(r_f)
    +
    a_1 \Vec{Y}_{\mathrm{hom},\mathrm{L},1}(r_f)
    +
    a_2 \Vec{Y}_{\mathrm{hom},\mathrm{L},2}(r_f)
    =
    \nonumber\\
    &\Vec{Y}_{\mathrm{right}, \mathrm{part}}(r_f)
    +
    b_1 \Vec{Y}_{\mathrm{hom},\mathrm{R},1}(r_f)
    +
    b_2 \Vec{Y}_{\mathrm{hom},\mathrm{R},2}(r_f)
    \nonumber\\
    &+
    b_3 \Vec{Y}_{\mathrm{hom},\mathrm{R},3}(r_f)
    \,.
\end{align}
To completely specify the solution, we need a relation between $H$ and $H'$ at the surface.
This is obtained by matching $H$ and its first derivative to the exterior vacuum metric solution of
Eq.~\eqref{eq:H0-ansatz-v1}
\begin{subequations}\label{eq:matching-surface}
\begin{align}
    &H_{\mathrm{right}, \mathrm{part}}(R)
    +
    b_1 H_{\mathrm{hom},\mathrm{R},1}(R)
    +
    b_2 H_{\mathrm{hom},\mathrm{R},2}(R)
    \nonumber\\
    &+
    b_3 H_{\mathrm{hom},\mathrm{R},3}(R)
    =
    \nonumber\\
    &
    \bigg.
    \frac{4 \pi d_{\ell m} (\omega) }{2 \ell +1} \frac{M^{\ell}}{r^{\ell}}
    \bigg[ \frac{2 \hat{K}_{\ell} (\omega)}{C^{2 \ell +1}} \left(\hat{Q}_{\ell}^2 + (M\omega)^2 \hat{N}\mathbb{H}_{Q,\ell}^{(2)}\right)  
    \nonumber\\
    &+ \hat{P}^2_{\ell} + (M\omega)^2 \hat{N}\mathbb{H}_{P,\ell}^{(2)}  \bigg]
    \bigg|_{r=R} \,,\\
%------------------------------------------------------------
    &\frac{d}{dr}H_{\mathrm{right}, \mathrm{part}}(R)
    +b_1 \frac{d}{dr}H_{\mathrm{hom},\mathrm{R},1}(R)
    \nonumber\\
    &+
    b_2 \frac{d}{dr}H_{\mathrm{hom},\mathrm{R},2}(R)
    +
    b_3 \frac{d}{dr}H_{\mathrm{hom},\mathrm{R},3}(R)
    =
    \nonumber\\
    &
    \frac{d}{dr}\bigg.\bigg\{
    \frac{4 \pi d_{\ell m} (\omega) }{2 \ell +1} \frac{M^{\ell}}{r^{\ell}}
    \bigg[ \frac{2 \hat{K}_{\ell} (\omega)}{C^{2 \ell +1}} \left(\hat{Q}_{\ell}^2 + (M\omega)^2 \hat{N}\mathbb{H}_{Q,\ell}^{(2)}\right)  
    \nonumber\\
    &+ \hat{P}^2_{\ell} + (M\omega)^2 \hat{N}\mathbb{H}_{P,\ell}^{(2)}  \bigg]
     \bigg\}\bigg|_{r=R}
     \,.
\end{align}
\end{subequations}
For a given value of the background TOV solution, the tidal field $d_{\ell m }(\omega)$, and the frequency $\omega$, Eqs.~\eqref{eq:matching-fiducial} and \eqref{eq:matching-surface} provide 6 linear equations for the unknowns $(a_1,a_2,b_1,b_2,b_3, \hat{K}_{\ell}(\omega))$, which can be solved to obtain these quantities, including the tidal response function $\hat{K}_{\ell}(\omega)$. 
%===================================================================================
\subsection{Summary of solutions in the body zone}\label{sec:summary-body-zone}

While the set of equations we provide above in principle are sufficient to describe the driven tidal response of a star, here we provide more details on the algorithm we used in practice to obtain the tidal response function $\hat{K}_{\ell}\left(\omega\right)$ for two special cases that we investigate in more detail later in this paper. 
First, we discuss how we obtain the tidal response for non-viscous (perfect fluid) stars, and then we describe how we obtain the low-frequency limit of the tidal response for viscous stars. 
We hope the summary below serves as a useful guide for numerical implementations.

%===================================================================================
\subsubsection{High-frequency conservative tidal response function for inviscid stars}\label{sec:conservative-tides-summary}
Here, we summarize the steps needed to obtain the high-frequency tidal response of an inviscid star with $S_{\mu\nu}=0$. 
The field equations for polar perturbations are governed by three master equations, which are written schematically as shown in Eq.~\eqref{eq:master-equation-matrix}
\begin{align}
    \Vec{Y}' &= \mathbf{A} \Vec{Y} \,.
\end{align}
To integrate these equations, we divide the domain $(0,R)$ into a left domain $(0,r_f)$ and a right domain $(r_f,R)$ separated by a fiducial point $r_f$.
The choice of the fiducial point depends on the problem but typically we choose $r_f \sim (0.5 - 0.8)R$.
For stratified stars, we choose the fiducial point around $r_f \sim 0.8 R$.
We solve the master equation in the left and right domains using the technique described in Sec.~\ref{sec:boundary-conditions}.
The right and left solutions are matched at the fiducial point to obtain a system of linear equations
\begin{align}\label{eq:match-zero-cont}
    &a_1 \Vec{Y}_{\mathrm{hom},\mathrm{L},1}(r_f)
    +
    a_2 \Vec{Y}_{\mathrm{hom},\mathrm{L},2}(r_f)
    =
    \nonumber\\
    &
    b_1 \Vec{Y}_{\mathrm{hom},\mathrm{R},1}(r_f)
    +
    b_2 \Vec{Y}_{\mathrm{hom},\mathrm{R},2}(r_f)
    \nonumber\\
    &+
    b_3 \Vec{Y}_{\mathrm{hom},\mathrm{R},3}(r_f)
    \,.
\end{align}
The boundary condition for the gravitational potential for the zeroth-order solution is given by
\begin{align}\label{eq:H_ext_summary}
    H&= 
    \frac{4 \pi d_{\ell m} (\omega) }{2 \ell +1} \frac{M^{\ell}}{r^{\ell}}
    \bigg[ \frac{2k_{\ell}}{C^{2 \ell +1}} \left(\hat{Q}_{\ell}^2 + (M\omega)^2 \hat{N}\mathbb{H}_{Q,\ell}^{(2)}\right)  \nonumber\\
    &+ \hat{P}^2_{\ell} + (M\omega)^2 \hat{N}\mathbb{H}_{P,\ell}^{(2)}  \bigg]
    \,,
\end{align}
where $k_{\ell}(\omega)$ is the conservative tidal response function.
Matching this solution and its derivative with the right domain solution at the surface, we obtain the following system of equations 
\begin{subequations}\label{eq:match-zero-boundary}
\begin{align}
    &b_1 H_{\mathrm{hom},\mathrm{R},1}(R)
    +
    b_2 H_{\mathrm{hom},\mathrm{R},2}(R)
    \nonumber\\
    &+
    b_3 H_{\mathrm{hom},\mathrm{R},3}(R)
    =
    \nonumber\\
    &
    \bigg.
    \frac{4 \pi d_{\ell m} (\omega) }{2 \ell +1} \frac{M^{\ell}}{r^{\ell}}
    \bigg[ \frac{2k_{\ell}}{C^{2 \ell +1}} \left(\hat{Q}_{\ell}^2 + (M\omega)^2 \hat{N}\mathbb{H}_{Q,\ell}^{(2)}\right)  
    \nonumber\\
    &+ \hat{P}^2_{\ell} + (M\omega)^2 \hat{N}\mathbb{H}_{P,\ell}^{(2)}  \bigg]
    \bigg|_{r=R} \,,\\
%------------------------------------------------------------
    &b_1\frac{d}{dr}H_{\mathrm{hom},\mathrm{R},1}(R)
    +
    b_2 \frac{d}{dr}H_{\mathrm{hom},\mathrm{R},2}(R)
    \nonumber\\
    &+
    b_3 \frac{d}{dr}H_{\mathrm{hom},\mathrm{R},3}(R)
    =
    \nonumber\\
    &
    \bigg.\frac{d}{dr}\bigg\{
    \frac{4 \pi d_{\ell m} (\omega) }{2 \ell +1} \frac{M^{\ell}}{r^{\ell}}
    \bigg[ \frac{2k_{\ell}}{C^{2 \ell +1}} \left(\hat{Q}_{\ell}^2 + (M\omega)^2 \hat{N}\mathbb{H}_{Q,\ell}^{(2)}\right)  
    \nonumber\\
    &+ \hat{P}^2_{\ell} + (M\omega)^2 \hat{N}\mathbb{H}_{P,\ell}^{(2)}  \bigg]
    \bigg\}\bigg|_{r=R}
    \,.
\end{align}
\end{subequations}
For a given value of frequency, Eqs.~\eqref{eq:match-zero-cont} and \eqref{eq:match-zero-boundary} provide us with six equations for the six variables $(a_1, a_2, b_1,b_2,b_3,k_{\ell} (\omega) )$. 
We solve these linear equations to obtain the solution and the conservative tidal response function $k_{\ell}(\omega)$ for inviscid stars. 
We note that we can always scale the linear system of equations by $d_{\ell m}(\omega)$, i.e.~the solution $H$ depends linearly on $d_{\ell m}$.
We use this freedom to set $d_{\ell m} (\omega)=1$.
\subsubsection{Low-frequency dissipative tidal response function}\label{sec:perturbation-theory-visc-summary}
To obtain the tidal lag parameter, we need to solve the master equations in a small viscosity and small frequency approximation.
We recall the small-viscosity expansion is given by Eqs.~\eqref{eq:perturbative-solution-interior-to-the-star} and \eqref{eq:perturbative-scheme-master-equation}.
We first obtain the conservative tidal solution $\Vec{Y}_{(0)}$ at a sufficiently small frequency, which, in practice, corresponds to $\stf{\Omega} \sim 0.01-0.05$. 
We then use this solution to compute the viscous source $\vec{S}$.
The explicit expression for the bulk and shear sources are provided in Appendix~\ref{appendix:source-functions}.
From this, we can compute the particular solution to $\vec{Y}_{(1)}$.
We then proceed to obtain the first-order solution to obtain the tidal lag parameter.
The basic steps are essentially the same as the ones we described in Sec.~\ref{sec:conservative-tides-summary}.
The difference is that now we need to obtain the particular solution corresponding to the presence of the viscous source term.

We now describe the local structure of the particular solution near the origin for bulk and shear viscous sources.
At the surface of the star we assume that the sources go to zero, so the local structure of the particular solution is the same as that of the homogeneous solution of Eq.~\eqref{eq:surface-solution-homogenous}.
The analytical solution near the origin depends on the zeroth-order solution, we schematically discuss the expansion in Appendix~\ref{appendix:source-functions} and the explicit expression are provided in the supplementary \texttt{Mathematica} notebook~\cite{github-code}.

With these boundary conditions, we integrate the first-order master equations and obtain the solutions in the left and right domains.
These solutions are matched at the fiducial point to obtain 
\begin{align}\label{eq:match-first-cont}
    &\Vec{Y}_{\mathrm{left}, \mathrm{part}}(r_f)
    +
    a_1^{(1)} \Vec{Y}_{(1),\mathrm{hom},\mathrm{L},1}(r_f)
    +
    a_2^{(1)} \Vec{Y}_{(1),\mathrm{hom},\mathrm{L},2}(r_f)
    =
    \nonumber\\
    &\Vec{Y}_{\mathrm{right}, \mathrm{part}}(r_f)
    +
    b_1^{(1)} \Vec{Y}_{(1),\mathrm{hom},\mathrm{R},1}(r_f)
    +
    b_2^{(1)} \Vec{Y}_{(1),\mathrm{hom},\mathrm{R},2}(r_f)
    \nonumber\\
    &+
    b_3^{(1)} \Vec{Y}_{(1),\mathrm{hom},\mathrm{R},3}(r_f)
    \,.
\end{align}
We also expand the tidal response function in $\epsilon$ assuming no g-mode resonances
\begin{align}
    \hat{K}_{\ell } (\omega) &= k_{\ell }(\omega)  + i \epsilon k_{\ell }(\omega) \tau_{d,\ell}(\omega) \omega  
    + \mathcal{O} \left(\epsilon^2\right) \,,
\end{align}
where $k_{\ell } (\omega)$ is the conservative tidal response function and $\tau_{d,\ell}(\omega)$ is the tidal lag function.
With this expansion, the exterior potential simplifies to
\begin{align}
    H &= H_{(0)} + i\epsilon H_{(1)} + \mathcal{O}\left(\epsilon^2\right)\,,
\end{align}
where (see Eq.~\eqref{eq:H0-boundary-v1})
\begin{subequations}
\begin{align}
    &H_{(0)}= \frac{4 \pi d_{\ell m} }{2 \ell +1} \frac{M^{\ell}}{r^{\ell}}
    \bigg[ \frac{2 k_{\ell } (\omega)}{C^{2 \ell +1}} \left(\hat{Q}_{\ell}^2 + (M\omega)^2 \hat{N}\mathbb{H}_{Q,\ell}^{(2)}\right)  \nonumber\\
    &+ \hat{P}^2_{\ell} + (M\omega)^2 \hat{N}\mathbb{H}_{P,\ell}^{(2)}  \bigg]
    \,,\\
    &H_{(1)}=
    \frac{4 \pi d_{\ell m} (\omega) }{2 \ell +1} \frac{M^{\ell}}{r^{\ell}}
    \bigg[ \frac{2 k_{\ell}(\omega) \tau_{d,\ell}(\omega) \omega }{C^{2 \ell +1}} \bigg(\hat{Q}_{\ell}^2 \nonumber\\
    &\hspace{3cm}
    + (M\omega)^2 \hat{N}\mathbb{H}_{Q,\ell}^{(2)}\bigg)
     \bigg]
    \,.
\end{align}
\end{subequations}
The function $k_{\ell}(\omega)$ is obtained by matching the zeroth order solution.
The boundary condition for the gravitational potential for the first-order solution at the stellar surface is given by
\begin{align}
    H_{(1)}&= 
    \frac{4 \pi d_{\ell m} (\omega) }{2 \ell +1} \frac{M^{\ell}}{r^{\ell}}
    \bigg[ \frac{2k_{\ell} \tau_{d,\ell} \omega}{C^{2 \ell +1}} \left(\hat{Q}_{\ell}^2 + (M\omega)^2 \hat{N}\mathbb{H}_{Q,\ell}^{(2)}\right)  \bigg]
    \,.
\end{align}
Matching this with the right domain solution at the surface, we obtain 
\begin{subequations}\label{eq:match-first-boundary}
\begin{align}
    &H_{\mathrm{right}, \mathrm{part}}(R)
    +
    b_1^{(1)} H_{(1),\mathrm{hom},\mathrm{R},1}(R)
    +
    b_2^{(1)} H_{(1),\mathrm{hom},\mathrm{R},2}(R)
    \nonumber\\
    &+
    b_3^{(1)} H_{(1),\mathrm{hom},\mathrm{R},3}(R)
    =
    \nonumber\\
    &
    \bigg.\frac{4 \pi d_{\ell m} (\omega) }{2 \ell +1} \frac{M^{\ell}}{r^{\ell}}
    \bigg[ \frac{2k_{\ell} \tau_{d,\ell} \omega}{C^{2 \ell +1}} \left(\hat{Q}_{\ell}^2 + (M\omega)^2 \hat{N}\mathbb{H}_{Q,\ell}^{(2)}\right)  \bigg]\bigg|_{r=R} \,,\\
%------------------------------------------------------------
    &b_1^{(1)} \frac{d}{dr}H_{(1),\mathrm{hom},\mathrm{R},1}(R)
    +
    b_2^{(1)} \frac{d}{dr}H_{(1),\mathrm{hom},\mathrm{R},2}(R)
    \nonumber\\
    &+
    b_3^{(1)} \frac{d}{dr}H_{(1),\mathrm{hom},\mathrm{R},3}(R)
    =
    \nonumber\\
    &
    \bigg.\frac{d}{dr}\bigg\{\frac{4 \pi d_{\ell m} (\omega) }{2 \ell +1} \frac{M^{\ell}}{r^{\ell}}
    \bigg[ \frac{2k_{\ell } \tau_{d,\ell} \omega}{C^{2 \ell +1}} \left(\hat{Q}_{\ell}^2 + (M\omega)^2 \hat{N}\mathbb{H}_{Q,\ell}^{(2)}\right)  \bigg]\bigg\}\bigg|_{r=R}
    \,,
\end{align}
\end{subequations}
which are the low-frequency expansions of Eq.~\eqref{eq:H0-ansatz-v1}.
Equations~\eqref{eq:match-first-cont} and \eqref{eq:match-first-boundary} provide 6 equations for the 6 unknowns $(a_1^{(1)}, a_2^{(1)}, b_1^{(1)},b_2^{(1)},b_3^{(1)},\tau_{d,\ell}(\omega) )$, which we solve to obtain the first-order-in-$\epsilon$ solution and the tidal lag parameter.

%-----------------------------------------------------
\section{Quadrupolar tidal response function of polytropic stars }\label{sec:Polytropic-Star-Dissipative-Love}
In this section, we apply the formalism we presented in Sec.~\ref{sec:non-radial-perturbations} to compute the $\ell=2$ tidal response function for relativistic stars with a polytropic EoS.
We consider an ``energy'' polytrope with the EoS
\begin{align}
    \label{eq:polytropic-eos}
    p(e) &= p_c \left(\frac{e}{e_c}\right)^{1+1/n} \,,
\end{align}
where $n$ is the polytropic index, while $p_c$ and $e_c$ are the central pressure and energy density.
\begin{figure*}[thp!]
    \centering
    \includegraphics[width = 0.95 \textwidth]{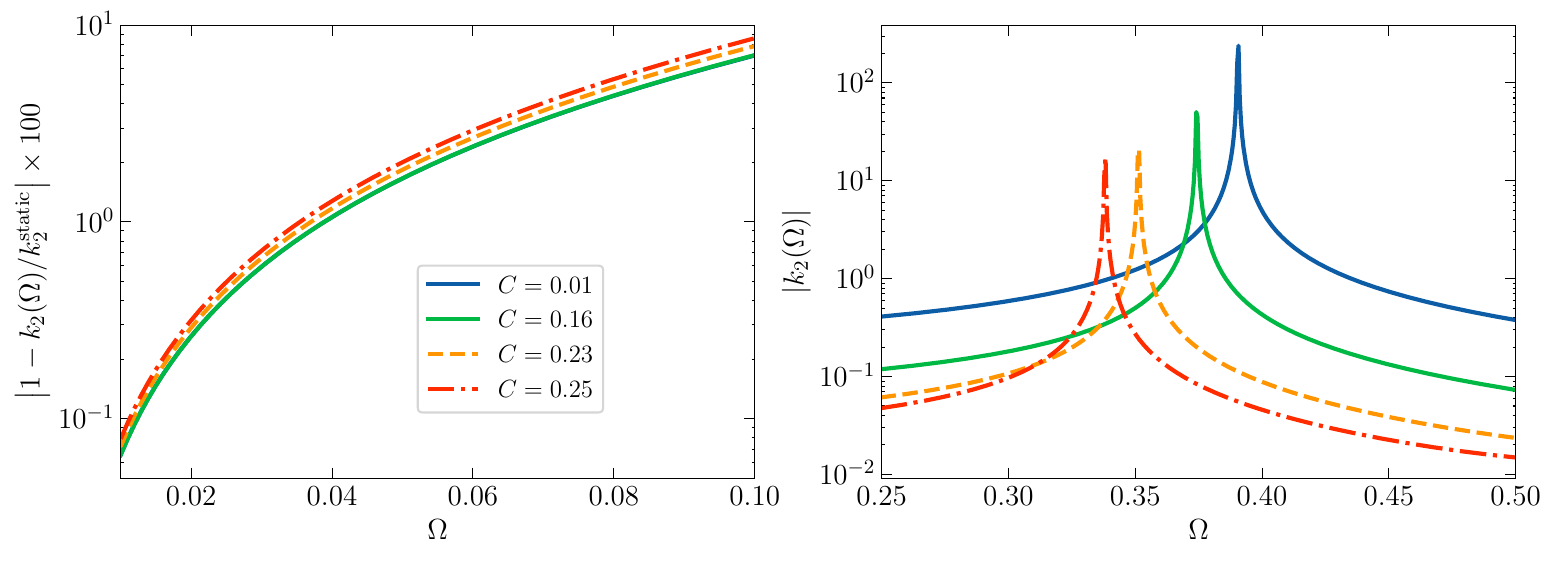}
    \caption{
    Dynamical conservative tidal response function for a convectively neutrally stable $n=1$ relativistic polytrope.
    \textbf{Left:} Fractional percent difference between the static and the dynamical tidal Love number as a function of the dimensionless frequency for different values of compactness.
    We see that the dynamical tidal Love number converges to the static Love number as $\Omega$ goes to zero.
    \textbf{Right:} high-frequency tidal response function for different values of compactness.
    We see resonances due to the $f$-mode contribution to the tidal response function. 
     }
    \label{fig:frac_diff_love_number}
\end{figure*}
For a polytropic star, it is easier to quote results based on dimensionless parameters.
We define dimensionless frequency $\Omega$ and space $\xi$ variables through 
\begin{align}
   r &= r_0 \; \xi \,, \qquad
   \omega = \Omega_s \; \Omega  \,.
\end{align}
where we have defined
\begin{align}
   \Omega_{s} &\equiv \sqrt{4 \pi e_c}\,,
\end{align}
and $r_0$ is a characteristic length scale.
The dimensionless variables are related to the relativistic Lane-Emden variables $\bar{\theta}$ and $\bar{\mu}$ via~\cite{Poisson:2009di} 
\begin{align}\label{eq:e-p-theta}
    e = e_c \; \bar{\theta}^n \,, \quad m= m_0 \; \bar{\mu} \,,
\end{align}
where $m_0$ is a characteristic mass scale.
The values of $m_0$ and $r_0$ are chosen to be 
\begin{align}
    m_0 \equiv 4 \pi r_0^3 e_c \,, \quad r_0^2 \equiv \frac{(n+1)p_c}{4 \pi e_c^2}\,,
\end{align}
because this will simplify the TOV equations later. We denote the dimensionless mass of the star and radius of the star by $\bar{\mu}_1$ and $\xi_1$, respectively.  Additionally, we introduce the relativistic factor 
\begin{align}
    b \equiv \frac{p_c}{e_c} \,,
\end{align}
which is a way of identifying relativistic corrections to the Newtonian limit.

The background spacetime is completely specified by the value of the polytropic index $n$ and the relativistic factor $b$. From the above definitions of dimensionless variables, it follows that the polytropic EoS [Eq.~\eqref{eq:polytropic-eos}] reduces to $p= p_c \bar{\theta}^{n+1}$.
We parameterize the adiabatic index $\gamma$ as
\begin{align}
    \label{eq:def-adiabatic-index}
    \gamma &= \left(1 + \frac{1}{N} \right)(1 + b \bar{\theta})
    \,,
\end{align}
the constant $N$ denotes the degree of stratification.
To understand when a star is stable against convection we
substitute the above expression into Eq.~\eqref{eq:Schwarzschild-criteria} and use Eq.~\eqref{eq:e-p-theta} and simplify to get 
\begin{align}
    \left(1 + \frac{1}{N} \right)(1 + b \bar{\theta})
    >
    \frac{e+p}{p} \frac{dp}{de} = 
    \left(1 + \frac{1}{n} \right)(1 + b \bar{\theta})
    \,.
\end{align}
Therefore, we see that a star is stable against convection if $n>N$.
Finally, we note that the parameter $\chi_0$ that appears in Eq.~\eqref{eq:chi_0_surface} is equal to 
\begin{align}
    \chi_0 &= \frac{1+ 1/N}{1+1/n} \,,
\end{align}
for the parameterization of the adiabatic index given in Eq.~\eqref{eq:def-adiabatic-index}.

For the viscosity profiles of the star, we use
\begin{subequations}\label{eq:viscous-profiles}
\begin{align}
    \zeta &= \frac{\bar{\zeta} p_c}{\Omega_s} \bar{\theta}^{n+3}\,,\\
    \eta &= \frac{\bar{\eta} p_c}{\Omega_s} \bar{\theta}^{n+3} \,,
\end{align}
\end{subequations}
this ensures that the viscous sources go to zero faster than the perfect fluid pressure near the surface (see Eq.~\eqref{eq:visc-to-zero-cond}), which holds for $n>0$.
While this is not the most general function that obeys Eq.~\eqref{eq:visc-to-zero-cond}, we use this simple choice to illustrate the difference in impact of the shear and bulk viscosity for the same fixed viscous profile.
We also rewrite the tidal lag function due to bulk viscosity and shear viscosity as~\cite{Ripley:2023qxo} 
\begin{align}\label{eq:taud-bulk-def}
    \tau_{d,\ell,\mathrm{bulk}}(\omega) &= \frac{p_{\ell,\mathrm{bulk}}(\omega) \stf{\zeta}}{ \stf{e} C}\,,\\
    \tau_{d,\ell,\mathrm{shear}}(\omega) &= \frac{p_{\ell,\mathrm{shear}}(\omega) \stf{\zeta}}{ \stf{e} C} \,,
\end{align}
where $C\equiv M/R$ is the compactness of the star, and where $p_{\ell,\mathrm{bulk/shear}}(\omega)$ now carries all the frequency dependence of the tidal lag function; it is easier to quote and plot the values of $p_{\ell,\mathrm{bulk/shear}}(\omega)$ instead of $\tau_{d,\ell,\mathrm{bulk/shear}}$, which is why we rewrite the latter above.

As we discuss in Sec.~\ref{sec:background-metric}, we solve for the background by numerically integrating the TOV equations given in Eq.~\eqref{eq:TOV-equations}.
We have checked that the values we obtain for the relativistic factor $b$, compactness $C$ match those given in Table V of~\cite{Binnington:2009bb}.
The method of integration of the fluid perturbation equations was described in Sec.~\ref{sec:summary-body-zone}. In what follows, we use these numerical solutions to present results for the conservative tidal response function $k_{\ell}(\omega)$ and the tidal lag function through $p_{\ell}(\omega)$ for the ${\ell}=2$ multipole, i.e.~the leading-order quadrupolar deformation.

Before we proceed further to present our results for the dynamical tidal response function, we note that a \texttt{C++} code to generate obtain the dynamical tidal response of a relativistic polytrope is available at~\cite{github-code}.
%-------------------------------------------------------
\subsection{Conservative Tidal Response Function}
We first present results for the full tidal response (non-perturbative in the forcing frequency $\omega$) for perfect fluid stars.
The sourceless master equations [Eq.\eqref{eq:master-equation-matrix}] can be integrated to obtain the conservative tidal response function for inviscid stars.
We emphasize that unlike in previous references~\cite{Hinderer:2007mb,Binnington:2009bb,Damour:2009vw}, we calculate the tidal response function $k_{\ell}$ \emph{without} assuming the forcing tidal frequency $\omega$ is small (see Sec.~\ref{sec:conservative-tides-summary}).
That is, we here do \textit{not} perform a small frequency expansion for the interior stellar solution.

Let us first discuss the results for a marginally convectively unstable star, i.e. $c_s^2 = c_e^2 \implies N = n$; this assumption is commonly used in modelling cold nuclear EoSs and in calculating the Love numbers in the adiabatic limit.
In this case, there are no $g$ modes in the mode spectrum of the stellar model.
In the left panel of Fig.~\ref{fig:frac_diff_love_number}, we compare our dynamical calculation of $k_2(\Omega)$ with the static value $k_2^{(\rm{static})}$ provided in Table V of~\cite{Binnington:2009bb} at low frequencies for an $n=1$ polytrope for different values of compactness.
As we can see from the plot, the dynamical tidal response function agrees well with the static value when $\Omega$ goes to zero. 
Differences larger than 10\%, however, begin to arise as soon as $\Omega > 0.1$.
To map this to a physical number, we need to pick a value for $e_c$.
Setting $e_c = 624 \mathrm{MeV}/\mathrm{fm}^3 $, we see that for a star with compactness $C = 0.16$ (bold green line) we obtain $M= 1.4 M_{\odot}$ and $R = 12.6 \mathrm{km} $.
For this star in an equal mass binary system, we see that differences larger than $10 \%$ in the value of $k_2(\Omega)$ arise for gravitational frequencies of $f_{\mathrm{GW}} >184 \mathrm{Hz} $.
Modelling dynamical effects is therefore important when inferring the tidal love numbers from gravitational wave data.
We leave a detailed analysis of the systematic effects arising from not modelling dynamical effects in the evolution of the tixal response to future work.

We now present results for the high-frequency conservative tidal response function and the resonance due to the $f$-mode frequency for different values of compactness in the right panel of Fig.~\ref{fig:frac_diff_love_number} for $n=N=1$.
We see that as compactness increases, relativistic effects make the $f$-mode resonance occur at a smaller value of $\Omega$ given the same polytropic EoS. 
We leave a detailed exploration of the behavior of the tidal response function of different nuclear EoS to future work.  
\begin{figure}[h!]
    \centering
    \includegraphics[width = 0.45 \textwidth]{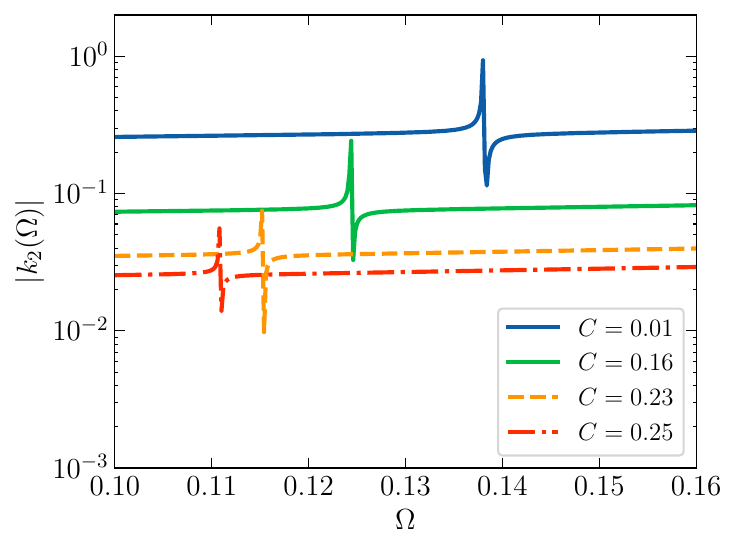}
    \caption{Low-frequency resonances due to $g$ modes in the dynamical conservative tidal response function for a relativistic polytrope with $n=1$ and $N=3/4$ for different values of compactness.
    }
    \label{fig:g_modes}
\end{figure}
\begin{figure*}[thp!]
    \centering
    \includegraphics[width = 0.95 \textwidth]{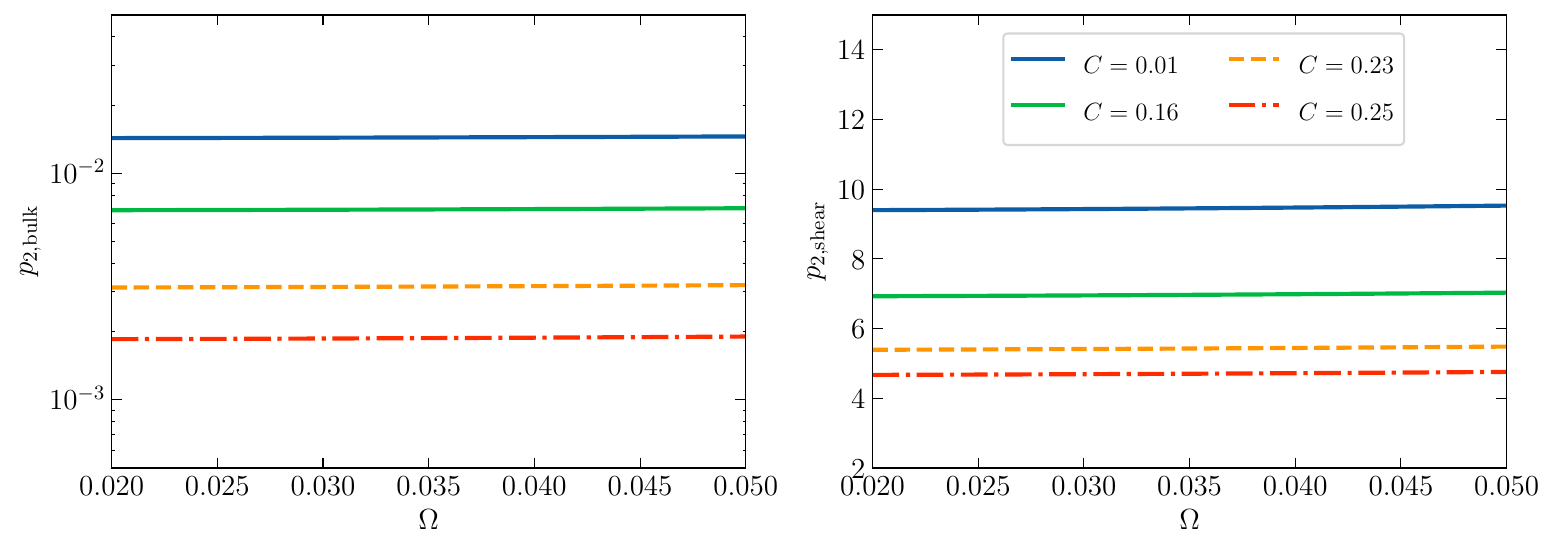}
    \caption{Dissipative tidal Love number $p_2$ for bulk (left) and shear (right) viscous interactions as a function of frequency for the same viscosity profile (see Eq.~\eqref{eq:viscous-profiles}) with $\bar{\eta}=1=\bar{\zeta}$.
    We see that at low frequencies the value of $p_2$ remains constant and that the value of $p_2$ for shear is 100-1000 larger than the bulk contribution.
    Relativistic corrections also lead to smaller values of $p_2(\omega)$.}
    \label{fig:shear_bulk_ratio}
\end{figure*}
Finally, in Fig.~\ref{fig:g_modes} we present the results for a convectively stable polytrope with $n=1$ and $N = 3/4$ for different values of compactness.
This is the first calculation of $g$-mode resonances for relativistic stars; for Newtonian calculations, see~\cite{Lai:1993di,Andersson_2020}.
As we can see from the figure, the presence of $g$ modes leads to resonances in the low-frequency tidal response.
Similar to the $f$-mode resonance structure, we see that the resonances introduced by $g$ modes is shifted to lower values of dimensionless frequency $\Omega$ due to relativistic corrections, as compared to the $g$-mode resonance of a less compact star with the same values of $N$ and $n$, i.e, the resonances are shifted to lower frequencies as the compactness $C$ increases. For example, we see that the first $g$-mode resonance is at $\Omega \sim 0.14$ for the blue solid curve with $C=0.01$, as compared to $\Omega \sim 0.125$ for green, dashed line with $C=0.15$.  
Resolving the $g$-mode resonances at extremely low frequencies requires the construction of a numerical code which has increases resolution near the surface to capture very low-frequency resonances which pile up near the surface of the star.
Sine our code currently does not have this feature, we cannot resolve the very low frequency $g$-mode resonances. 
We leave the construction of a numerical code which resolves the low-frequency $g$ modes for future work.
%=============================================================================
\subsection{Tidal lag parameter}

Let us now consider an example calculation of the tidal lag time $\tau_{\ell m}$ for a polytropic star, with viscous corrections.
To obtain the tidal lag due to viscosity, we need to solve the master equations perturbatively, as summarized in Sec.~\ref{sec:perturbation-theory-visc-summary}.
Figure~\ref{fig:shear_bulk_ratio} shows $p_{2}(\omega)$ induced by bulk (left panel) and shear viscosity (right panel), assuming the bulk and shear viscous profiles provided in Eq.~\eqref{eq:viscous-profiles} with $\bar{\zeta} = 1 = \bar{\eta}$ and different values of compactness, and assuming a marginally convectively unstable star with $n=1=N$.
Observe that the tidal lag function induced by shear viscosity is 1000 times larger than that induced by bulk viscosity.
Observe also that $p_2\left(\omega\right)$ does not depend sensitively on $\omega$ in the regime of small frequencies.
We also see that relativistic effects (higher compactness) leads to smaller values of $p_2\left(\omega\right)$.
This is qualitatively similar to the case of $k_{2}\left(\omega\right)$.
We have checked that we obtain qualitatively similar behavior for different polytropic indices $n$ (assuming $n=N$), and thus, what we have discussed here seems to be generic for polytropic EoSs. 

As we mentioned above, for the same viscosity profile and numerically \textit{equal} value of $\bar{\zeta}$ and $\bar{\eta}$, the dissipative tidal response is almost two orders of magnitude larger when it is induced by shear viscosities.
The physical reason behind this is that shear viscous effects are traverse, traceless and ``tensorial'' in nature (i.e.~arising through a nonzero shear tensor $\sigma_{\mu\nu}$), while bulk viscous excitations are ``scalar'' in nature (i.e.~arising from a nonzero fluid expansion scalar $\theta$).
Thus, the shear viscosity couples more strongly to the tensorial gravitational perturbations of the star. 

We warn the reader that the expected physical values of bulk viscosity is order $4-8$ orders of magnitude larger than those of the shear viscosity for neutron star mergers (see Sec.~\ref{sec:dissipation-review}). We here chose to focus our calculation on a simple example, where we used the same profile for the shear and bulk viscosity to compare the nature of the dissipative tidal response.
A physically realistic calculation would use a more realistic equation of state, bulk viscosity and shear viscosity profiles provided by nuclear physics calculations.
%------------------------------------------------------
%------------------------------------------------------
\section{Conclusion}\label{sec:conclusions}
%{\AH{Please read the conclusions and Appendix D and E.}}
Accurately modeling the tidal response of neutron stars is crucial to obtain information about the internal structure of these objects from GW observations. 
In this article, we introduced a formalism for calculating the dynamical tidal response of non-rotating neutron stars in general relativity.
Our formalism builds on~\cite{Poisson:2020vap,Lindblom-Mendell-Ipser-1997}, and reduces the internal problem of sourced polar perturbations into a set of three master equations: one for the gravitational potential, and two from the polar and radial components of the Lagrangian displacement vector.
We solved for the external, perturbed gravitational field in a small frequency approximation.
We then re-summed the external small-frequency expansion to obtain the tidal response function in the high-frequency regime.
We also proved that this re-summation procedure can be extended to all orders in the frequency expansion.
Finally, we  showed how to incorporate shear and bulk viscous effects in the perturbative tidal response of a star within a low-frequency expansion.
With this addition, we calculated the dissipative tidal response function of a spherical polytropic star.

To demonstrate the power of our approach, we applied our method to study the tidal response of relativistic polytropes in both the high-frequency and low-frequency regime.
We first showed that for marginally convectively unstable inviscid stars, the low-frequency limit of our method to calculate the tidal Love number agrees with the methods used to calculate the static tidal response~\cite{Binnington:2009bb,Damour:2009vw,Hinderer:2009ca}.
We then showed how our formalism enables the computation of the location of the $f$ and $g$-mode tidal resonances (for relativistic perfect fluid stars) of highly relativistic stars.
Finally, we presented results for the dissipative tidal response function due to shear and bulk viscous sources.
We showed that, given the same profile and numerical value for both viscosities, shear viscosity couples more strongly to the perturbing gravitational field than bulk viscosity does.
The upshot of this is that, for a fixed viscosity profile, the tidal response function is a more sensitive function of the shear viscosity than it is of the bulk viscosity.

Our work enables several directions for future work.
First, it would be interesting to compare our method of calculating the tidal response function to the Newtonian-inspired phenomenological models currently used in EoB framework to see how to systematically improve the latter. 
One could also see how numerical relativity simulations compare to our method of calculating the tidal response function, both with and without viscosity.
Another direction for future work is to extend the formalism presented here to slowly- and rapidly-rotating neutron stars.
We expect our approach to readily extend to slowly-rotating neutron stars (by employing a Hartle-Thorne-like expansion to the background and field perturbations).
Extending our method to rapidly-rotating neutron stars will likely be much more challenging, as spin could no longer be added perturbatively about a fiducial, spherically symmetric background.

A third direction for future work concerns the high-frequency behavior of the tidal response. While one can easily obtain analytical expressions for the external problem in a small frequency expansion to $\mathcal{O}\left(M\omega\right)^2$, obtaining analytical expressions at higher order appears difficult.
Nonetheless, if one could obtain such analytical solutions, one could then also explore how the addition of higher-order terms enters the re-summing of the external solution and changes or improves the high-frequency resonant behavior.
A fourth direction for future work concerns extensions beyond the simple EoSs considered in our paper. Here, we only explicitly calculated the conservative and dissipative tidal response function for polytropic stars.
Calculating the conservative and dissipative tidal response functions for more physically realistic EoSs will be necessary to apply our results to gravitational waveform models to constrain the equilibrium and out-of-equilibrium physics of neutron stars ~\cite{Hinderer_2010,Ripley:2023qxo}.
A final direction for future work would be to extend the computation of the dynamical black hole tidal Love number of non-spinning/slowly spinning black holes and to compare the results obtained with calculations from~\cite{Poisson:2020vap,Saketh:2023bul}. 
%-----------------------------------------------------
\begin{acknowledgements}
We are grateful to Elias Most for simulating the binary neutron star mergers to confirm our predictions about the tidal response function, which helped us look at the normalization of the particular solution in more detail.
We thank Eric Poisson for reading the paper, providing helpful comments and for discussing the calculation of the tidal response function in full general relativity.
We thank Rohit Chandramouli for discussions about the paper.
We acknowledge support from the Simons Foundation through Award No. 896696, the National Science Foundation (NSF) Grant No. PHY-2207650, and the National Aeronautics and Space Agency (NASA) Award No. 80NSSC22K0806.
\end{acknowledgements}
%-------------------------------------------------------------------------------------------------------
\appendix
\input{appendix}
%---------------------------------------------------
%\bibliographystyle{apsrev4-1}
\bibliography{ref}
\end{document}

%% file: appendix.tex
\section{Scalar, vector, and tensor spherical harmonics
\label{appendix:scalar_vector_tensor_spherical_harmonics}}
We work on the unit two-sphere $\mathbb{S}^2$, with metric $\Omega_{AB}$,
Levi-Cevita tensor $\varepsilon_{AB}$, 
and metric compatible derivative $D_A$. 
The Ricci tensor is $R=+2$.
Our notation for the spherical harmonics follows that of
\cite{Nagar:2005ea}.

%---------------------------------------------------------------------------
\subsection{Scalar spherical harmonics}
The scalar spherical harmonics satisfy
\begin{align}
    \left(
        \Omega^{AB}D_AD_B
        +
        \ell\left(\ell+1\right)
    \right)
    Y^m_{\ell}
    =
    0
    ,
\end{align}
along with the following orthogonality relation
\begin{align}
    \int d\Omega Y^m_{\ell}Y^{m^{\prime}}_{\ell^{\prime}}
    =
    \delta_{\ell\ell^{\prime}}\delta_{mm^{\prime}}
    .
\end{align}
%---------------------------------------------------------------------------
\subsection{Vector spherical harmonics}
The polar and axial
vector spherical harmonics respectively are
\begin{subequations}
\begin{align}
    \left[E^m_{\ell}\right]_A
    &\equiv
    D_AY^m_{\ell}
    ,\\
    \left[S^m_{\ell}\right]_A
    &\equiv
    \varepsilon_{BA}D^BY^m_{\ell}
    .
\end{align}
\end{subequations}
The vector spherical harmonics satisfy
\begin{align}
    \left(
        \Omega^{AB}D_AD_B
        +
        \left(-1+\ell\left(\ell+1\right)\right)
    \right)
    \left[V^m_{\ell}\right]_C
    =
    0
    ,
\end{align}
along with the following orthogonality relation
\begin{align}
    \int d\Omega 
        \left[V^m_{\ell}\right]_A
        \left[V^{m^{\prime}}_{\ell^{\prime}}\right]^A
    =
    \ell\left(\ell+1\right)
    \delta_{\ell\ell^{\prime}}\delta_{mm^{\prime}}
    .
\end{align}
The divergence of the polar and axial vector spherical harmonics
respectively are
\begin{subequations}
\begin{align}
    D_A\left[E^m_{\ell}\right]^A
    =&
    D_AD^AY^m_{\ell}
    \nonumber\\
    =&
    -
    \ell\left(\ell+1\right)Y^m_{\ell}
    .\\
    D_A\left[S^m_{\ell}\right]^A
    =&
    \varepsilon_{BA}D^BD^AY^m_{\ell}
    \nonumber\\
    =&
    0
    .
\end{align}
\end{subequations}
%---------------------------------------------------------------------------
\subsection{Tensor spherical harmonics}
The polar and axial tensor spherical harmonics respectively are
\begin{subequations}
\begin{align}
    \left[Z^m_{\ell}\right]_{AB}
    &\equiv
    D_AD_BY^m_{\ell}
    +
    \frac{1}{2}\ell\left(\ell+1\right)
    \Omega_{AB}Y^m_{\ell}
    ,\\
    \left[S^m_{\ell}\right]_{AB}
    &\equiv
    D_{(A}\left[S^m_{\ell}\right]_{B)}
    .
\end{align}
\end{subequations}
The tensor spherical harmonics satisfy
\begin{align}
    \left(
        \Omega^{AB}D_AD_B
        +
        \left(-2+\ell\left(\ell+1\right)\right)
    \right)
    \left[T^m_{\ell}\right]_{CD}
    =
    0
    ,
\end{align}
along with the following orthogonality relation
\begin{align}
    &\int d\Omega 
        \left[T^m_{\ell}\right]_{AB}
        \left[T^{m^{\prime}}_{\ell^{\prime}}\right]^{AB}
    \nonumber\\
    &\qquad=
    \frac{1}{2}
    \left(\ell-1\right)\ell\left(\ell+1\right)\left(\ell+2\right)
    \delta_{\ell\ell^{\prime}}\delta_{mm^{\prime}}
    .
\end{align}
The polar and axial tensor spherical harmonics are both traceless
\begin{subequations}
\begin{align}
    \Omega^{AB}\left[Z^m_{\ell}\right]_{AB}
    =&
    D_AD^AY^m_{\ell}
    +
    \ell\left(\ell+1\right)Y^m_{\ell}
    \nonumber\\
    =&
    0
    ,\\
    \Omega^{AB}\left[S^m_{\ell}\right]_{AB}
    =&
    D_A\left[S^m_{\ell}\right]^A
    \nonumber\\
    =&
    0
    .
\end{align}
\end{subequations}
The trace is captured by the scalar spherical harmonic $\Omega_{AB}Y^m_{\ell}$.
The divergence of the polar and axial tensor spherical harmonics
respectively are
\begin{subequations}
\begin{align}
    D_A\left[Z^m_{\ell}\right]^{AB}
    =&
    D_AD^AD^BY^m_{\ell}
    +
    \frac{1}{2}\ell\left(\ell+1\right)D^BY^m_{\ell}
    \nonumber\\
    =&
    \left(
        1
        -
        \frac{1}{2}\ell\left(\ell+1\right)
    \right)
    \left[E^m_{\ell}\right]^B
    ,\\
    D_A\left[S^m_{\ell}\right]^{AB}
    =&
    \frac{1}{2}D_A\left(
        D^A\left[S^m_{\ell}\right]^B
        +
        D^B\left[S^m_{\ell}\right]^A
    \right)
    \nonumber\\
    =&
    \left(
        1
        -
        \frac{1}{2}\ell\left(\ell+1\right)
    \right)
    \left[S^m_{\ell}\right]^B
    .
\end{align}
\end{subequations}
We have used that $D_AD_BV^C-D_BD_AV^C=R^C_{DAB}V^D$, 
$R_{ABCD}=(1/2)(\Omega_{AC}\Omega_{BD}-\Omega_{AD}\Omega_{BC})R$ and $R=2$.
Similarly,
\begin{subequations}
\begin{align}
    D^AD_A E_B &= \left(1 - \llp \right)E_B \\
    D^AD_A S_B &= \left(1 - \llp \right)S_B \,.
\end{align}
\end{subequations}
%=======================================
\section{Relativistic theory of Lagrangian fluid perturbations}
\label{appendix:relativistic-perturbation-theory}
Any field $Q$ can be described in two ways: through the Eulerian variation
$\delta Q$, which is the change in the quantity at a fixed point in spacetime,
or the Lagrangian perturbation $\Delta Q$, which is the variation of the fluid
quantity with respect to a frame that is dragged along with the fluid
along a \emph{displacement vector} $\xi^{\alpha}$
(for a review, see \cite{1978CMaPh..62..247F}).
The displacement vector $\xi^{\alpha}$ connects fluid elements
in the unperturbed fluid configuration to the elements in the perturbed fluid.
In the language of differential geometry, $\xi^{\alpha}$ is the generator
of diffeomorphisms that take world lines in the unperturbed fluid into 
world lines of the perturbed fluid.
For general fields one has
\begin{align}
\label{eq:lagrangian_variation}
    \Delta
    =
    \delta
    +
    \mathcal{L}_{\xi}
    ,
\end{align}
where $\mathcal{L}_{\xi}$ is the Lie derivative along $\xi^{\alpha}$.
Unlike the Eulerian variation,
the Lagrangian variation is invariant under infinitesimal changes in the coordinates.
The Eulerian and Lagrangian variations are both invariant under the change $\xi^{\alpha}\to\xi^{\alpha}+f u^{\alpha}$, where $f$ is a scalar field. 
We use this freedom to make $\xi^{\alpha}$ purely spacelike, i.e.~$u_{\alpha}\xi^{\alpha}=0$.
From $u^{\alpha}u^{\beta}g_{\alpha\beta}=-1$, 
and $\Delta u^{\alpha} \propto u^{\alpha}$, we then have
\begin{align}
\label{eq:lagrangian_variation_fluid_velocity}
    \Delta u^{\alpha}
    =
    \frac{1}{2}u^{\alpha}u^{\beta}u^{\gamma}\Delta g_{\beta\gamma}
    .
\end{align}

We now list thermodynamic identities that are valid in the comoving Lagrangian frame of the fluid.
A detailed derivation of these equations is provided in~\cite{1978CMaPh..62..247F}.
For adiabatic perturbations, the Lagrangian perturbation of the baryon number density $n$ is given by solving the baryon conservation equation $\nabla_{\alpha} \left(u^{\alpha} n \right) = 0$, and taking the Lagrangian perturbation of that expression
\begin{align}\label{eq:Deltanbyn}
    \Delta \nabla_{\mu}\left(n u^{\mu}\right)
    &= 0
    \nonumber\\
    \implies
    \frac{\Delta n}{n} 
    &= 
    -
    \frac{1}{2} \left(g^{\alpha \beta} + u^{\alpha} u^{\beta} \right) \Delta g_{\alpha \beta}
    \,.
\end{align}
From the first law of thermodynamics we have thats
\begin{align}\label{eq:Deltaebyeplusp}
    \frac{\Delta e}{e +p} &= \frac{\Delta n}{n} 
    \,.
\end{align}
Note that the above relations only hold when Eq.~\eqref{eq:S-perp-condition} holds.
We parameterize the Lagrangian perturbation of the pressure with
\begin{align}\label{eq:Deltapbyp}
    \frac{\Delta p}{p} &= \gamma \frac{\Delta n}{n} \,.
\end{align}
%--------------------------------------------------
\section{Relationship between $H(r)$ and the Regge-Wheeler function}\label{appendix:H-to-ZM-RW}
One can show that if $X_{RW,\ell}$ is a solution to the Regge-Wheeler equation
\begin{align}
    &\frac{d^2 X_{RW,\ell}}{d r_{\star}^2} + \left[\omega^2 - V_{RW}(r)  \right] X_{RW,\ell} =0\,,
\end{align}
where $r_{\star} = r + 2M \log(r/2M-1)$ is the tortoise coordinate and $V_{RW}(r)$ is the potential
\begin{align}
    V_{RW}(r) = \left(1-\frac{2M}{r}\right)\left(\frac{\ell(\ell+1)}{r^2} - \frac{6M}{r^3} \right)\,,
\end{align}
then 
\begin{align}\label{eq:}
    H &= \frac{J_0}{r^2 (r-2M)} X_{RW,\ell} + \frac{J_1}{r^2 } \frac{d}{d r}X_{RW,\ell}   \,,
\end{align}
is a solution to the master equation [Eq.~\eqref{eq:master-equation-matrix}] in the region exterior to the star.
The functions $J_0$ and $J_1$ are given by 
\begin{subequations}
\begin{align}
    J_0 &= \ell \left(\ell^3+2 \ell^2-\ell-2\right) M (2 M-r) (\ell (\ell+1) r-6 M)\nonumber\\
    &-2 M r^2 \omega ^2 (6 M-\ell (\ell+1) r) \left(\left(\ell^2+\ell-2\right) r+6 M\right) \,,\\
    J_1 &= 
    2 \ell \left(\ell^3+2 \ell^2-\ell-2\right) M r (3 M-r)+24 M^2 r^3 \omega ^2
    \,.
\end{align}
\end{subequations}
Relating $H$ to the Zerilli-Moncrief function is straightforward using a Chandrasekhar transformation~\cite{1975RSPSA.343..289C}.
%------------------------------------------------------------
\section{Source functions for viscous perturbations}\label{appendix:source-functions}
In this section we provide the expressions for the spherical harmonic decomposition (Eq.~\eqref{eq:spherical-decomposition-source-term}) of the source tensor for bulk and shear viscosity (see Eq.~\eqref{eq:source-tensors-visc}).

\subsection{Bulk viscosity}
The nonzero source functions for the bulk viscous source function [Eq.~\eqref{eq:bulk-viscous-source-tensor}] are given by 
\begin{subequations}\label{eq:sources-spherical-decomp-bulk}
\begin{align}
    \frac{-i S_{\Omega}}{\zeta(p)}
    &=
    -\frac{e^{-\nu /2} r^2 \omega  H_{(0)} (e+p)}{p \gamma}
    +\frac{e^{-3 \nu /2} r^2 \omega ^3
   (e+p) V_{(0)}}{p \gamma}
   \nonumber\\
   &-\frac{e^{-\frac{\lambda }{2}-\frac{\nu }{2}} r \omega  W_{(0)} p'}{p
   \gamma}
   \,,\\
   S_0 &= S_{\Omega} e^{\lambda} \,.
\end{align}
\end{subequations}
We now note the following identities
\begin{subequations}\label{eq:identities-alpha-V-H-8}
\begin{align}
    \alpha_{V,8} e^{\lambda} + \alpha_{V,11} &=0\,,\\
    \alpha_{H,8} e^{\lambda} + \alpha_{H,11} &=0\,.
\end{align}
\end{subequations}
We can use Eqs.~\eqref{eq:sources-spherical-decomp-bulk} and \eqref{eq:identities-alpha-V-H-8} to simplify the first order master equation given in Eq.~\eqref{eq:master-equation-1-order} for bulk viscous perturbations to the following
\begin{subequations}\label{eq:master-equation-simplified-bulk}
\begin{align}
%----------------------------------------------------
W_{(1)}' &=  H_{(1)} \alpha_{W,0}
+W_{(1)} \alpha_{W,1}
+V_{(1)} \alpha_{W,2}
+\alpha_{W,3} H_{(1)}'
\nonumber\\
&
+ 
\alpha_{W,4} S_0 + \alpha_{W,7} S_{\Omega}
\,,\\
%----------------------------------------------------
%----------------------------------------------------
V_{(1)}' &=H_{(1)} \alpha_{V,0}
+W_{(1)} \alpha_{V,1}
+V_{(1)} \alpha_{V,2}
+\alpha_{V,3} H_{(1)}'
\nonumber \\
&
+ \alpha_{V,4} S_0
+ \alpha_{V,7} S_{\Omega}
+ \alpha_{V,8} S_0 \lambda'
\,,\\
%----------------------------------------------------
H_{(1)}'' &= H_{(1)} \alpha_{H,0}
+W_{(1)} \alpha_{H,1}
+V_{(1)} \alpha_{H,2}
+\alpha_{H,3} H_{(1)}'
\nonumber\\
&
+ \alpha_{H,4} S_0
+ \alpha_{H,7} S_{\Omega}
+ \alpha_{H,8} S_0 \lambda'
\,.
\end{align}
\end{subequations}
That is, we are able to eliminate $S_0^{\prime}$ and $S_{\Omega}^{\prime}$ from the master equations. 
Thus for bulk viscous perturbations, we only need $\zeta(p)$ to calculate the solutions to the first order master equations (and not any derivatives of $\zeta(p)$).
Here the subscripts $(1)$ correspond to the solution at first order in the viscosity ($\epsilon$) expansion.

The schematic series solution near the origin for a bulk viscous source ($\eta=0$) depends on the inviscid solution obtained at zeroth order in perturbation theory
\begin{subequations}
\begin{align}
    W_{(1)}(r) &= w_{2,\mathrm{bulk}}(h_0^{(0)}, w_0^{(0)})r^2 + \mathcal{O}(r^4) \,,\\
    V_{(1)}(r) &= v_{2,\mathrm{bulk}}(h_0^{(0)}, w_0^{(0)})r^2 + \mathcal{O}(r^4) \,,\\
    H_{(1)}(r) &= h_{2,\mathrm{bulk}}(h_0^{(0)}, w_0^{(0)})r^2 + \mathcal{O}(r^4) \,.
\end{align}
\end{subequations}
The expressions for $w_{2,\mathrm{bulk}},v_{2,\mathrm{bulk}}$ and $h_{2,\mathrm{bulk}}$ are available in the supplementary mathematica notebook.
%-------------------------------------------------
\subsection{Shear viscosity}
The nonzero source functions for the shear viscous source function [Eq.~\eqref{eq:shear-viscous-source-tensor}] are given by
\begin{subequations}
\begin{align}
    &-\frac{iS_{\Omega}}{\eta(p)}
    = 
    -\frac{1}{3} e^{-\frac{\lambda }{2}-\frac{\nu }{2}} r \omega  \left(2 \alpha_{W,0}+e^{\lambda
   /2} r (2+\alpha_0)\right) H_{(0)}
   \nonumber\\
   &-\frac{1}{3} e^{-\frac{\lambda }{2}-\frac{\nu }{2}} \omega  \left(-4+2 \ell+2 r \alpha_{W,1}+e^{\lambda /2} r^2 \alpha_{1}\right)W_{(0)}
   \nonumber\\
   &-\frac{1}{3} e^{-\frac{\lambda }{2}-\frac{\nu }{2}} \omega  \left(2 r \alpha_{W,2}-e^{\lambda
   /2} \left(\ell+\ell^2-r^2 \alpha_2\right)\right) V_{(0)}
   \nonumber\\
   &-\frac{1}{3} e^{-\frac{\lambda }{2}-\frac{\nu }{2}} \omega  \left(2 r \alpha_{W,3}+e^{\lambda
   /2} r^2 \alpha_3\right) H_{(0)}'
    \,,\\
%-------------------------------------------------------
    &-\frac{iS_Z}{\eta(p)}
    = -2 e^{-\nu /2} \omega  V_{(0)}
    \,,\\
%-------------------------------------------------------
    &-\frac{iS_1}{\eta(p)}
    = 
    -e^{-\nu /2} r \omega  H_{(0)} \alpha_{V,0}
    \nonumber\\
    &+e^{-\nu /2} \omega  W_{(0)} \left(e^{\lambda /2}-r
    \alpha_{V,1}\right)
    -e^{-\nu /2} r \omega 
    \alpha_{V,3} H_{(0)}' 
    \nonumber\\
    &-e^{-\nu /2} \omega  V_{(0)} (-2+\ell+r \alpha_{V,2})
    \,,\\
%-------------------------------------------------------
    &S_{0} = -2 S_{\Omega} e^{\lambda} \,.
\end{align}
\end{subequations}
Unlike the bulk viscous case, the master equations~\eqref{eq:master-equation-1-order} depend on $\eta(p)$, $d \eta/d p$ and $d^2 \eta/d p^2$ because of the appearance of second derivatives of $S_1$ in the master equations \eqref{eq:master-equation-1-order}.

The schematic series solution near the origin for shear viscous source depends on the inviscid solution obtained at zeroth order in perturbation theory
\begin{subequations}
\begin{align}
    W_{(1)}(r) &= w_{2,\mathrm{shear}}(h_0^{(0)}, w_0^{(0)})r^2 + \mathcal{O}(r^4) \,,\\
    V_{(1)}(r) &= v_{2,\mathrm{shear}}(h_0^{(0)}, w_0^{(0)})r^2 + \mathcal{O}(r^4) \,,\\
    H_{(1)}(r) &= h_{2,\mathrm{shear}}(h_0^{(0)}, w_0^{(0)})r^2 + \mathcal{O}(r^4) \,.
\end{align}
\end{subequations}
The expressions for $w_{2,\mathrm{shear}},v_{2,\mathrm{shear}}$ and $h_{2,\mathrm{shear}}$ are available in the supplementary mathematica notebook.
%===========================================================================
%===========================================================================
\section{Method of re-summing the external solution}\label{appendix:resum}
We here present a proof that the re-summation of the external potential performed in Sec.~\ref{sec:tidal-bcs} can be extended to all orders in perturbation theory.
To describe the re-summation procedure, it is useful to switch to a more compact notation.
The master equation for the external gravitational potential can be written in the following schematic form 
\begin{align}
    \mathcal{L}(\varepsilon) H(\varepsilon) &=0\,,
\end{align}
where $\varepsilon = M \omega$, $H(\varepsilon)$ is the gravitational potential, and $\mathcal{L}(\varepsilon)$ is a second-order differential operator.
We next expand the latter two as follows: 
\begin{subequations}
\begin{align}
    \mathcal{L}(\varepsilon) &= \sum_{j=0}^{n} \varepsilon^{2j} \mathcal{L}_{2j} + \mathcal{O}(\varepsilon^{2n+2}) \,,\\
    H(\varepsilon) &= \sum_{j=0}^{n} \varepsilon^{2j} H_{2j} + \mathcal{O}(\varepsilon^{2n+2})\,.
\end{align}
\end{subequations}
We have only included even terms in the expansion of $\mathcal{L}$ because it is even in $\varepsilon$, and, as we argued below Eq.~\eqref{eq:H-odd-even-small-frequency}, the functional form for the gravitational potential with even and odd powers of $\varepsilon$ is the same (so we only include the even powers here for our analysis).  
The arguments given below apply equally well to the terms in the gravitational potential that are odd in $\varepsilon$. 

The perturbative expansion yields the following tower of equations 
\begin{subequations}
\begin{align}
    \mathcal{L}_0 H_0 &= 0\,,\\
    \mathcal{L}_0 H_{2} +  \mathcal{L}_2 H_{0} &= 0\,,\\
    \mathcal{L}_0 H_{2j} + \mathcal{L}_{2j} H_0 + \sum_{i=1}^{j-1} \mathcal{L}_{2i} H_{2j - 2i} &=0\,, j>1\,.
\end{align}
\end{subequations}
We define a set of solutions with the following properties 
\begin{subequations}
\begin{align}
    &\mathcal{L}_0 P = 0 \,,\\
    &\mathcal{L}_0 Q = 0\,, \\
    &\mathcal{L}_0 P_{2} + \mathcal{L}_2 P = 0 \,,\\
    &\mathcal{L}_0 Q_{2} + \mathcal{L}_2 Q = 0 \,,\\ 
    &\mathcal{L}_0 P_{2a_1 2 a_2 \cdots 2 a_j} + \mathcal{L}_{2 a_1} P_{2 a_2 \cdots 2 a_j} = 0\,, j>1\\
    &\mathcal{L}_0 Q_{2a_1 2 a_2 \cdots 2 a_j} + \mathcal{L}_{2 a_1} Q_{2 a_2 \cdots 2 a_j} = 0\,, j>1 \,.
\end{align}
\end{subequations}
The functions $P$, $Q$, $P_2$ and $Q_2$ were defined in the main text as $P=M^{\ell} r^{-\ell} \hat{P}_{\ell}^2$, $Q=M^{\ell} r^{-\ell} \hat{Q}_{\ell}^2$, $P_2 = M^{\ell} r^{-\ell} \mathbb{H}_{P,\ell}^{(2)}$, and $Q_2 = M^{\ell} r^{-\ell} \mathbb{H}_{Q,\ell}^{(2)}$.
We use this compact notation to keep our expressions succinct.
We note that each of the basis solutions $P_{{2a_1 2 a_2 \cdots 2 a_j}}$ is not unique because we can always add a term proportional to $P$ or $Q$, which solves the homogeneous equation.
Similarly, we can add solutions that are proportional to $P_{{2a_1 2 a_2 \cdots 2 a_k}}$ with $k<j$.
One can make a unique choice by demanding additional restriction of the form of $P_{{2a_1 2 a_2 \cdots 2 a_j}}$ and $Q_{{2a_1 2 a_2 \cdots 2 a_j}}$ near spatial infinity.
We make the normalization choice that the asymptotic expansion of $P_{{2a_1 2 a_2 \cdots 2 a_j}} (1-2 M/r) r^{\ell}$ does not contain any term proportional to $\bar{r}^{\ell}$ or $\bar{r}^{-\ell-1}$.
Normalization choices beyond the leading-order condition are difficult to impose without understanding the particular form of the solutions. 
It is well known that only polynomial or logarithmic terms appear in the near zone expansion of the gravitational potential~\cite{Blanchet:2013haa,Sasaki_2003}. Therefore, implementing the normalization procedure for terms which scale as $\log(\bar{r}/M)^{n} r^{m}$ is straightforward in principle but understanding the exact structure would require explicit calculation.
We assume that such sub-leading normalization conditions can be chosen as well in the following discussion.
We demand the same normalization condition for $Q_{{2a_1 2 a_2 \cdots 2 a_j}}$ near spatial infinity.
We denote the normalized solutions by $\hat{N}P_{{2a_1 2 a_2 \cdots 2 a_j}}$ and $\hat{N}Q_{{2a_1 2 a_2 \cdots 2 a_j}}$.
Note that this normalization condition is equivalent to that made in the main body of the paper when $j=2$.

With this unique basis of solutions, we can define the solution at order $j$ in perturbation theory.
For example,
\begin{subequations}
\begin{align}
    H_0 &= a_{p,0} P + a_{Q,0} Q \,,\\
    H_2 &= a_{p,2} P + a_{Q,2} Q + a_{p,0} \hat{N}P_{2} + a_{Q,0} \hat{N}Q_2 \,,\\
    H_4 &= a_{p,4} P + a_{Q,4} Q \nonumber\\
    &+ a_{p,2} \hat{N}P_{2} + a_{Q,2} \hat{N}Q_2 + a_{p,0} \left(\hat{N}P_{2 2} + \hat{N}P_4\right) \nonumber\\
    &+ a_{Q,0} \left( \hat{N}Q_{22} + \hat{N}Q_4 \right) \,,\\
    H_6 &= a_{p,0} \left(\hat{N}P_{6} + \hat{N}P_{42} + \hat{N}P_{24} + \hat{N}P_{222}\right) \nonumber\\
    &+ 
    a_{p,2} \left(\hat{N}P_{4} + \hat{N}P_{22} \right)
    \nonumber\\
    &+a_{p,4} \hat{N}P_{2}
    + a_{p,6} \hat{N}P
    + \left(P \leftrightarrow Q\right) \,.
\end{align}
\end{subequations}
The general solution at $j$th order in perturbation theory is given by 
\begin{align}\label{eq:H2j-sol}
    H_{2j}
    &=
    a_{p,2j} P 
    +
    \sum_{k=0}^{j-1}
    a_{p,2k} \left(\sum_{m=1}^{j-k} \sum_{a_1 + \cdots a_m = j-k} \hat{N} P_{2a_1 \cdots 2 a_m} \right)
    \nonumber\\
    &+
    \left(P \leftrightarrow Q \right) \,.
\end{align}
We introduce 
\begin{subequations}
\begin{align}
    \mathcal{F}(j-k) &\equiv \sum_{m=1}^{j-k} \sum_{a_1 + \cdots a_m = j-k} \hat{N} P_{2a_1 \cdots 2 a_m} \,,\\
    \mathcal{G}(j-k) &\equiv  \sum_{m=1}^{j-k} \sum_{a_1 + \cdots a_m = j-k} \hat{N} Q_{2a_1 \cdots 2 a_m}\,.
\end{align}
\end{subequations}
With these definitions, we can rewrite Eq.~\eqref{eq:H2j-sol} as
\begin{align}
    H_{2j} &= a_{p,2j} P + a_{Q,2j} Q \nonumber\\
    &+ \sum_{k=0}^{j-1} a_{p,2k} \mathcal{F}(j-k) + a_{Q,2k} \mathcal{G}(j-k) \,.
\end{align}
The solution at $n$th-order in perturbation theory is then given by 
\begin{widetext}
\begin{align}\label{eq:Hpert-v1}
    &H(\varepsilon) = \sum_{j=0}^{n} \varepsilon^{2j} H_{2j}
    +
    \mathcal{O}\left(\varepsilon^{2n+2}\right) \,,\nonumber\\
    &= 
    \sum_{j=0}^{n} \left( a_{p,2j} \varepsilon^{2j} \right) P
    +
    \sum_{j=0}^{n} \left( a_{Q,2j} \varepsilon^{2j} \right) Q
    +
    \sum_{j=1}^{n} \varepsilon^{2j} \sum_{k=0}^{j-1} a_{p,2k} \mathcal{F}\left(j-k\right) 
    +
    \sum_{j=1}^{n} \varepsilon^{2j} \sum_{k=0}^{j-1} a_{Q,2k} \mathcal{G}\left(j-k\right)
    +
    \mathcal{O}\left(\varepsilon^{2n+2}\right)
    \,,\nonumber \\
    &=
    \sum_{j=0}^{n} \left( a_{p,2j} \varepsilon^{2j} \right) P
    +
    \sum_{j=0}^{n} \left( a_{Q,2j} \varepsilon^{2j} \right) Q
    +
    \sum_{j=1}^{n} \varepsilon^{2j} \mathcal{F}\left(j\right) 
    \left(\sum_{k=0}^{n-j} a_{p,2k} \varepsilon^{2k} \right)
    +
    \sum_{j=1}^{n} \varepsilon^{2j} \mathcal{G}\left(j\right) 
    \left(\sum_{k=0}^{n-j} a_{Q,2k} \varepsilon^{2k} \right)
    +
    \mathcal{O}\left(\varepsilon^{2n+2}\right)
    \,,\nonumber \\
    &=
    \sum_{j=0}^{n} \left( a_{p,2j} \varepsilon^{2j} \right) P
    +
    \sum_{j=0}^{n} \left( a_{Q,2j} \varepsilon^{2j} \right) Q
    +
    \sum_{j=1}^{n} \varepsilon^{2j} \mathcal{F}\left(j\right) 
    \left(\sum_{k=0}^{n-j} a_{p,2k} \varepsilon^{2k} \right)
    +
    \sum_{j=1}^{n} \varepsilon^{2j} \mathcal{G}\left(j\right) 
    \left(\sum_{k=0}^{n-j} a_{Q,2k} \varepsilon^{2k} \right)
    +
    \mathcal{O}\left(\varepsilon^{2n+2}\right)
    \,.
\end{align}
We now note that adding
\begin{align}
    \sum_{j=1}^{n} \varepsilon^{2j} \mathcal{F}\left(j\right) 
    \left(\sum_{k=n-j+1}^{n} a_{p,2k} \varepsilon^{2k} \right)
    +
    \sum_{j=1}^{n} \varepsilon^{2j} \mathcal{G}\left(j\right) 
    \left(\sum_{k=n-j+1}^{n} a_{Q,2k} \varepsilon^{2k} \right)
    =
    \mathcal{O}\left(\varepsilon^{2n+2}\right) 
\end{align}
does not change the uncontrolled remainder.
Moreover, note that these terms are actually present if we go to higher-order in perturbation theory.
Therefore, we rewrite Eq.~\eqref{eq:Hpert-v1} as 
\begin{align}
    &H(\varepsilon) =
    \sum_{j=0}^{n} \left( a_{p,2j} \varepsilon^{2j} \right) P
    +
    \sum_{j=0}^{n} \left( a_{Q,2j} \varepsilon^{2j} \right) Q
    +
    \sum_{j=1}^{n} \varepsilon^{2j} \mathcal{F}\left(j\right) 
    \sum_{k=0}^{n} \left( a_{p,2k} \varepsilon^{2j} \right)
    +
    \sum_{j=1}^{n} \varepsilon^{2j} \mathcal{G}\left(j\right) 
    \sum_{k=0}^{n} \left( a_{Q,2k} \varepsilon^{2j} \right)
    +
    \mathcal{O}\left(\varepsilon^{2n+2}\right)
    \,,\nonumber\\
    &=
    \left(\sum_{k=0}^{n} a_{p,2k} \varepsilon^{2k}\right)
    \left( P + \sum_{j=1}^{n} \varepsilon^{2j} \mathcal{F}(j) \right)
    +
    \left(\sum_{k=0}^{n} a_{Q,2k} \varepsilon^{2k}\right)
    \left( Q + \sum_{j=1}^{n} \varepsilon^{2j} \mathcal{G}(j) \right)
    +
    \mathcal{O}\left(\varepsilon^{2n+2}\right) 
    \,.
\end{align}
Using the same PN matching procedure as in Sec.~\ref{sec:Matching-PN}, we can identify
\begin{align}
    \sum_{k=0}^{n} \left( a_{p,2j} \varepsilon^{2j} \right)
    &\equiv 
    \frac{4 \pi d_{\ell m}(\omega)}{2\ell + 1} \,,\\
    \sum_{k=0}^{n} \left( a_{Q,2j} \varepsilon^{2j} \right)
    &\equiv 
    \frac{8 \pi d_{\ell m}(\omega) \hat{K}_{\ell}}{(2\ell + 1) C^{2\ell + 1}}\,,
\end{align}
to obtain 
\begin{align}
    H(\varepsilon) 
    &=
    \frac{4 \pi d_{\ell m}}{2\ell + 1}
    \bigg[ 
    \frac{2 \hat{K}_{\ell}(\omega)}{C^{2\ell + 1}}
    \left(Q + \sum_{j=1}^{n} \varepsilon^{2j} \mathcal{G}(j) \right)
    +
    \left(P + \sum_{j=1}^{n} \varepsilon^{2j} \mathcal{F}(j)\right) 
    \bigg]
    +
    \mathcal{O}
    \left( \varepsilon^{2n+2} \right)
    \,.
\end{align}
This matches the expression given in Eq.~\eqref{eq:H_ext_summary} when $n=2$.
We note that we have assumed that the logarithmic corrections are not important when performing this re-summation. 
This assumption is valid because we are matching to a Newtonian potential in the buffer-zone.
For a binary system, this approximation works, provided the stars are not too close to contact.
\end{widetext}
%-------------------------------------------------------
\section{Comparison to previous works}\label{appendix:compare-Eric-to-us}
In this appendix, we compare our prescription for computing the tidal response function with that of Pitre and Poisson (PP)~\cite{Pitre:2023xsr,Poisson:2020vap}.
Let us first note that PP only calculated the conservative tidal response of relativistic stars in the low-frequency regime.
However, the overall approach followed by PP remains is essentially the approach taken in this article except, for the differences which we outline here.

The first difference arises because the particular solution at $\mathcal{O}(\varepsilon^2)$ used by PP is not the same as the normalized particular solution used in this article, as we described in detail in Sec.~\ref{sec:tidal-bcs}.
The particular solution used by PP ($\mathcal{H}^{(2)}_{P/Q,\ell}$) differs from the normalized particular solution ($\hat{N}\mathbb{H}^{(2)}_{P/Q,\ell}$) by a linear combination of the homogeneous solutions
\begin{subequations}\label{eq:PP-potentials}
\begin{align}
    \mathbb{H}_{P,\ell}^{(2)}(r) &= \hat{N}\mathbb{H}_{P,\ell}^{(2)}(r) + a_{1,\ell} \hat{Q}_{\ell}^2\left(\frac{r}{M}-1\right) 
    \nonumber\\
    &+ a_{2,\ell} \hat{P}_{\ell}^2 \left(\frac{r}{M}-1\right)  \,\\
%--------------------------
    \mathbb{H}_{Q,\ell}^{(2)}(r) &= \hat{N}\mathbb{H}_{Q,\ell}^{(2)}(r) + a_{3,\ell} \hat{Q}_{\ell}^2\left(\frac{r}{M}-1\right)  \nonumber\\
    &+ a_{4,\ell} \hat{P}_{\ell}^2\left(\frac{r}{M}-1\right)  \,,
\end{align}
\end{subequations}
where $a_{i,\ell}$ are constants.
The value of the constants $a_{i,2}$ are provided in Eq.~\eqref{eq:ai-vals-PP-2}.

The second difference arises in the definition of the tidal response.
Let us denote the metric potential used by PP by $\mathcal{H}$, the tidal moments by $\mathcal{D}_{\ell m}(\omega)$ and the conservative tidal response function by $k_{\ell,\mathrm{PP}}(\omega)$.
The metric potential used by PP written in our notation is then
\begin{align}\label{eq:H-PP}
    \mathcal{H} &= \frac{4 \pi \mathcal{D}_{\ell m} (\omega) }{2 \ell +1} \frac{M^{\ell}}{r^{\ell}}
    \bigg[ \frac{2k_{\ell,\mathrm{PP}}}{C^{2 \ell +1}} \left(\hat{Q}_{\ell}^2 + (M\omega)^2 \mathbb{H}_{Q,\ell}^{(2)}\right)  \nonumber\\
    &+ \hat{P}^2_{\ell} + (M\omega)^2 \mathbb{H}_{P,\ell}^{(2)}  \bigg]
    \,.
\end{align}
Our metric potential, given in Eq.~\eqref{eq:H_ext_summary}, will be exactly equal to the above expression, but the value of $D_{\ell m}(\omega)$ and $k_{\ell,\mathrm{PP}}(\omega)$ will differ from that of $d_{\ell m}(\omega)$ and $k_{\ell}(\omega)$ at a sufficiently high frequency.
This difference is due to the normalization of the particular solution, the interpretation of the tidal moments and the response used by PP.

Let us now compare the metrics obtained by PP to that found in this paper. Let us denote the $(t,t)$ component of the metric obtained using Eq.~\eqref{eq:H-PP} by $g_{tt,\mathrm{PP}}$.
Outside the star, the metric is given by
\begin{align}
    g_{tt,\mathrm{PP}} &= -\left(1-\frac{2M}{r}\right)
    \left[ 
    1- 2 \mathcal{H}(r) r^{\ell}e^{-i \omega t} Y_{\ell m}
    \right]\,.
\end{align}
Performing an asymptotic expansion of the above metric in the buffer zone, using Eq.~\eqref{eq:H-PP}, we find
\begin{align}\label{eq:gtt-asymptotic-PP}
    &g_{tt,\mathrm{PP}} = 
    -1 + \frac{2 M}{\bar{r}} \nonumber\\
    &+ \frac{8 \pi \mathcal{D}_{\ell m} e^{-i\omega t} \bar{r}^{\ell}}{(2\ell+1)} \left[1 + 
    a_{2,\ell} \varepsilon^2 + a_{4,\ell} \varepsilon^2 \left(\frac{2 k_{\ell, \mathrm{PP}}}{C^{2\ell + 1}} \right)
    \right] \nonumber\\
    &+
    \frac{8 \pi \mathcal{D}_{\ell m}  e^{-i\omega t} M^{2\ell +1} }{\bar{r}^{\ell+1} (2\ell+1) }
    \left[\frac{2 k_{\ell, \mathrm{PP}}}{C^{2\ell + 1}}\left(1+ a_{3,\ell} \varepsilon^2 \right) + a_{1,\ell} \varepsilon^2 \right]
    \nonumber\\
    &+
    \mathcal{O}\left(c^{-4}\right)\,.
\end{align}
The asymptotic expansion of the $(t,t)$ component of the metric obtained by us is given by 
\begin{align}\label{eq:gtt-asymptotic-us}
    g_{tt} &= 
    -1 + \frac{2 M}{\bar{r}} + \frac{8 \pi d_{\ell m}  e^{-i\omega t} \bar{r}^{\ell}}{(2\ell+1)} \nonumber\\
    &+
    \frac{8 \pi d_{\ell m}  e^{-i\omega t} M^{2\ell +1} }{\bar{r}^{\ell+1} (2\ell+1) }
    \left(\frac{2 k_{\ell}}{C^{2\ell + 1}}\right)
    +
    \mathcal{O}\left(\mathrm{1 PN}\right)\,.
\end{align}
Since the metrics must be the same, we equate Eq.~\eqref{eq:gtt-asymptotic-PP} to Eq.~\eqref{eq:gtt-asymptotic-us} to obtain
\begin{align}\label{eq:matching-PP-to-us}
    &\mathcal{D}_{\ell m}(\omega)\left[1 + 
    a_{2,\ell} \varepsilon^2 + a_{4,\ell} \varepsilon^2 \left(\frac{2 k_{\ell, \mathrm{PP}}}{C^{2\ell + 1}} \right)
    \right]
    = 
    d_{\ell m}(\omega) \,,\\
    &\mathcal{D}_{\ell m} \left[\frac{2 k_{\ell, \mathrm{PP}}(\omega)}{C^{2\ell + 1}}\left(1+ a_{3,\ell} \varepsilon^2 \right) + a_{1,\ell} \varepsilon^2 \right]
    =
    \nonumber\\
    &
    \hspace{4cm}
    \frac{2 d_{\ell m}(\omega) k_{\ell}(\omega)}{C^{2\ell + 1}}
    \,.
\end{align}

To invert these expressions we note that PP only defined the tidal moments and the tidal response functions in the low-frequency regime, assuming that a small $\omega$ expansion is different from a small $\varepsilon$ expansion.
Physically, this corresponds to saying that the tidal moments and the tidal response function are slowly-varying quantities on an external time-scale $\omega^{-1}$, which is different from the PN time scale $M^{-1}$.
We also note that PP did not perform any re-summation. We thus invert the above expression first in a small $\varepsilon$ expansion to obtain 
\begin{align}
\label{eq:Delm-dlm-eq-v1}
    &D_{\ell m }(\omega)
    =
    d_{\ell m }(\omega)\left(1 - a_{2,\ell} \varepsilon^2 - \frac{2 k_{\ell}(\omega) a_{4,\ell} \varepsilon^2}{C^{2\ell + 1}} \right) \nonumber\\
    &+ \mathcal{O}\left(\varepsilon^4\right)\,,\\
\label{eq:KlmPP-klm-eq-v1}
    &k_{\ell, PP}(\omega)
    = 
    k_{\ell}(\omega) \left(1 + \left[a_{2,\ell} - a_{3,\ell} + \frac{2 a_{4,\ell} k_{\ell}(\omega)}{C^{2\ell + 1}}\right]\varepsilon^2\right)
    \nonumber\\
    &- \frac{C^{2\ell + 1} a_{1,\ell} \varepsilon^2}{2}
    + \mathcal{O}\left(\varepsilon^4\right)
    \,.
\end{align}

Let us first look at the first equation in the expression above.
To interpret $d_{\ell m}(\omega)$ as the tidal moment we had to match the body metric [Eq.~\eqref{eq:gtt-asymptotic-us}] to a PN metric [Eq.~\eqref{eq:gtt-PN-BZ}].
This matching was done at leading, Newtonian order in this paper, and it was done to 1PN order in~\cite{Poisson:2020vap}.
If one requires that Eq.~\eqref{eq:Delm-dlm-eq-v1} only hold to Newtonian order, then the terms inside the bracket, which are proportional to $\varepsilon^2$, can be said to be of 3PN order, when $a_{2,\ell}$ and $2 k_{\ell}(\omega) a_{4,\ell} \varepsilon^2C^{-2\ell - 1}$ are assumed to scale as 0PN quantities. This assumption was made by PP; we therefore obtain that
\begin{align}\label{eq:matching-PP-to-us-D}
    D_{\ell m }(\omega)
    =
    d_{\ell m }(\omega) + \mathcal{O}\left(\mathrm{3 PN }\right) \,.
\end{align}

Now we apply the same logic to Eq.~\eqref{eq:KlmPP-klm-eq-v1}.
Expanding in small $\omega$, we find
\begin{align}\label{eq:matching-PP-to-us-PN-kell}
    &k_{\ell,\mathrm{PP}}(0) = k_{\ell}(0) \,,\\
    &k_{\ell,\mathrm{PP}}''(0) = k_{\ell}''(0)
    +2 k_{\ell}(0) M^2\left[a_{2,\ell} - a_{3,\ell} + \frac{2 a_{4,\ell} k_{\ell}(\omega)}{C^{2\ell + 1}} \right]\nonumber\\
    &\hspace{2cm}
    -C^{2\ell + 1} a_{1,\ell} M^2
    \,.
\end{align}
We see that, while the adiabatic Love number computed by us agrees with that of PP, the low-frequency dynamical tidal Love number satisfies 
\begin{align}
    k_{\ell,\mathrm{PP}}''(0) = k_{\ell}''(0) + \mathcal{O}\left(\mathrm{3 PN }\right) \,.
\end{align}

Is this disagreement in the low-frequency dynamical tidal Love number a problem? No. 
From a physical standpoint, the invariant function that we must all agree on is the metric potentials $H(r)$ and $\mathcal{H}(r)$ because these functions directly impact observable quantities. As we mentioned above, these two functions are equal to each other irrespective of the definition of the tidal response function.
From a practical point of view, if the star is not extremely compact, the 3PN corrections should be small and the two definitions of the Love numbers will be practically equivalent in the low-frequency regime. 
Since our approach normalizes the particular solution, one can interpret the $\bar{r}^{-\ell-1}$ piece as a tidal response \textit{irrespective} of the order in $\varepsilon$ of the exterior solution.

Moreover, we stress that choosing arbitrary non-zero values for $a_{i,\ell}$ and ignoring the values of these constants as higher PN corrections when matching in the buffer zone [Eqs.~\eqref{eq:matching-PP-to-us}, \eqref{eq:matching-PP-to-us-D} and \eqref{eq:matching-PP-to-us-PN-kell} ] can lead to unphysical solutions even for small frequencies for highly compact objects.
To show this, let us suppose that the values $a_{i,2}$ in Eq.~\eqref{eq:PP-potentials} are given by
\begin{subequations}
\begin{align}
    a_{1,2} &= -\frac{3029}{1575}-\frac{1}{15} 64 \log (2) \,,\\
    a_{2,2} &= \frac{214 \log (2)}{105}-\frac{1037}{45} \,,\\
    a_{3,2} &= \frac{239453}{11025}-\frac{214 \log (2)}{105} \,,\\
    a_{4,2} &=0\,.
\end{align}
\end{subequations}
\begin{figure}
    \centering
    \includegraphics[width = 1\columnwidth]{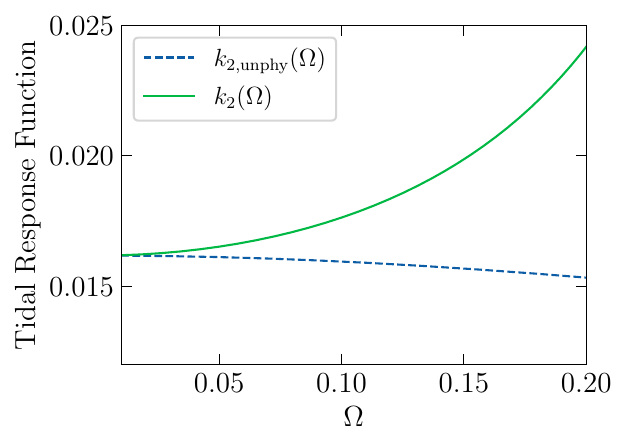}
    \caption{Comparing the tidal response of an non-normalized solution $k_{2,\mathrm{unphy}}(\Omega)$ with that of the normalized solution $k_{2}(\Omega)$ for an extremely relativistic polytrope.
    We see that the tidal response decreases as a function of the dimensionless frequency for the non-normalized solution (dashed blue line) while the normalized solution (solid green line) provides the right physical behavior.
    The unphysical behavior can be traced back to the artificial truncation PN truncation of the tidal moments and the tidal response function in the buffer zone [Eqs.~\eqref{eq:matching-PP-to-us}, \eqref{eq:matching-PP-to-us-D} and \eqref{eq:matching-PP-to-us-PN-kell}].  }
    \label{fig:unphysical-vs-physical}
\end{figure}
\noindent Let us denote the value of the metric potential for these values of $a_{i,\ell}$ by $\mathcal{H}_{\mathrm{unphy}}$, the tidal response function by $k_{\ell,\mathrm{unphy}}$ and the tidal moments by $\mathcal{D}_{\ell m,\mathrm{unphy}} $.
The metric potential is then given by
\begin{align}\label{eq:H-UP}
    \mathcal{H}_{\mathrm{unphy}} &= \frac{4 \pi D_{\ell m,\mathrm{unphy}} (\omega) }{2 \ell +1} \frac{M^{\ell}}{r^{\ell}}
    \bigg[ \nonumber\\
    &\frac{2k_{\ell,\mathrm{unphy}}}{C^{2 \ell +1}} \left(\hat{Q}_{\ell}^2 + (M\omega)^2 \mathbb{H}_{Q,\ell}^{(2)}\right)  \nonumber\\
    &+ \hat{P}^2_{\ell} + (M\omega)^2 \mathbb{H}_{P,\ell}^{(2)}  \bigg]
    \,.
\end{align}
We compare the value of $k_{2,\mathrm{unphy}}(\Omega)$ with the value of the tidal response function obtained using the normalized solution $k_{2}(\Omega)$, for an extremely relativistic polytrope with $n=0.75$, $b = 0.58$ and $C = 0.304$ in Fig.~\ref{fig:unphysical-vs-physical}.
As we can see from the plot, the value of $k_{2,\mathrm{unphy}}(\Omega)$ decreases with $\Omega$ while the normalized (physical) tidal response function $k_{2}(\Omega)$ increases as a function of frequency.
From the value of $k_{2,\mathrm{unphy}}(\Omega)$ one could incorrectly interpret that the star is unstable to tidal interactions in a binary system.
We have verified that this is indeed not true~\cite{talk_with_Elias}.
The unphysical behavior is an artifact of not normalizing the particular solution correctly.
The above example highlights the importance of normalizing the particular solutions to extract the right physical behavior. 
In conclusion, the logic used by PP when matching to the buffer zone PN metric works very well for moderately relativistic stars at small frequencies.
Normalization is essential when extending the method to higher frequencies and to extremely relativistic stars.

Let us conclude this section with a short comparison of our approach and that of~\cite{chakrabarti2013new}.
The authors in~\cite{chakrabarti2013new} also used the master equation approach of~\cite{Lindblom-Mendell-Ipser-1997}, but did not consider stratification due to the presence of $g$ modes or include effects from viscous dissipation.
The treatment of the external problem additionally is different from the approach we followed.
The authors of~\cite{chakrabarti2013new} adopted an approach based on effective field theory and perturbations of the Regge-Wheeler equation using singular sources.
They then solved the Regge-Wheeler equation using analytical method of Mano, Suzuki and Takasugi~\cite{Mano_1996}.
The method used by the authors required the matching and fitting of a re-normalization constant, in addition to the matching of solutions at the surface of the star.
The re-normalization constant appeared due to the presence of a logarithmic term in the near zone expansion. In our work, we fixed a normalization condition from the beginning which simplified the formalism considerably.
Because of these differences, it is not straightforward to do a direct comparison of their work to ours.

%% file: main.bbl
\begin{thebibliography}{85}%
\makeatletter
\providecommand \@ifxundefined [1]{%
 \@ifx{#1\undefined}
}%
\providecommand \@ifnum [1]{%
 \ifnum #1\expandafter \@firstoftwo
 \else \expandafter \@secondoftwo
 \fi
}%
\providecommand \@ifx [1]{%
 \ifx #1\expandafter \@firstoftwo
 \else \expandafter \@secondoftwo
 \fi
}%
\providecommand \natexlab [1]{#1}%
\providecommand \enquote  [1]{``#1''}%
\providecommand \bibnamefont  [1]{#1}%
\providecommand \bibfnamefont [1]{#1}%
\providecommand \citenamefont [1]{#1}%
\providecommand \href@noop [0]{\@secondoftwo}%
\providecommand \href [0]{\begingroup \@sanitize@url \@href}%
\providecommand \@href[1]{\@@startlink{#1}\@@href}%
\providecommand \@@href[1]{\endgroup#1\@@endlink}%
\providecommand \@sanitize@url [0]{\catcode `\\12\catcode `\$12\catcode
  `\&12\catcode `\#12\catcode `\^12\catcode `\_12\catcode `\%12\relax}%
\providecommand \@@startlink[1]{}%
\providecommand \@@endlink[0]{}%
\providecommand \url  [0]{\begingroup\@sanitize@url \@url }%
\providecommand \@url [1]{\endgroup\@href {#1}{\urlprefix }}%
\providecommand \urlprefix  [0]{URL }%
\providecommand \Eprint [0]{\href }%
\providecommand \doibase [0]{http://dx.doi.org/}%
\providecommand \selectlanguage [0]{\@gobble}%
\providecommand \bibinfo  [0]{\@secondoftwo}%
\providecommand \bibfield  [0]{\@secondoftwo}%
\providecommand \translation [1]{[#1]}%
\providecommand \BibitemOpen [0]{}%
\providecommand \bibitemStop [0]{}%
\providecommand \bibitemNoStop [0]{.\EOS\space}%
\providecommand \EOS [0]{\spacefactor3000\relax}%
\providecommand \BibitemShut  [1]{\csname bibitem#1\endcsname}%
\let\auto@bib@innerbib\@empty
%</preamble>
\bibitem [{\citenamefont {\"Ozel}\ and\ \citenamefont
  {Freire}(2016)}]{Ozel:2016oaf}%
  \BibitemOpen
  \bibfield  {author} {\bibinfo {author} {\bibfnamefont {Feryal}\ \bibnamefont
  {\"Ozel}}\ and\ \bibinfo {author} {\bibfnamefont {Paulo}\ \bibnamefont
  {Freire}},\ }\bibfield  {title} {\enquote {\bibinfo {title} {{Masses, Radii,
  and the Equation of State of Neutron Stars}},}\ }\href {\doibase
  10.1146/annurev-astro-081915-023322} {\bibfield  {journal} {\bibinfo
  {journal} {Ann. Rev. Astron. Astrophys.}\ }\textbf {\bibinfo {volume} {54}},\
  \bibinfo {pages} {401--440} (\bibinfo {year} {2016})},\ \Eprint
  {http://arxiv.org/abs/1603.02698} {arXiv:1603.02698 [astro-ph.HE]}
  \BibitemShut {NoStop}%
\bibitem [{\citenamefont {Lattimer}(2021)}]{Lattimer:2021emm}%
  \BibitemOpen
  \bibfield  {author} {\bibinfo {author} {\bibfnamefont {J.~M.}\ \bibnamefont
  {Lattimer}},\ }\bibfield  {title} {\enquote {\bibinfo {title} {{Neutron Stars
  and the Nuclear Matter Equation of State}},}\ }\href {\doibase
  10.1146/annurev-nucl-102419-124827} {\bibfield  {journal} {\bibinfo
  {journal} {Ann. Rev. Nucl. Part. Sci.}\ }\textbf {\bibinfo {volume} {71}},\
  \bibinfo {pages} {433--464} (\bibinfo {year} {2021})}\BibitemShut {NoStop}%
\bibitem [{\citenamefont {Burgio}\ \emph {et~al.}(2021)\citenamefont {Burgio},
  \citenamefont {Schulze}, \citenamefont {Vidana},\ and\ \citenamefont
  {Wei}}]{Burgio:2021vgk}%
  \BibitemOpen
  \bibfield  {author} {\bibinfo {author} {\bibfnamefont {G.~F.}\ \bibnamefont
  {Burgio}}, \bibinfo {author} {\bibfnamefont {H.~J.}\ \bibnamefont {Schulze}},
  \bibinfo {author} {\bibfnamefont {I.}~\bibnamefont {Vidana}}, \ and\ \bibinfo
  {author} {\bibfnamefont {J.~B.}\ \bibnamefont {Wei}},\ }\bibfield  {title}
  {\enquote {\bibinfo {title} {{Neutron stars and the nuclear equation of
  state}},}\ }\href {\doibase 10.1016/j.ppnp.2021.103879} {\bibfield  {journal}
  {\bibinfo  {journal} {Prog. Part. Nucl. Phys.}\ }\textbf {\bibinfo {volume}
  {120}},\ \bibinfo {pages} {103879} (\bibinfo {year} {2021})},\ \Eprint
  {http://arxiv.org/abs/2105.03747} {arXiv:2105.03747 [nucl-th]} \BibitemShut
  {NoStop}%
\bibitem [{\citenamefont {Flanagan}\ and\ \citenamefont
  {Hinderer}(2008)}]{Flanagan:2007ix}%
  \BibitemOpen
  \bibfield  {author} {\bibinfo {author} {\bibfnamefont {Eanna~E.}\
  \bibnamefont {Flanagan}}\ and\ \bibinfo {author} {\bibfnamefont {Tanja}\
  \bibnamefont {Hinderer}},\ }\bibfield  {title} {\enquote {\bibinfo {title}
  {{Constraining neutron star tidal Love numbers with gravitational wave
  detectors}},}\ }\href {\doibase 10.1103/PhysRevD.77.021502} {\bibfield
  {journal} {\bibinfo  {journal} {Phys. Rev. D}\ }\textbf {\bibinfo {volume}
  {77}},\ \bibinfo {pages} {021502} (\bibinfo {year} {2008})},\ \Eprint
  {http://arxiv.org/abs/0709.1915} {arXiv:0709.1915 [astro-ph]} \BibitemShut
  {NoStop}%
\bibitem [{\citenamefont {Andersson}\ and\ \citenamefont
  {Pnigouras}(2021)}]{andersson2021phenomenology}%
  \BibitemOpen
  \bibfield  {author} {\bibinfo {author} {\bibfnamefont {N.}~\bibnamefont
  {Andersson}}\ and\ \bibinfo {author} {\bibfnamefont {P.}~\bibnamefont
  {Pnigouras}},\ }\href@noop {} {\enquote {\bibinfo {title} {The phenomenology
  of dynamical neutron star tides},}\ } (\bibinfo {year} {2021}),\ \Eprint
  {http://arxiv.org/abs/1905.00012} {arXiv:1905.00012 [gr-qc]} \BibitemShut
  {NoStop}%
\bibitem [{\citenamefont {Hinderer}\ \emph {et~al.}(2016)\citenamefont
  {Hinderer}, \citenamefont {Taracchini}, \citenamefont {Foucart},
  \citenamefont {Buonanno}, \citenamefont {Steinhoff}, \citenamefont {Duez},
  \citenamefont {Kidder}, \citenamefont {Pfeiffer}, \citenamefont {Scheel},
  \citenamefont {Szilagyi}, \citenamefont {Hotokezaka}, \citenamefont
  {Kyutoku}, \citenamefont {Shibata},\ and\ \citenamefont
  {Carpenter}}]{Hinderer_2016}%
  \BibitemOpen
  \bibfield  {author} {\bibinfo {author} {\bibfnamefont {Tanja}\ \bibnamefont
  {Hinderer}}, \bibinfo {author} {\bibfnamefont {Andrea}\ \bibnamefont
  {Taracchini}}, \bibinfo {author} {\bibfnamefont {Francois}\ \bibnamefont
  {Foucart}}, \bibinfo {author} {\bibfnamefont {Alessandra}\ \bibnamefont
  {Buonanno}}, \bibinfo {author} {\bibfnamefont {Jan}\ \bibnamefont
  {Steinhoff}}, \bibinfo {author} {\bibfnamefont {Matthew}\ \bibnamefont
  {Duez}}, \bibinfo {author} {\bibfnamefont {Lawrence~E.}\ \bibnamefont
  {Kidder}}, \bibinfo {author} {\bibfnamefont {Harald~P.}\ \bibnamefont
  {Pfeiffer}}, \bibinfo {author} {\bibfnamefont {Mark~A.}\ \bibnamefont
  {Scheel}}, \bibinfo {author} {\bibfnamefont {Bela}\ \bibnamefont {Szilagyi}},
  \bibinfo {author} {\bibfnamefont {Kenta}\ \bibnamefont {Hotokezaka}},
  \bibinfo {author} {\bibfnamefont {Koutarou}\ \bibnamefont {Kyutoku}},
  \bibinfo {author} {\bibfnamefont {Masaru}\ \bibnamefont {Shibata}}, \ and\
  \bibinfo {author} {\bibfnamefont {Cory~W.}\ \bibnamefont {Carpenter}},\
  }\bibfield  {title} {\enquote {\bibinfo {title} {Effects of neutron-star
  dynamic tides on gravitational waveforms within the effective-one-body
  approach},}\ }\href {\doibase 10.1103/physrevlett.116.181101} {\bibfield
  {journal} {\bibinfo  {journal} {Physical Review Letters}\ }\textbf {\bibinfo
  {volume} {116}} (\bibinfo {year} {2016}),\
  10.1103/physrevlett.116.181101}\BibitemShut {NoStop}%
\bibitem [{\citenamefont {Steinhoff}\ \emph
  {et~al.}(2016{\natexlab{a}})\citenamefont {Steinhoff}, \citenamefont
  {Hinderer}, \citenamefont {Buonanno},\ and\ \citenamefont
  {Taracchini}}]{Steinhoff_2016}%
  \BibitemOpen
  \bibfield  {author} {\bibinfo {author} {\bibfnamefont {Jan}\ \bibnamefont
  {Steinhoff}}, \bibinfo {author} {\bibfnamefont {Tanja}\ \bibnamefont
  {Hinderer}}, \bibinfo {author} {\bibfnamefont {Alessandra}\ \bibnamefont
  {Buonanno}}, \ and\ \bibinfo {author} {\bibfnamefont {Andrea}\ \bibnamefont
  {Taracchini}},\ }\bibfield  {title} {\enquote {\bibinfo {title} {Dynamical
  tides in general relativity: Effective action and effective-one-body
  hamiltonian},}\ }\href {\doibase 10.1103/physrevd.94.104028} {\bibfield
  {journal} {\bibinfo  {journal} {Physical Review D}\ }\textbf {\bibinfo
  {volume} {94}} (\bibinfo {year} {2016}{\natexlab{a}}),\
  10.1103/physrevd.94.104028}\BibitemShut {NoStop}%
\bibitem [{\citenamefont {Gupta}\ \emph {et~al.}(2021)\citenamefont {Gupta},
  \citenamefont {Steinhoff},\ and\ \citenamefont {Hinderer}}]{Gupta_2021}%
  \BibitemOpen
  \bibfield  {author} {\bibinfo {author} {\bibfnamefont {Pawan~Kumar}\
  \bibnamefont {Gupta}}, \bibinfo {author} {\bibfnamefont {Jan}\ \bibnamefont
  {Steinhoff}}, \ and\ \bibinfo {author} {\bibfnamefont {Tanja}\ \bibnamefont
  {Hinderer}},\ }\bibfield  {title} {\enquote {\bibinfo {title} {Relativistic
  effective action of dynamical gravitomagnetic tides for slowly rotating
  neutron stars},}\ }\href {\doibase 10.1103/physrevresearch.3.013147}
  {\bibfield  {journal} {\bibinfo  {journal} {Physical Review Research}\
  }\textbf {\bibinfo {volume} {3}} (\bibinfo {year} {2021}),\
  10.1103/physrevresearch.3.013147}\BibitemShut {NoStop}%
\bibitem [{\citenamefont {Abac}\ \emph {et~al.}(2023)\citenamefont {Abac},
  \citenamefont {Dietrich}, \citenamefont {Buonanno}, \citenamefont
  {Steinhoff},\ and\ \citenamefont {Ujevic}}]{abac2023nrtidalv3}%
  \BibitemOpen
  \bibfield  {author} {\bibinfo {author} {\bibfnamefont {Adrian}\ \bibnamefont
  {Abac}}, \bibinfo {author} {\bibfnamefont {Tim}\ \bibnamefont {Dietrich}},
  \bibinfo {author} {\bibfnamefont {Alessandra}\ \bibnamefont {Buonanno}},
  \bibinfo {author} {\bibfnamefont {Jan}\ \bibnamefont {Steinhoff}}, \ and\
  \bibinfo {author} {\bibfnamefont {Maximiliano}\ \bibnamefont {Ujevic}},\
  }\href@noop {} {\enquote {\bibinfo {title} {Nrtidalv3: A new and robust
  gravitational-waveform model for high-mass-ratio binary neutron star systems
  with dynamical tidal effects},}\ } (\bibinfo {year} {2023}),\ \Eprint
  {http://arxiv.org/abs/2311.07456} {arXiv:2311.07456 [gr-qc]} \BibitemShut
  {NoStop}%
\bibitem [{\citenamefont {Schmidt}\ and\ \citenamefont
  {Hinderer}(2019)}]{PhysRevD.100.021501}%
  \BibitemOpen
  \bibfield  {author} {\bibinfo {author} {\bibfnamefont {Patricia}\
  \bibnamefont {Schmidt}}\ and\ \bibinfo {author} {\bibfnamefont {Tanja}\
  \bibnamefont {Hinderer}},\ }\bibfield  {title} {\enquote {\bibinfo {title}
  {Frequency domain model of $f$-mode dynamic tides in gravitational waveforms
  from compact binary inspirals},}\ }\href {\doibase
  10.1103/PhysRevD.100.021501} {\bibfield  {journal} {\bibinfo  {journal}
  {Phys. Rev. D}\ }\textbf {\bibinfo {volume} {100}},\ \bibinfo {pages}
  {021501} (\bibinfo {year} {2019})}\BibitemShut {NoStop}%
\bibitem [{\citenamefont {Pratten}\ \emph {et~al.}(2020)\citenamefont
  {Pratten}, \citenamefont {Schmidt},\ and\ \citenamefont
  {Hinderer}}]{Pratten_2020}%
  \BibitemOpen
  \bibfield  {author} {\bibinfo {author} {\bibfnamefont {Geraint}\ \bibnamefont
  {Pratten}}, \bibinfo {author} {\bibfnamefont {Patricia}\ \bibnamefont
  {Schmidt}}, \ and\ \bibinfo {author} {\bibfnamefont {Tanja}\ \bibnamefont
  {Hinderer}},\ }\bibfield  {title} {\enquote {\bibinfo {title}
  {Gravitational-wave asteroseismology with fundamental modes from compact
  binary inspirals},}\ }\href {\doibase 10.1038/s41467-020-15984-5} {\bibfield
  {journal} {\bibinfo  {journal} {Nature Communications}\ }\textbf {\bibinfo
  {volume} {11}} (\bibinfo {year} {2020}),\
  10.1038/s41467-020-15984-5}\BibitemShut {NoStop}%
\bibitem [{\citenamefont {Alford}\ \emph {et~al.}(2018)\citenamefont {Alford},
  \citenamefont {Bovard}, \citenamefont {Hanauske}, \citenamefont {Rezzolla},\
  and\ \citenamefont {Schwenzer}}]{Alford:2017rxf}%
  \BibitemOpen
  \bibfield  {author} {\bibinfo {author} {\bibfnamefont {Mark~G.}\ \bibnamefont
  {Alford}}, \bibinfo {author} {\bibfnamefont {Luke}\ \bibnamefont {Bovard}},
  \bibinfo {author} {\bibfnamefont {Matthias}\ \bibnamefont {Hanauske}},
  \bibinfo {author} {\bibfnamefont {Luciano}\ \bibnamefont {Rezzolla}}, \ and\
  \bibinfo {author} {\bibfnamefont {Kai}\ \bibnamefont {Schwenzer}},\
  }\bibfield  {title} {\enquote {\bibinfo {title} {{Viscous Dissipation and
  Heat Conduction in Binary Neutron-Star Mergers}},}\ }\href {\doibase
  10.1103/PhysRevLett.120.041101} {\bibfield  {journal} {\bibinfo  {journal}
  {Phys. Rev. Lett.}\ }\textbf {\bibinfo {volume} {120}},\ \bibinfo {pages}
  {041101} (\bibinfo {year} {2018})},\ \Eprint
  {http://arxiv.org/abs/1707.09475} {arXiv:1707.09475 [gr-qc]} \BibitemShut
  {NoStop}%
\bibitem [{\citenamefont {Alford}\ and\ \citenamefont
  {Harris}(2019)}]{Alford:2019qtm}%
  \BibitemOpen
  \bibfield  {author} {\bibinfo {author} {\bibfnamefont {Mark~G.}\ \bibnamefont
  {Alford}}\ and\ \bibinfo {author} {\bibfnamefont {Steven~P.}\ \bibnamefont
  {Harris}},\ }\bibfield  {title} {\enquote {\bibinfo {title} {{Damping of
  density oscillations in neutrino-transparent nuclear matter}},}\ }\href
  {\doibase 10.1103/PhysRevC.100.035803} {\bibfield  {journal} {\bibinfo
  {journal} {Phys. Rev. C}\ }\textbf {\bibinfo {volume} {100}},\ \bibinfo
  {pages} {035803} (\bibinfo {year} {2019})},\ \Eprint
  {http://arxiv.org/abs/1907.03795} {arXiv:1907.03795 [nucl-th]} \BibitemShut
  {NoStop}%
\bibitem [{\citenamefont {Most}\ \emph {et~al.}(2021)\citenamefont {Most},
  \citenamefont {Harris}, \citenamefont {Plumberg}, \citenamefont {Alford},
  \citenamefont {Noronha}, \citenamefont {Noronha-Hostler}, \citenamefont
  {Pretorius}, \citenamefont {Witek},\ and\ \citenamefont
  {Yunes}}]{Most:2021zvc}%
  \BibitemOpen
  \bibfield  {author} {\bibinfo {author} {\bibfnamefont {Elias~R.}\
  \bibnamefont {Most}}, \bibinfo {author} {\bibfnamefont {Steven~P.}\
  \bibnamefont {Harris}}, \bibinfo {author} {\bibfnamefont {Christopher}\
  \bibnamefont {Plumberg}}, \bibinfo {author} {\bibfnamefont {Mark~G.}\
  \bibnamefont {Alford}}, \bibinfo {author} {\bibfnamefont {Jorge}\
  \bibnamefont {Noronha}}, \bibinfo {author} {\bibfnamefont {Jacquelyn}\
  \bibnamefont {Noronha-Hostler}}, \bibinfo {author} {\bibfnamefont {Frans}\
  \bibnamefont {Pretorius}}, \bibinfo {author} {\bibfnamefont {Helvi}\
  \bibnamefont {Witek}}, \ and\ \bibinfo {author} {\bibfnamefont {Nicol\'as}\
  \bibnamefont {Yunes}},\ }\bibfield  {title} {\enquote {\bibinfo {title}
  {{Projecting the likely importance of weak-interaction-driven bulk viscosity
  in neutron star mergers}},}\ }\href {\doibase 10.1093/mnras/stab2793}
  {\bibfield  {journal} {\bibinfo  {journal} {Mon. Not. Roy. Astron. Soc.}\
  }\textbf {\bibinfo {volume} {509}},\ \bibinfo {pages} {1096--1108} (\bibinfo
  {year} {2021})},\ \Eprint {http://arxiv.org/abs/2107.05094} {arXiv:2107.05094
  [astro-ph.HE]} \BibitemShut {NoStop}%
\bibitem [{\citenamefont {Most}\ \emph {et~al.}(2022)\citenamefont {Most},
  \citenamefont {Haber}, \citenamefont {Harris}, \citenamefont {Zhang},
  \citenamefont {Alford},\ and\ \citenamefont {Noronha}}]{Most:2022yhe}%
  \BibitemOpen
  \bibfield  {author} {\bibinfo {author} {\bibfnamefont {Elias~R.}\
  \bibnamefont {Most}}, \bibinfo {author} {\bibfnamefont {Alexander}\
  \bibnamefont {Haber}}, \bibinfo {author} {\bibfnamefont {Steven~P.}\
  \bibnamefont {Harris}}, \bibinfo {author} {\bibfnamefont {Ziyuan}\
  \bibnamefont {Zhang}}, \bibinfo {author} {\bibfnamefont {Mark~G.}\
  \bibnamefont {Alford}}, \ and\ \bibinfo {author} {\bibfnamefont {Jorge}\
  \bibnamefont {Noronha}},\ }\bibfield  {title} {\enquote {\bibinfo {title}
  {{Emergence of microphysical viscosity in binary neutron star post-merger
  dynamics}},}\ }\href@noop {} {\  (\bibinfo {year} {2022})},\ \Eprint
  {http://arxiv.org/abs/2207.00442} {arXiv:2207.00442 [astro-ph.HE]}
  \BibitemShut {NoStop}%
\bibitem [{\citenamefont {Chabanov}\ and\ \citenamefont
  {Rezzolla}(2023{\natexlab{a}})}]{Chabanov:2023blf}%
  \BibitemOpen
  \bibfield  {author} {\bibinfo {author} {\bibfnamefont {Michail}\ \bibnamefont
  {Chabanov}}\ and\ \bibinfo {author} {\bibfnamefont {Luciano}\ \bibnamefont
  {Rezzolla}},\ }\bibfield  {title} {\enquote {\bibinfo {title} {{Impact of
  bulk viscosity on the post-merger gravitational-wave signal from merging
  neutron stars}},}\ }\href@noop {} {\  (\bibinfo {year}
  {2023}{\natexlab{a}})},\ \Eprint {http://arxiv.org/abs/2307.10464}
  {arXiv:2307.10464 [gr-qc]} \BibitemShut {NoStop}%
\bibitem [{\citenamefont {Jones}(2001)}]{Jones:2001ya}%
  \BibitemOpen
  \bibfield  {author} {\bibinfo {author} {\bibfnamefont {P.~B.}\ \bibnamefont
  {Jones}},\ }\bibfield  {title} {\enquote {\bibinfo {title} {{Bulk viscosity
  of neutron star matter}},}\ }\href {\doibase 10.1103/PhysRevD.64.084003}
  {\bibfield  {journal} {\bibinfo  {journal} {Phys. Rev. D}\ }\textbf {\bibinfo
  {volume} {64}},\ \bibinfo {pages} {084003} (\bibinfo {year}
  {2001})}\BibitemShut {NoStop}%
\bibitem [{\citenamefont {Lindblom}\ and\ \citenamefont
  {Owen}(2002)}]{Lindblom:2001hd}%
  \BibitemOpen
  \bibfield  {author} {\bibinfo {author} {\bibfnamefont {Lee}\ \bibnamefont
  {Lindblom}}\ and\ \bibinfo {author} {\bibfnamefont {Benjamin~J.}\
  \bibnamefont {Owen}},\ }\bibfield  {title} {\enquote {\bibinfo {title}
  {{Effect of hyperon bulk viscosity on neutron star r modes}},}\ }\href
  {\doibase 10.1103/PhysRevD.65.063006} {\bibfield  {journal} {\bibinfo
  {journal} {Phys. Rev. D}\ }\textbf {\bibinfo {volume} {65}},\ \bibinfo
  {pages} {063006} (\bibinfo {year} {2002})},\ \Eprint
  {http://arxiv.org/abs/astro-ph/0110558} {arXiv:astro-ph/0110558} \BibitemShut
  {NoStop}%
\bibitem [{\citenamefont {Gusakov}\ and\ \citenamefont
  {Kantor}(2008)}]{Gusakov:2008hv}%
  \BibitemOpen
  \bibfield  {author} {\bibinfo {author} {\bibfnamefont {M.~E.}\ \bibnamefont
  {Gusakov}}\ and\ \bibinfo {author} {\bibfnamefont {E.~M.}\ \bibnamefont
  {Kantor}},\ }\bibfield  {title} {\enquote {\bibinfo {title} {{Bulk viscosity
  of superfluid hyperon stars}},}\ }\href {\doibase 10.1103/PhysRevD.78.083006}
  {\bibfield  {journal} {\bibinfo  {journal} {Phys. Rev. D}\ }\textbf {\bibinfo
  {volume} {78}},\ \bibinfo {pages} {083006} (\bibinfo {year} {2008})},\
  \Eprint {http://arxiv.org/abs/0806.4914} {arXiv:0806.4914 [astro-ph]}
  \BibitemShut {NoStop}%
\bibitem [{\citenamefont {Alford}\ and\ \citenamefont
  {Haber}(2021)}]{Alford:2020pld}%
  \BibitemOpen
  \bibfield  {author} {\bibinfo {author} {\bibfnamefont {Mark~G.}\ \bibnamefont
  {Alford}}\ and\ \bibinfo {author} {\bibfnamefont {Alexander}\ \bibnamefont
  {Haber}},\ }\bibfield  {title} {\enquote {\bibinfo {title}
  {{Strangeness-changing Rates and Hyperonic Bulk Viscosity in Neutron Star
  Mergers}},}\ }\href {\doibase 10.1103/PhysRevC.103.045810} {\bibfield
  {journal} {\bibinfo  {journal} {Phys. Rev. C}\ }\textbf {\bibinfo {volume}
  {103}},\ \bibinfo {pages} {045810} (\bibinfo {year} {2021})},\ \Eprint
  {http://arxiv.org/abs/2009.05181} {arXiv:2009.05181 [nucl-th]} \BibitemShut
  {NoStop}%
\bibitem [{\citenamefont {Lai}(1994)}]{Lai:1993di}%
  \BibitemOpen
  \bibfield  {author} {\bibinfo {author} {\bibfnamefont {Dong}\ \bibnamefont
  {Lai}},\ }\bibfield  {title} {\enquote {\bibinfo {title} {{Resonant
  oscillations and tidal heating in coalescing binary neutron stars}},}\ }\href
  {\doibase 10.1093/mnras/270.3.611} {\bibfield  {journal} {\bibinfo  {journal}
  {Mon. Not. Roy. Astron. Soc.}\ }\textbf {\bibinfo {volume} {270}},\ \bibinfo
  {pages} {611} (\bibinfo {year} {1994})},\ \Eprint
  {http://arxiv.org/abs/astro-ph/9404062} {arXiv:astro-ph/9404062} \BibitemShut
  {NoStop}%
\bibitem [{\citenamefont {Andersson}\ and\ \citenamefont
  {Pnigouras}(2020)}]{Andersson_2020}%
  \BibitemOpen
  \bibfield  {author} {\bibinfo {author} {\bibfnamefont {N.}~\bibnamefont
  {Andersson}}\ and\ \bibinfo {author} {\bibfnamefont {P.}~\bibnamefont
  {Pnigouras}},\ }\bibfield  {title} {\enquote {\bibinfo {title} {Exploring the
  effective tidal deformability of neutron stars},}\ }\href {\doibase
  10.1103/physrevd.101.083001} {\bibfield  {journal} {\bibinfo  {journal}
  {Physical Review D}\ }\textbf {\bibinfo {volume} {101}} (\bibinfo {year}
  {2020}),\ 10.1103/physrevd.101.083001}\BibitemShut {NoStop}%
\bibitem [{\citenamefont {Ogilvie}(2014)}]{Ogilvie-review}%
  \BibitemOpen
  \bibfield  {author} {\bibinfo {author} {\bibfnamefont {Gordon~I.}\
  \bibnamefont {Ogilvie}},\ }\bibfield  {title} {\enquote {\bibinfo {title}
  {Tidal dissipation in stars and giant planets},}\ }\href {\doibase
  10.1146/annurev-astro-081913-035941} {\bibfield  {journal} {\bibinfo
  {journal} {Annual Review of Astronomy and Astrophysics}\ }\textbf {\bibinfo
  {volume} {52}},\ \bibinfo {pages} {171--210} (\bibinfo {year}
  {2014})}\BibitemShut {NoStop}%
\bibitem [{\citenamefont {Hinderer}(2008)}]{Hinderer:2007mb}%
  \BibitemOpen
  \bibfield  {author} {\bibinfo {author} {\bibfnamefont {Tanja}\ \bibnamefont
  {Hinderer}},\ }\bibfield  {title} {\enquote {\bibinfo {title} {{Tidal Love
  numbers of neutron stars}},}\ }\href {\doibase 10.1086/533487} {\bibfield
  {journal} {\bibinfo  {journal} {Astrophys. J.}\ }\textbf {\bibinfo {volume}
  {677}},\ \bibinfo {pages} {1216--1220} (\bibinfo {year} {2008})},\ \Eprint
  {http://arxiv.org/abs/0711.2420} {arXiv:0711.2420 [astro-ph]} \BibitemShut
  {NoStop}%
\bibitem [{\citenamefont {Binnington}\ and\ \citenamefont
  {Poisson}(2009)}]{Binnington:2009bb}%
  \BibitemOpen
  \bibfield  {author} {\bibinfo {author} {\bibfnamefont {Taylor}\ \bibnamefont
  {Binnington}}\ and\ \bibinfo {author} {\bibfnamefont {Eric}\ \bibnamefont
  {Poisson}},\ }\bibfield  {title} {\enquote {\bibinfo {title} {{Relativistic
  theory of tidal Love numbers}},}\ }\href {\doibase
  10.1103/PhysRevD.80.084018} {\bibfield  {journal} {\bibinfo  {journal} {Phys.
  Rev. D}\ }\textbf {\bibinfo {volume} {80}},\ \bibinfo {pages} {084018}
  (\bibinfo {year} {2009})},\ \Eprint {http://arxiv.org/abs/0906.1366}
  {arXiv:0906.1366 [gr-qc]} \BibitemShut {NoStop}%
\bibitem [{\citenamefont {Damour}\ and\ \citenamefont
  {Nagar}(2009)}]{Damour:2009vw}%
  \BibitemOpen
  \bibfield  {author} {\bibinfo {author} {\bibfnamefont {Thibault}\
  \bibnamefont {Damour}}\ and\ \bibinfo {author} {\bibfnamefont {Alessandro}\
  \bibnamefont {Nagar}},\ }\bibfield  {title} {\enquote {\bibinfo {title}
  {{Relativistic tidal properties of neutron stars}},}\ }\href {\doibase
  10.1103/PhysRevD.80.084035} {\bibfield  {journal} {\bibinfo  {journal} {Phys.
  Rev. D}\ }\textbf {\bibinfo {volume} {80}},\ \bibinfo {pages} {084035}
  (\bibinfo {year} {2009})},\ \Eprint {http://arxiv.org/abs/0906.0096}
  {arXiv:0906.0096 [gr-qc]} \BibitemShut {NoStop}%
\bibitem [{\citenamefont {Ripley}\ \emph
  {et~al.}(2023{\natexlab{a}})\citenamefont {Ripley}, \citenamefont {Hegade
  K.~R.},\ and\ \citenamefont {Yunes}}]{Ripley:2023qxo}%
  \BibitemOpen
  \bibfield  {author} {\bibinfo {author} {\bibfnamefont {Justin~L.}\
  \bibnamefont {Ripley}}, \bibinfo {author} {\bibfnamefont {Abhishek}\
  \bibnamefont {Hegade K.~R.}}, \ and\ \bibinfo {author} {\bibfnamefont
  {Nicolas}\ \bibnamefont {Yunes}},\ }\bibfield  {title} {\enquote {\bibinfo
  {title} {{Probing internal dissipative processes of neutron stars with
  gravitational waves during the inspiral of neutron star binaries}},}\
  }\href@noop {} {\  (\bibinfo {year} {2023}{\natexlab{a}})},\ \Eprint
  {http://arxiv.org/abs/2306.15633} {arXiv:2306.15633 [gr-qc]} \BibitemShut
  {NoStop}%
\bibitem [{\citenamefont {{Cox}}(1980)}]{Cox-book}%
  \BibitemOpen
  \bibfield  {author} {\bibinfo {author} {\bibfnamefont {John~P.}\ \bibnamefont
  {{Cox}}},\ }\href@noop {} {\emph {\bibinfo {title} {{Theory of Stellar
  Pulsation. (PSA-2), Volume 2}}}},\ Vol.~\bibinfo {volume} {2}\ (\bibinfo
  {year} {1980})\BibitemShut {NoStop}%
\bibitem [{\citenamefont {Lindblom}\ \emph {et~al.}(1997)\citenamefont
  {Lindblom}, \citenamefont {Mendell},\ and\ \citenamefont
  {Ipser}}]{Lindblom-Mendell-Ipser-1997}%
  \BibitemOpen
  \bibfield  {author} {\bibinfo {author} {\bibfnamefont {Lee}\ \bibnamefont
  {Lindblom}}, \bibinfo {author} {\bibfnamefont {Gregory}\ \bibnamefont
  {Mendell}}, \ and\ \bibinfo {author} {\bibfnamefont {James~R.}\ \bibnamefont
  {Ipser}},\ }\bibfield  {title} {\enquote {\bibinfo {title} {Relativistic
  stellar pulsations with near-zone boundary conditions},}\ }\href {\doibase
  10.1103/PhysRevD.56.2118} {\bibfield  {journal} {\bibinfo  {journal} {Phys.
  Rev. D}\ }\textbf {\bibinfo {volume} {56}},\ \bibinfo {pages} {2118--2126}
  (\bibinfo {year} {1997})}\BibitemShut {NoStop}%
\bibitem [{\citenamefont {Poisson}(2021)}]{Poisson:2020vap}%
  \BibitemOpen
  \bibfield  {author} {\bibinfo {author} {\bibfnamefont {Eric}\ \bibnamefont
  {Poisson}},\ }\bibfield  {title} {\enquote {\bibinfo {title} {{Compact body
  in a tidal environment: New types of relativistic Love numbers, and a
  post-Newtonian operational definition for tidally induced multipole
  moments}},}\ }\href {\doibase 10.1103/PhysRevD.103.064023} {\bibfield
  {journal} {\bibinfo  {journal} {Phys. Rev. D}\ }\textbf {\bibinfo {volume}
  {103}},\ \bibinfo {pages} {064023} (\bibinfo {year} {2021})},\ \Eprint
  {http://arxiv.org/abs/2012.10184} {arXiv:2012.10184 [gr-qc]} \BibitemShut
  {NoStop}%
\bibitem [{\citenamefont {{Terquem}}\ \emph {et~al.}(1998)\citenamefont
  {{Terquem}}, \citenamefont {{Papaloizou}}, \citenamefont {{Nelson}},\ and\
  \citenamefont {{Lin}}}]{Terquem_1998}%
  \BibitemOpen
  \bibfield  {author} {\bibinfo {author} {\bibfnamefont {C.}~\bibnamefont
  {{Terquem}}}, \bibinfo {author} {\bibfnamefont {J.~C.~B.}\ \bibnamefont
  {{Papaloizou}}}, \bibinfo {author} {\bibfnamefont {R.~P.}\ \bibnamefont
  {{Nelson}}}, \ and\ \bibinfo {author} {\bibfnamefont {D.~N.~C.}\ \bibnamefont
  {{Lin}}},\ }\bibfield  {title} {\enquote {\bibinfo {title} {{On the Tidal
  Interaction of a Solar-Type Star with an Orbiting Companion: Excitation of
  g-Mode Oscillation and Orbital Evolution}},}\ }\href {\doibase
  10.1086/305927} {\bibfield  {journal} {\bibinfo  {journal} {\apj}\ }\textbf
  {\bibinfo {volume} {502}},\ \bibinfo {pages} {788--801} (\bibinfo {year}
  {1998})},\ \Eprint {http://arxiv.org/abs/astro-ph/9801280}
  {arXiv:astro-ph/9801280 [astro-ph]} \BibitemShut {NoStop}%
\bibitem [{\citenamefont {Pitre}\ and\ \citenamefont
  {Poisson}(2023)}]{Pitre:2023xsr}%
  \BibitemOpen
  \bibfield  {author} {\bibinfo {author} {\bibfnamefont {Tristan}\ \bibnamefont
  {Pitre}}\ and\ \bibinfo {author} {\bibfnamefont {Eric}\ \bibnamefont
  {Poisson}},\ }\bibfield  {title} {\enquote {\bibinfo {title} {{General
  relativistic dynamical tides in binary inspirals, without modes}},}\
  }\href@noop {} {\  (\bibinfo {year} {2023})},\ \Eprint
  {http://arxiv.org/abs/2311.04075} {arXiv:2311.04075 [gr-qc]} \BibitemShut
  {NoStop}%
\bibitem [{\citenamefont {Chakrabarti}\ \emph
  {et~al.}(2013{\natexlab{a}})\citenamefont {Chakrabarti}, \citenamefont
  {Delsate},\ and\ \citenamefont {Steinhoff}}]{chakrabarti2013new}%
  \BibitemOpen
  \bibfield  {author} {\bibinfo {author} {\bibfnamefont {Sayan}\ \bibnamefont
  {Chakrabarti}}, \bibinfo {author} {\bibfnamefont {Térence}\ \bibnamefont
  {Delsate}}, \ and\ \bibinfo {author} {\bibfnamefont {Jan}\ \bibnamefont
  {Steinhoff}},\ }\href@noop {} {\enquote {\bibinfo {title} {New perspectives
  on neutron star and black hole spectroscopy and dynamic tides},}\ } (\bibinfo
  {year} {2013}{\natexlab{a}}),\ \Eprint {http://arxiv.org/abs/1304.2228}
  {arXiv:1304.2228 [gr-qc]} \BibitemShut {NoStop}%
\bibitem [{\citenamefont {Zahn}(2008)}]{Zahn:2008fk}%
  \BibitemOpen
  \bibfield  {author} {\bibinfo {author} {\bibfnamefont {Jean-Paul}\
  \bibnamefont {Zahn}},\ }\bibfield  {title} {\enquote {\bibinfo {title}
  {{Tidal dissipation in binary systems}},}\ }\href {\doibase
  10.1051/eas:0829002} {\bibfield  {journal} {\bibinfo  {journal} {EAS Publ.
  Ser.}\ }\textbf {\bibinfo {volume} {29}},\ \bibinfo {pages} {67} (\bibinfo
  {year} {2008})},\ \Eprint {http://arxiv.org/abs/0807.4870} {arXiv:0807.4870
  [astro-ph]} \BibitemShut {NoStop}%
\bibitem [{\citenamefont {{Kochanek}}(1992)}]{1992ApJ...398..234K}%
  \BibitemOpen
  \bibfield  {author} {\bibinfo {author} {\bibfnamefont {Christopher~S.}\
  \bibnamefont {{Kochanek}}},\ }\bibfield  {title} {\enquote {\bibinfo {title}
  {{Coalescing Binary Neutron Stars}},}\ }\href {\doibase 10.1086/171851}
  {\bibfield  {journal} {\bibinfo  {journal} {\apj}\ }\textbf {\bibinfo
  {volume} {398}},\ \bibinfo {pages} {234} (\bibinfo {year}
  {1992})}\BibitemShut {NoStop}%
\bibitem [{\citenamefont {Shternin}\ and\ \citenamefont
  {Yakovlev}(2008)}]{Shternin:2008es}%
  \BibitemOpen
  \bibfield  {author} {\bibinfo {author} {\bibfnamefont {P.~S.}\ \bibnamefont
  {Shternin}}\ and\ \bibinfo {author} {\bibfnamefont {D.~G.}\ \bibnamefont
  {Yakovlev}},\ }\bibfield  {title} {\enquote {\bibinfo {title} {{Shear
  viscosity in neutron star cores}},}\ }\href {\doibase
  10.1103/PhysRevD.78.063006} {\bibfield  {journal} {\bibinfo  {journal} {Phys.
  Rev. D}\ }\textbf {\bibinfo {volume} {78}},\ \bibinfo {pages} {063006}
  (\bibinfo {year} {2008})},\ \Eprint {http://arxiv.org/abs/0808.2018}
  {arXiv:0808.2018 [astro-ph]} \BibitemShut {NoStop}%
\bibitem [{\citenamefont {Sawyer}(1989)}]{physrevd.39.3804}%
  \BibitemOpen
  \bibfield  {author} {\bibinfo {author} {\bibfnamefont {Raymond~F.}\
  \bibnamefont {Sawyer}},\ }\bibfield  {title} {\enquote {\bibinfo {title}
  {Bulk viscosity of hot neutron-star matter and the maximum rotation rates of
  neutron stars},}\ }\href {\doibase 10.1103/PhysRevD.39.3804} {\bibfield
  {journal} {\bibinfo  {journal} {Phys. Rev. D}\ }\textbf {\bibinfo {volume}
  {39}},\ \bibinfo {pages} {3804--3806} (\bibinfo {year} {1989})}\BibitemShut
  {NoStop}%
\bibitem [{\citenamefont {Alford}\ \emph {et~al.}(2023)\citenamefont {Alford},
  \citenamefont {Haber},\ and\ \citenamefont {Zhang}}]{Alford:2023gxq}%
  \BibitemOpen
  \bibfield  {author} {\bibinfo {author} {\bibfnamefont {Mark~G.}\ \bibnamefont
  {Alford}}, \bibinfo {author} {\bibfnamefont {Alexander}\ \bibnamefont
  {Haber}}, \ and\ \bibinfo {author} {\bibfnamefont {Ziyuan}\ \bibnamefont
  {Zhang}},\ }\bibfield  {title} {\enquote {\bibinfo {title} {{Isospin
  Equilibration in Neutron Star Mergers}},}\ }\href@noop {} {\  (\bibinfo
  {year} {2023})},\ \Eprint {http://arxiv.org/abs/2306.06180} {arXiv:2306.06180
  [nucl-th]} \BibitemShut {NoStop}%
\bibitem [{\citenamefont {Yang}\ \emph {et~al.}(2023)\citenamefont {Yang},
  \citenamefont {Hippert}, \citenamefont {Speranza},\ and\ \citenamefont
  {Noronha}}]{Yang:2023ogo}%
  \BibitemOpen
  \bibfield  {author} {\bibinfo {author} {\bibfnamefont {Yumu}\ \bibnamefont
  {Yang}}, \bibinfo {author} {\bibfnamefont {Mauricio}\ \bibnamefont
  {Hippert}}, \bibinfo {author} {\bibfnamefont {Enrico}\ \bibnamefont
  {Speranza}}, \ and\ \bibinfo {author} {\bibfnamefont {Jorge}\ \bibnamefont
  {Noronha}},\ }\bibfield  {title} {\enquote {\bibinfo {title}
  {{Far-from-equilibrium bulk-viscous transport coefficients in neutron star
  mergers}},}\ }\href@noop {} {\  (\bibinfo {year} {2023})},\ \Eprint
  {http://arxiv.org/abs/2309.01864} {arXiv:2309.01864 [nucl-th]} \BibitemShut
  {NoStop}%
\bibitem [{\citenamefont {Chabanov}\ and\ \citenamefont
  {Rezzolla}(2023{\natexlab{b}})}]{Chabanov:2023abq}%
  \BibitemOpen
  \bibfield  {author} {\bibinfo {author} {\bibfnamefont {Michail}\ \bibnamefont
  {Chabanov}}\ and\ \bibinfo {author} {\bibfnamefont {Luciano}\ \bibnamefont
  {Rezzolla}},\ }\bibfield  {title} {\enquote {\bibinfo {title} {{Numerical
  modelling of bulk viscosity in neutron stars}},}\ }\href@noop {} {\
  (\bibinfo {year} {2023}{\natexlab{b}})},\ \Eprint
  {http://arxiv.org/abs/2311.13027} {arXiv:2311.13027 [gr-qc]} \BibitemShut
  {NoStop}%
\bibitem [{\citenamefont {Chabanov}\ and\ \citenamefont
  {Rezzolla}(2023{\natexlab{c}})}]{chabanov2023impact}%
  \BibitemOpen
  \bibfield  {author} {\bibinfo {author} {\bibfnamefont {Michail}\ \bibnamefont
  {Chabanov}}\ and\ \bibinfo {author} {\bibfnamefont {Luciano}\ \bibnamefont
  {Rezzolla}},\ }\href@noop {} {\enquote {\bibinfo {title} {Impact of bulk
  viscosity on the post-merger gravitational-wave signal from merging neutron
  stars},}\ } (\bibinfo {year} {2023}{\natexlab{c}}),\ \Eprint
  {http://arxiv.org/abs/2307.10464} {arXiv:2307.10464 [gr-qc]} \BibitemShut
  {NoStop}%
\bibitem [{\citenamefont {Radice}\ \emph {et~al.}(2022)\citenamefont {Radice},
  \citenamefont {Bernuzzi}, \citenamefont {Perego},\ and\ \citenamefont
  {Haas}}]{Radice_2022}%
  \BibitemOpen
  \bibfield  {author} {\bibinfo {author} {\bibfnamefont {David}\ \bibnamefont
  {Radice}}, \bibinfo {author} {\bibfnamefont {Sebastiano}\ \bibnamefont
  {Bernuzzi}}, \bibinfo {author} {\bibfnamefont {Albino}\ \bibnamefont
  {Perego}}, \ and\ \bibinfo {author} {\bibfnamefont {Roland}\ \bibnamefont
  {Haas}},\ }\bibfield  {title} {\enquote {\bibinfo {title} {A new moment-based
  general-relativistic neutrino-radiation transport code: Methods and first
  applications to neutron star mergers},}\ }\href {\doibase
  10.1093/mnras/stac589} {\bibfield  {journal} {\bibinfo  {journal} {Monthly
  Notices of the Royal Astronomical Society}\ }\textbf {\bibinfo {volume}
  {512}},\ \bibinfo {pages} {1499–1521} (\bibinfo {year} {2022})}\BibitemShut
  {NoStop}%
\bibitem [{\citenamefont {Espino}\ \emph {et~al.}(2023)\citenamefont {Espino},
  \citenamefont {Hammond}, \citenamefont {Radice}, \citenamefont {Bernuzzi},
  \citenamefont {Gamba}, \citenamefont {Zappa}, \citenamefont {Micchi},\ and\
  \citenamefont {Perego}}]{espino2023neutrino}%
  \BibitemOpen
  \bibfield  {author} {\bibinfo {author} {\bibfnamefont {Pedro~Luis}\
  \bibnamefont {Espino}}, \bibinfo {author} {\bibfnamefont {Peter}\
  \bibnamefont {Hammond}}, \bibinfo {author} {\bibfnamefont {David}\
  \bibnamefont {Radice}}, \bibinfo {author} {\bibfnamefont {Sebastiano}\
  \bibnamefont {Bernuzzi}}, \bibinfo {author} {\bibfnamefont {Rossella}\
  \bibnamefont {Gamba}}, \bibinfo {author} {\bibfnamefont {Francesco}\
  \bibnamefont {Zappa}}, \bibinfo {author} {\bibfnamefont {Luis Felipe~Longo}\
  \bibnamefont {Micchi}}, \ and\ \bibinfo {author} {\bibfnamefont {Albino}\
  \bibnamefont {Perego}},\ }\href@noop {} {\enquote {\bibinfo {title} {Neutrino
  trapping and out-of-equilibrium effects in binary neutron star merger
  remnants},}\ } (\bibinfo {year} {2023}),\ \Eprint
  {http://arxiv.org/abs/2311.00031} {arXiv:2311.00031 [astro-ph.HE]}
  \BibitemShut {NoStop}%
\bibitem [{\citenamefont {Ripley}\ \emph
  {et~al.}(2023{\natexlab{b}})\citenamefont {Ripley}, \citenamefont {Hegade
  K.~R.}, \citenamefont {Chandramouli},\ and\ \citenamefont
  {Yunes}}]{Ripley:2023lsq}%
  \BibitemOpen
  \bibfield  {author} {\bibinfo {author} {\bibfnamefont {Justin~L.}\
  \bibnamefont {Ripley}}, \bibinfo {author} {\bibfnamefont {Abhishek}\
  \bibnamefont {Hegade K.~R.}}, \bibinfo {author} {\bibfnamefont {Rohit~S.}\
  \bibnamefont {Chandramouli}}, \ and\ \bibinfo {author} {\bibfnamefont
  {Nicolas}\ \bibnamefont {Yunes}},\ }\bibfield  {title} {\enquote {\bibinfo
  {title} {{First constraint on the dissipative tidal deformability of neutron
  stars}},}\ }\href@noop {} {\  (\bibinfo {year} {2023}{\natexlab{b}})},\
  \Eprint {http://arxiv.org/abs/2312.11659} {arXiv:2312.11659 [gr-qc]}
  \BibitemShut {NoStop}%
\bibitem [{\citenamefont {Arras}\ and\ \citenamefont
  {Weinberg}(2019)}]{Arras:2018fxj}%
  \BibitemOpen
  \bibfield  {author} {\bibinfo {author} {\bibfnamefont {Phil}\ \bibnamefont
  {Arras}}\ and\ \bibinfo {author} {\bibfnamefont {Nevin~N.}\ \bibnamefont
  {Weinberg}},\ }\bibfield  {title} {\enquote {\bibinfo {title} {{Urca
  reactions during neutron star inspiral}},}\ }\href {\doibase
  10.1093/mnras/stz880} {\bibfield  {journal} {\bibinfo  {journal} {Mon. Not.
  Roy. Astron. Soc.}\ }\textbf {\bibinfo {volume} {486}},\ \bibinfo {pages}
  {1424--1436} (\bibinfo {year} {2019})},\ \Eprint
  {http://arxiv.org/abs/1806.04163} {arXiv:1806.04163 [astro-ph.HE]}
  \BibitemShut {NoStop}%
\bibitem [{\citenamefont {Perego}\ \emph {et~al.}(2019)\citenamefont {Perego},
  \citenamefont {Bernuzzi},\ and\ \citenamefont {Radice}}]{Perego:2019adq}%
  \BibitemOpen
  \bibfield  {author} {\bibinfo {author} {\bibfnamefont {Albino}\ \bibnamefont
  {Perego}}, \bibinfo {author} {\bibfnamefont {Sebastiano}\ \bibnamefont
  {Bernuzzi}}, \ and\ \bibinfo {author} {\bibfnamefont {David}\ \bibnamefont
  {Radice}},\ }\bibfield  {title} {\enquote {\bibinfo {title} {{Thermodynamics
  conditions of matter in neutron star mergers}},}\ }\href {\doibase
  10.1140/epja/i2019-12810-7} {\bibfield  {journal} {\bibinfo  {journal} {Eur.
  Phys. J. A}\ }\textbf {\bibinfo {volume} {55}},\ \bibinfo {pages} {124}
  (\bibinfo {year} {2019})},\ \Eprint {http://arxiv.org/abs/1903.07898}
  {arXiv:1903.07898 [gr-qc]} \BibitemShut {NoStop}%
\bibitem [{\citenamefont {Poisson}\ and\ \citenamefont
  {Will}(2014)}]{Poisson-Will}%
  \BibitemOpen
  \bibfield  {author} {\bibinfo {author} {\bibfnamefont {Eric}\ \bibnamefont
  {Poisson}}\ and\ \bibinfo {author} {\bibfnamefont {Clifford~M.}\ \bibnamefont
  {Will}},\ }\href {\doibase 10.1017/CBO9781139507486} {\emph {\bibinfo {title}
  {Gravity: Newtonian, Post-Newtonian, Relativistic}}}\ (\bibinfo  {publisher}
  {Cambridge University Press},\ \bibinfo {year} {2014})\BibitemShut {NoStop}%
\bibitem [{\citenamefont {Hinderer}\ \emph
  {et~al.}(2010{\natexlab{a}})\citenamefont {Hinderer}, \citenamefont {Lackey},
  \citenamefont {Lang},\ and\ \citenamefont {Read}}]{Hinderer_2010}%
  \BibitemOpen
  \bibfield  {author} {\bibinfo {author} {\bibfnamefont {Tanja}\ \bibnamefont
  {Hinderer}}, \bibinfo {author} {\bibfnamefont {Benjamin~D.}\ \bibnamefont
  {Lackey}}, \bibinfo {author} {\bibfnamefont {Ryan~N.}\ \bibnamefont {Lang}},
  \ and\ \bibinfo {author} {\bibfnamefont {Jocelyn~S.}\ \bibnamefont {Read}},\
  }\bibfield  {title} {\enquote {\bibinfo {title} {Tidal deformability of
  neutron stars with realistic equations of state and their gravitational wave
  signatures in binary inspiral},}\ }\href {\doibase
  10.1103/physrevd.81.123016} {\bibfield  {journal} {\bibinfo  {journal}
  {Physical Review D}\ }\textbf {\bibinfo {volume} {81}} (\bibinfo {year}
  {2010}{\natexlab{a}}),\ 10.1103/physrevd.81.123016}\BibitemShut {NoStop}%
\bibitem [{\citenamefont {{Smeyers}}\ and\ \citenamefont {{van
  Hoolst}}(2010)}]{Smeyers-Book}%
  \BibitemOpen
  \bibfield  {author} {\bibinfo {author} {\bibfnamefont {Paul}\ \bibnamefont
  {{Smeyers}}}\ and\ \bibinfo {author} {\bibfnamefont {Tim}\ \bibnamefont {{van
  Hoolst}}},\ }\href {\doibase 10.1007/978-3-642-13030-4} {\emph {\bibinfo
  {title} {{Linear Isentropic Oscillations of Stars: Theoretical
  Foundations}}}},\ Vol.\ \bibinfo {volume} {371}\ (\bibinfo {year}
  {2010})\BibitemShut {NoStop}%
\bibitem [{\citenamefont {{Press}}\ and\ \citenamefont
  {{Teukolsky}}(1977)}]{Press-Teukolsky}%
  \BibitemOpen
  \bibfield  {author} {\bibinfo {author} {\bibfnamefont {W.~H.}\ \bibnamefont
  {{Press}}}\ and\ \bibinfo {author} {\bibfnamefont {S.~A.}\ \bibnamefont
  {{Teukolsky}}},\ }\bibfield  {title} {\enquote {\bibinfo {title} {{On
  formation of close binaries by two-body tidal capture.}}}\ }\href {\doibase
  10.1086/155143} {\bibfield  {journal} {\bibinfo  {journal} {\apj}\ }\textbf
  {\bibinfo {volume} {213}},\ \bibinfo {pages} {183--192} (\bibinfo {year}
  {1977})}\BibitemShut {NoStop}%
\bibitem [{\citenamefont {{Cowling}}(1941)}]{1941MNRAS.101..367C}%
  \BibitemOpen
  \bibfield  {author} {\bibinfo {author} {\bibfnamefont {T.~G.}\ \bibnamefont
  {{Cowling}}},\ }\bibfield  {title} {\enquote {\bibinfo {title} {{The
  non-radial oscillations of polytropic stars}},}\ }\href {\doibase
  10.1093/mnras/101.8.367} {\bibfield  {journal} {\bibinfo  {journal} {Mon.
  Not. Roy. Astron. Soc.}\ }\textbf {\bibinfo {volume} {101}},\ \bibinfo
  {pages} {367} (\bibinfo {year} {1941})}\BibitemShut {NoStop}%
\bibitem [{\citenamefont {{Schwarzschild}}(1958)}]{1958ses..book.....S}%
  \BibitemOpen
  \bibfield  {author} {\bibinfo {author} {\bibfnamefont {Martin}\ \bibnamefont
  {{Schwarzschild}}},\ }\href@noop {} {\emph {\bibinfo {title} {{Structure and
  evolution of the stars.}}}}\ (\bibinfo {year} {1958})\BibitemShut {NoStop}%
\bibitem [{\citenamefont {Lai}(1997)}]{Lai_1997}%
  \BibitemOpen
  \bibfield  {author} {\bibinfo {author} {\bibfnamefont {Dong}\ \bibnamefont
  {Lai}},\ }\bibfield  {title} {\enquote {\bibinfo {title} {Dynamical tides in
  rotating binary stars},}\ }\href {\doibase 10.1086/304899} {\bibfield
  {journal} {\bibinfo  {journal} {The Astrophysical Journal}\ }\textbf
  {\bibinfo {volume} {490}},\ \bibinfo {pages} {847} (\bibinfo {year}
  {1997})}\BibitemShut {NoStop}%
\bibitem [{\citenamefont {Chakrabarti}\ \emph
  {et~al.}(2013{\natexlab{b}})\citenamefont {Chakrabarti}, \citenamefont
  {Delsate},\ and\ \citenamefont {Steinhoff}}]{Chakrabarti:2013lua}%
  \BibitemOpen
  \bibfield  {author} {\bibinfo {author} {\bibfnamefont {Sayan}\ \bibnamefont
  {Chakrabarti}}, \bibinfo {author} {\bibfnamefont {T\'erence}\ \bibnamefont
  {Delsate}}, \ and\ \bibinfo {author} {\bibfnamefont {Jan}\ \bibnamefont
  {Steinhoff}},\ }\bibfield  {title} {\enquote {\bibinfo {title} {{New
  perspectives on neutron star and black hole spectroscopy and dynamic
  tides}},}\ }\href@noop {} {\  (\bibinfo {year} {2013}{\natexlab{b}})},\
  \Eprint {http://arxiv.org/abs/1304.2228} {arXiv:1304.2228 [gr-qc]}
  \BibitemShut {NoStop}%
\bibitem [{\citenamefont {Steinhoff}\ \emph
  {et~al.}(2016{\natexlab{b}})\citenamefont {Steinhoff}, \citenamefont
  {Hinderer}, \citenamefont {Buonanno},\ and\ \citenamefont
  {Taracchini}}]{Steinhoff:2016rfi}%
  \BibitemOpen
  \bibfield  {author} {\bibinfo {author} {\bibfnamefont {Jan}\ \bibnamefont
  {Steinhoff}}, \bibinfo {author} {\bibfnamefont {Tanja}\ \bibnamefont
  {Hinderer}}, \bibinfo {author} {\bibfnamefont {Alessandra}\ \bibnamefont
  {Buonanno}}, \ and\ \bibinfo {author} {\bibfnamefont {Andrea}\ \bibnamefont
  {Taracchini}},\ }\bibfield  {title} {\enquote {\bibinfo {title} {{Dynamical
  Tides in General Relativity: Effective Action and Effective-One-Body
  Hamiltonian}},}\ }\href {\doibase 10.1103/PhysRevD.94.104028} {\bibfield
  {journal} {\bibinfo  {journal} {Phys. Rev. D}\ }\textbf {\bibinfo {volume}
  {94}},\ \bibinfo {pages} {104028} (\bibinfo {year} {2016}{\natexlab{b}})},\
  \Eprint {http://arxiv.org/abs/1608.01907} {arXiv:1608.01907 [gr-qc]}
  \BibitemShut {NoStop}%
\bibitem [{\citenamefont {Chatziioannou}(2020)}]{Chatziioannou_2020}%
  \BibitemOpen
  \bibfield  {author} {\bibinfo {author} {\bibfnamefont {Katerina}\
  \bibnamefont {Chatziioannou}},\ }\bibfield  {title} {\enquote {\bibinfo
  {title} {Neutron-star tidal deformability and equation-of-state
  constraints},}\ }\href {\doibase 10.1007/s10714-020-02754-3} {\bibfield
  {journal} {\bibinfo  {journal} {General Relativity and Gravitation}\ }\textbf
  {\bibinfo {volume} {52}} (\bibinfo {year} {2020}),\
  10.1007/s10714-020-02754-3}\BibitemShut {NoStop}%
\bibitem [{\citenamefont {Pratten}\ \emph {et~al.}(2022)\citenamefont
  {Pratten}, \citenamefont {Schmidt},\ and\ \citenamefont
  {Williams}}]{Pratten:2021pro}%
  \BibitemOpen
  \bibfield  {author} {\bibinfo {author} {\bibfnamefont {Geraint}\ \bibnamefont
  {Pratten}}, \bibinfo {author} {\bibfnamefont {Patricia}\ \bibnamefont
  {Schmidt}}, \ and\ \bibinfo {author} {\bibfnamefont {Natalie}\ \bibnamefont
  {Williams}},\ }\bibfield  {title} {\enquote {\bibinfo {title} {{Impact of
  Dynamical Tides on the Reconstruction of the Neutron Star Equation of
  State}},}\ }\href {\doibase 10.1103/PhysRevLett.129.081102} {\bibfield
  {journal} {\bibinfo  {journal} {Phys. Rev. Lett.}\ }\textbf {\bibinfo
  {volume} {129}},\ \bibinfo {pages} {081102} (\bibinfo {year} {2022})},\
  \Eprint {http://arxiv.org/abs/2109.07566} {arXiv:2109.07566 [astro-ph.HE]}
  \BibitemShut {NoStop}%
\bibitem [{\citenamefont {Kokkotas}\ and\ \citenamefont
  {Schmidt}(1999)}]{Kokkotas:1999bd}%
  \BibitemOpen
  \bibfield  {author} {\bibinfo {author} {\bibfnamefont {Kostas~D.}\
  \bibnamefont {Kokkotas}}\ and\ \bibinfo {author} {\bibfnamefont {Bernd~G.}\
  \bibnamefont {Schmidt}},\ }\bibfield  {title} {\enquote {\bibinfo {title}
  {{Quasinormal modes of stars and black holes}},}\ }\href {\doibase
  10.12942/lrr-1999-2} {\bibfield  {journal} {\bibinfo  {journal} {Living Rev.
  Rel.}\ }\textbf {\bibinfo {volume} {2}},\ \bibinfo {pages} {2} (\bibinfo
  {year} {1999})},\ \Eprint {http://arxiv.org/abs/gr-qc/9909058}
  {arXiv:gr-qc/9909058} \BibitemShut {NoStop}%
\bibitem [{\citenamefont {Yunes}\ \emph {et~al.}(2006)\citenamefont {Yunes},
  \citenamefont {Tichy}, \citenamefont {Owen},\ and\ \citenamefont
  {Bruegmann}}]{Yunes:2005nn}%
  \BibitemOpen
  \bibfield  {author} {\bibinfo {author} {\bibfnamefont {Nicolas}\ \bibnamefont
  {Yunes}}, \bibinfo {author} {\bibfnamefont {Wolfgang}\ \bibnamefont {Tichy}},
  \bibinfo {author} {\bibfnamefont {Benjamin~J.}\ \bibnamefont {Owen}}, \ and\
  \bibinfo {author} {\bibfnamefont {Bernd}\ \bibnamefont {Bruegmann}},\
  }\bibfield  {title} {\enquote {\bibinfo {title} {{Binary black hole initial
  data from matched asymptotic expansions}},}\ }\href {\doibase
  10.1103/PhysRevD.74.104011} {\bibfield  {journal} {\bibinfo  {journal} {Phys.
  Rev. D}\ }\textbf {\bibinfo {volume} {74}},\ \bibinfo {pages} {104011}
  (\bibinfo {year} {2006})},\ \Eprint {http://arxiv.org/abs/gr-qc/0503011}
  {arXiv:gr-qc/0503011} \BibitemShut {NoStop}%
\bibitem [{\citenamefont {Johnson-McDaniel}\ \emph {et~al.}(2009)\citenamefont
  {Johnson-McDaniel}, \citenamefont {Yunes}, \citenamefont {Tichy},\ and\
  \citenamefont {Owen}}]{Johnson-McDaniel:2009tvj}%
  \BibitemOpen
  \bibfield  {author} {\bibinfo {author} {\bibfnamefont {Nathan~K.}\
  \bibnamefont {Johnson-McDaniel}}, \bibinfo {author} {\bibfnamefont {Nicolas}\
  \bibnamefont {Yunes}}, \bibinfo {author} {\bibfnamefont {Wolfgang}\
  \bibnamefont {Tichy}}, \ and\ \bibinfo {author} {\bibfnamefont {Benjamin~J.}\
  \bibnamefont {Owen}},\ }\bibfield  {title} {\enquote {\bibinfo {title}
  {{Conformally curved binary black hole initial data including tidal
  deformations and outgoing radiation}},}\ }\href {\doibase
  10.1103/PhysRevD.80.124039} {\bibfield  {journal} {\bibinfo  {journal} {Phys.
  Rev. D}\ }\textbf {\bibinfo {volume} {80}},\ \bibinfo {pages} {124039}
  (\bibinfo {year} {2009})},\ \Eprint {http://arxiv.org/abs/0907.0891}
  {arXiv:0907.0891 [gr-qc]} \BibitemShut {NoStop}%
\bibitem [{\citenamefont {{Thorne}}\ and\ \citenamefont
  {{Campolattaro}}(1967)}]{1967ApJ...149..591T}%
  \BibitemOpen
  \bibfield  {author} {\bibinfo {author} {\bibfnamefont {Kip~S.}\ \bibnamefont
  {{Thorne}}}\ and\ \bibinfo {author} {\bibfnamefont {Alfonso}\ \bibnamefont
  {{Campolattaro}}},\ }\href {\doibase 10.1086/149288} {\enquote {\bibinfo
  {title} {{Non-Radial Pulsation of General-Relativistic Stellar Models. I.
  Analytic Analysis for L >= 2}},}\ }\bibinfo {howpublished} {Astrophysical
  Journal, vol. 149, p.591} (\bibinfo {year} {1967})\BibitemShut {NoStop}%
\bibitem [{\citenamefont {{Lindblom}}\ and\ \citenamefont
  {{Detweiler}}(1983)}]{Lindblom-Detweiler-1983}%
  \BibitemOpen
  \bibfield  {author} {\bibinfo {author} {\bibfnamefont {L.}~\bibnamefont
  {{Lindblom}}}\ and\ \bibinfo {author} {\bibfnamefont {S.~L.}\ \bibnamefont
  {{Detweiler}}},\ }\bibfield  {title} {\enquote {\bibinfo {title} {{The
  quadrupole oscillations of neutron stars.}}}\ }\href {\doibase
  10.1086/190884} {\bibfield  {journal} {\bibinfo  {journal} {APJs}\ }\textbf
  {\bibinfo {volume} {53}},\ \bibinfo {pages} {73--92} (\bibinfo {year}
  {1983})}\BibitemShut {NoStop}%
\bibitem [{\citenamefont {{Detweiler}}\ and\ \citenamefont
  {{Lindblom}}(1985)}]{Detweiler-Lindblom-1985}%
  \BibitemOpen
  \bibfield  {author} {\bibinfo {author} {\bibfnamefont {S.}~\bibnamefont
  {{Detweiler}}}\ and\ \bibinfo {author} {\bibfnamefont {L.}~\bibnamefont
  {{Lindblom}}},\ }\bibfield  {title} {\enquote {\bibinfo {title} {{On the
  nonradial pulsations of general relativistic stellar models}},}\ }\href
  {\doibase 10.1086/163127} {\bibfield  {journal} {\bibinfo  {journal} {\apj}\
  }\textbf {\bibinfo {volume} {292}},\ \bibinfo {pages} {12--15} (\bibinfo
  {year} {1985})}\BibitemShut {NoStop}%
\bibitem [{\citenamefont {{Ipser}}\ and\ \citenamefont
  {{Lindblom}}(1990)}]{Ipser-Lindblom-Newtonian-1990}%
  \BibitemOpen
  \bibfield  {author} {\bibinfo {author} {\bibfnamefont {James~R.}\
  \bibnamefont {{Ipser}}}\ and\ \bibinfo {author} {\bibfnamefont {Lee}\
  \bibnamefont {{Lindblom}}},\ }\bibfield  {title} {\enquote {\bibinfo {title}
  {{The Oscillations of Rapidly Rotating Newtonian Stellar Models}},}\ }\href
  {\doibase 10.1086/168757} {\bibfield  {journal} {\bibinfo  {journal} {\apj}\
  }\textbf {\bibinfo {volume} {355}},\ \bibinfo {pages} {226} (\bibinfo {year}
  {1990})}\BibitemShut {NoStop}%
\bibitem [{\citenamefont {Hegade~K.R.}(2024)}]{github-code}%
  \BibitemOpen
  \bibfield  {author} {\bibinfo {author} {\bibfnamefont {Abhishek}\
  \bibnamefont {Hegade~K.R.}},\ }\href@noop {} {\enquote {\bibinfo {title}
  {Supplementary code},}\ }\bibinfo {howpublished}
  {\url{https://github.com/AbhiHegade/Dynamical-Tidal-Response-of-Polytropic-Stars}}
  (\bibinfo {year} {2024})\BibitemShut {NoStop}%
\bibitem [{\citenamefont {Sasaki}\ and\ \citenamefont
  {Tagoshi}(2003)}]{Sasaki_2003}%
  \BibitemOpen
  \bibfield  {author} {\bibinfo {author} {\bibfnamefont {Misao}\ \bibnamefont
  {Sasaki}}\ and\ \bibinfo {author} {\bibfnamefont {Hideyuki}\ \bibnamefont
  {Tagoshi}},\ }\bibfield  {title} {\enquote {\bibinfo {title} {Analytic black
  hole perturbation approach to gravitational radiation},}\ }\href {\doibase
  10.12942/lrr-2003-6} {\bibfield  {journal} {\bibinfo  {journal} {Living
  Reviews in Relativity}\ }\textbf {\bibinfo {volume} {6}} (\bibinfo {year}
  {2003}),\ 10.12942/lrr-2003-6}\BibitemShut {NoStop}%
\bibitem [{\citenamefont {{Hiscock}}\ and\ \citenamefont
  {{Lindblom}}(1983)}]{1983AnPhy.151..466H}%
  \BibitemOpen
  \bibfield  {author} {\bibinfo {author} {\bibfnamefont {W.~A.}\ \bibnamefont
  {{Hiscock}}}\ and\ \bibinfo {author} {\bibfnamefont {L.}~\bibnamefont
  {{Lindblom}}},\ }\bibfield  {title} {\enquote {\bibinfo {title} {{Stability
  and causality in dissipative relativistic fluids.}}}\ }\href {\doibase
  10.1016/0003-4916(83)90288-9} {\bibfield  {journal} {\bibinfo  {journal}
  {Annals of Physics}\ }\textbf {\bibinfo {volume} {151}},\ \bibinfo {pages}
  {466--496} (\bibinfo {year} {1983})}\BibitemShut {NoStop}%
\bibitem [{\citenamefont {{Hiscock}}\ and\ \citenamefont
  {{Lindblom}}(1985)}]{1985PhRvD..31..725H}%
  \BibitemOpen
  \bibfield  {author} {\bibinfo {author} {\bibfnamefont {William~A.}\
  \bibnamefont {{Hiscock}}}\ and\ \bibinfo {author} {\bibfnamefont {Lee}\
  \bibnamefont {{Lindblom}}},\ }\bibfield  {title} {\enquote {\bibinfo {title}
  {{Generic instabilities in first-order dissipative relativistic fluid
  theories}},}\ }\href {\doibase 10.1103/PhysRevD.31.725} {\bibfield  {journal}
  {\bibinfo  {journal} {\prd}\ }\textbf {\bibinfo {volume} {31}},\ \bibinfo
  {pages} {725--733} (\bibinfo {year} {1985})}\BibitemShut {NoStop}%
\bibitem [{\citenamefont {{Israel}}(1976)}]{1976AnPhy.100..310I}%
  \BibitemOpen
  \bibfield  {author} {\bibinfo {author} {\bibfnamefont {Werner}\ \bibnamefont
  {{Israel}}},\ }\bibfield  {title} {\enquote {\bibinfo {title} {{Nonstationary
  irreversible thermodynamics: A causal relativistic theory}},}\ }\href
  {\doibase 10.1016/0003-4916(76)90064-6} {\bibfield  {journal} {\bibinfo
  {journal} {Annals of Physics}\ }\textbf {\bibinfo {volume} {100}},\ \bibinfo
  {pages} {310--331} (\bibinfo {year} {1976})}\BibitemShut {NoStop}%
\bibitem [{\citenamefont {{Israel}}\ and\ \citenamefont
  {{Stewart}}(1979)}]{1979RSPSA.365...43I}%
  \BibitemOpen
  \bibfield  {author} {\bibinfo {author} {\bibfnamefont {W.}~\bibnamefont
  {{Israel}}}\ and\ \bibinfo {author} {\bibfnamefont {J.~M.}\ \bibnamefont
  {{Stewart}}},\ }\bibfield  {title} {\enquote {\bibinfo {title} {{On transient
  relativistic thermodynamics and kinetic theory. II}},}\ }\href {\doibase
  10.1098/rspa.1979.0005} {\bibfield  {journal} {\bibinfo  {journal}
  {Proceedings of the Royal Society of London Series A}\ }\textbf {\bibinfo
  {volume} {365}},\ \bibinfo {pages} {43--52} (\bibinfo {year}
  {1979})}\BibitemShut {NoStop}%
\bibitem [{\citenamefont {{M\"{u}ller}}\ and\ \citenamefont
  {{Ruggeri}}(1993)}]{Muller-book}%
  \BibitemOpen
  \bibfield  {author} {\bibinfo {author} {\bibfnamefont {Ingo}\ \bibnamefont
  {{M\"{u}ller}}}\ and\ \bibinfo {author} {\bibfnamefont {Tommaso}\
  \bibnamefont {{Ruggeri}}},\ }\href {\doibase
  https://doi.org/10.1007/978-1-4612-2210-1} {\emph {\bibinfo {title}
  {{Rational Extended Thermodynamics}}}}\ (\bibinfo  {publisher} {Springer},\
  \bibinfo {year} {1993})\BibitemShut {NoStop}%
\bibitem [{\citenamefont {Baier}\ \emph {et~al.}(2008)\citenamefont {Baier},
  \citenamefont {Romatschke}, \citenamefont {Son}, \citenamefont {Starinets},\
  and\ \citenamefont {Stephanov}}]{Baier:2007ix}%
  \BibitemOpen
  \bibfield  {author} {\bibinfo {author} {\bibfnamefont {Rudolf}\ \bibnamefont
  {Baier}}, \bibinfo {author} {\bibfnamefont {Paul}\ \bibnamefont
  {Romatschke}}, \bibinfo {author} {\bibfnamefont {Dam~Thanh}\ \bibnamefont
  {Son}}, \bibinfo {author} {\bibfnamefont {Andrei~O.}\ \bibnamefont
  {Starinets}}, \ and\ \bibinfo {author} {\bibfnamefont {Mikhail~A.}\
  \bibnamefont {Stephanov}},\ }\bibfield  {title} {\enquote {\bibinfo {title}
  {{Relativistic viscous hydrodynamics, conformal invariance, and
  holography}},}\ }\href {\doibase 10.1088/1126-6708/2008/04/100} {\bibfield
  {journal} {\bibinfo  {journal} {JHEP}\ }\textbf {\bibinfo {volume} {04}},\
  \bibinfo {pages} {100} (\bibinfo {year} {2008})},\ \Eprint
  {http://arxiv.org/abs/0712.2451} {arXiv:0712.2451 [hep-th]} \BibitemShut
  {NoStop}%
\bibitem [{\citenamefont {Denicol}\ \emph {et~al.}(2012)\citenamefont
  {Denicol}, \citenamefont {Niemi}, \citenamefont {Moln{\'{a} }r},\ and\
  \citenamefont {Rischke}}]{Denicol_2012}%
  \BibitemOpen
  \bibfield  {author} {\bibinfo {author} {\bibfnamefont {G.~S.}\ \bibnamefont
  {Denicol}}, \bibinfo {author} {\bibfnamefont {H.}~\bibnamefont {Niemi}},
  \bibinfo {author} {\bibfnamefont {E.}~\bibnamefont {Moln{\'{a} }r}}, \ and\
  \bibinfo {author} {\bibfnamefont {D.~H.}\ \bibnamefont {Rischke}},\
  }\bibfield  {title} {\enquote {\bibinfo {title} {Derivation of transient
  relativistic fluid dynamics from the boltzmann equation},}\ }\href {\doibase
  10.1103/physrevd.85.114047} {\bibfield  {journal} {\bibinfo  {journal}
  {Physical Review D}\ }\textbf {\bibinfo {volume} {85}} (\bibinfo {year}
  {2012}),\ 10.1103/physrevd.85.114047}\BibitemShut {NoStop}%
\bibitem [{\citenamefont {Bemfica}\ \emph {et~al.}(2022)\citenamefont
  {Bemfica}, \citenamefont {Disconzi},\ and\ \citenamefont
  {Noronha}}]{Bemfica:2020zjp}%
  \BibitemOpen
  \bibfield  {author} {\bibinfo {author} {\bibfnamefont {Fabio~S.}\
  \bibnamefont {Bemfica}}, \bibinfo {author} {\bibfnamefont {Marcelo~M.}\
  \bibnamefont {Disconzi}}, \ and\ \bibinfo {author} {\bibfnamefont {Jorge}\
  \bibnamefont {Noronha}},\ }\bibfield  {title} {\enquote {\bibinfo {title}
  {{First-Order General-Relativistic Viscous Fluid Dynamics}},}\ }\href
  {\doibase 10.1103/PhysRevX.12.021044} {\bibfield  {journal} {\bibinfo
  {journal} {Phys. Rev. X}\ }\textbf {\bibinfo {volume} {12}},\ \bibinfo
  {pages} {021044} (\bibinfo {year} {2022})},\ \Eprint
  {http://arxiv.org/abs/2009.11388} {arXiv:2009.11388 [gr-qc]} \BibitemShut
  {NoStop}%
\bibitem [{\citenamefont {Kovtun}(2019)}]{Kovtun_2019}%
  \BibitemOpen
  \bibfield  {author} {\bibinfo {author} {\bibfnamefont {Pavel}\ \bibnamefont
  {Kovtun}},\ }\bibfield  {title} {\enquote {\bibinfo {title} {First-order
  relativistic hydrodynamics is stable},}\ }\href {\doibase
  10.1007/jhep10(2019)034} {\bibfield  {journal} {\bibinfo  {journal} {Journal
  of High Energy Physics}\ }\textbf {\bibinfo {volume} {2019}} (\bibinfo {year}
  {2019}),\ 10.1007/jhep10(2019)034}\BibitemShut {NoStop}%
\bibitem [{\citenamefont {Hegade K.~R.}\ \emph {et~al.}(2023)\citenamefont
  {Hegade K.~R.}, \citenamefont {Ripley},\ and\ \citenamefont
  {Yunes}}]{HegadeKR:2023glb}%
  \BibitemOpen
  \bibfield  {author} {\bibinfo {author} {\bibfnamefont {Abhishek}\
  \bibnamefont {Hegade K.~R.}}, \bibinfo {author} {\bibfnamefont {Justin~L.}\
  \bibnamefont {Ripley}}, \ and\ \bibinfo {author} {\bibfnamefont {Nicol\'as}\
  \bibnamefont {Yunes}},\ }\bibfield  {title} {\enquote {\bibinfo {title}
  {{Nonrelativistic limit of first-order relativistic viscous fluids}},}\
  }\href {\doibase 10.1103/PhysRevD.107.124029} {\bibfield  {journal} {\bibinfo
   {journal} {Phys. Rev. D}\ }\textbf {\bibinfo {volume} {107}},\ \bibinfo
  {pages} {124029} (\bibinfo {year} {2023})},\ \Eprint
  {http://arxiv.org/abs/2305.09725} {arXiv:2305.09725 [gr-qc]} \BibitemShut
  {NoStop}%
\bibitem [{\citenamefont {Poisson}(2009)}]{Poisson:2009di}%
  \BibitemOpen
  \bibfield  {author} {\bibinfo {author} {\bibfnamefont {Eric}\ \bibnamefont
  {Poisson}},\ }\bibfield  {title} {\enquote {\bibinfo {title} {{Tidal
  interaction of black holes and Newtonian viscous bodies}},}\ }\href {\doibase
  10.1103/PhysRevD.80.064029} {\bibfield  {journal} {\bibinfo  {journal} {Phys.
  Rev. D}\ }\textbf {\bibinfo {volume} {80}},\ \bibinfo {pages} {064029}
  (\bibinfo {year} {2009})},\ \Eprint {http://arxiv.org/abs/0907.0874}
  {arXiv:0907.0874 [gr-qc]} \BibitemShut {NoStop}%
\bibitem [{\citenamefont {Hinderer}\ \emph
  {et~al.}(2010{\natexlab{b}})\citenamefont {Hinderer}, \citenamefont {Lackey},
  \citenamefont {Lang},\ and\ \citenamefont {Read}}]{Hinderer:2009ca}%
  \BibitemOpen
  \bibfield  {author} {\bibinfo {author} {\bibfnamefont {Tanja}\ \bibnamefont
  {Hinderer}}, \bibinfo {author} {\bibfnamefont {Benjamin~D.}\ \bibnamefont
  {Lackey}}, \bibinfo {author} {\bibfnamefont {Ryan~N.}\ \bibnamefont {Lang}},
  \ and\ \bibinfo {author} {\bibfnamefont {Jocelyn~S.}\ \bibnamefont {Read}},\
  }\bibfield  {title} {\enquote {\bibinfo {title} {{Tidal deformability of
  neutron stars with realistic equations of state and their gravitational wave
  signatures in binary inspiral}},}\ }\href {\doibase
  10.1103/PhysRevD.81.123016} {\bibfield  {journal} {\bibinfo  {journal} {Phys.
  Rev. D}\ }\textbf {\bibinfo {volume} {81}},\ \bibinfo {pages} {123016}
  (\bibinfo {year} {2010}{\natexlab{b}})},\ \Eprint
  {http://arxiv.org/abs/0911.3535} {arXiv:0911.3535 [astro-ph.HE]} \BibitemShut
  {NoStop}%
\bibitem [{\citenamefont {Saketh}\ \emph {et~al.}(2023)\citenamefont {Saketh},
  \citenamefont {Zhou},\ and\ \citenamefont {Ivanov}}]{Saketh:2023bul}%
  \BibitemOpen
  \bibfield  {author} {\bibinfo {author} {\bibfnamefont {M.~V.~S.}\
  \bibnamefont {Saketh}}, \bibinfo {author} {\bibfnamefont {Zihan}\
  \bibnamefont {Zhou}}, \ and\ \bibinfo {author} {\bibfnamefont {Mikhail~M.}\
  \bibnamefont {Ivanov}},\ }\bibfield  {title} {\enquote {\bibinfo {title}
  {{Dynamical Tidal Response of Kerr Black Holes from Scattering
  Amplitudes}},}\ }\href@noop {} {\  (\bibinfo {year} {2023})},\ \Eprint
  {http://arxiv.org/abs/2307.10391} {arXiv:2307.10391 [hep-th]} \BibitemShut
  {NoStop}%
\bibitem [{\citenamefont {Nagar}\ and\ \citenamefont
  {Rezzolla}(2005)}]{Nagar:2005ea}%
  \BibitemOpen
  \bibfield  {author} {\bibinfo {author} {\bibfnamefont {Alessandro}\
  \bibnamefont {Nagar}}\ and\ \bibinfo {author} {\bibfnamefont {Luciano}\
  \bibnamefont {Rezzolla}},\ }\bibfield  {title} {\enquote {\bibinfo {title}
  {{Gauge-invariant non-spherical metric perturbations of Schwarzschild
  black-hole spacetimes}},}\ }\href {\doibase 10.1088/0264-9381/22/16/R01}
  {\bibfield  {journal} {\bibinfo  {journal} {Class. Quant. Grav.}\ }\textbf
  {\bibinfo {volume} {22}},\ \bibinfo {pages} {R167} (\bibinfo {year}
  {2005})},\ \bibinfo {note} {[Erratum: Class.Quant.Grav. 23, 4297 (2006)]},\
  \Eprint {http://arxiv.org/abs/gr-qc/0502064} {arXiv:gr-qc/0502064}
  \BibitemShut {NoStop}%
\bibitem [{\citenamefont {{Friedman}}(1978)}]{1978CMaPh..62..247F}%
  \BibitemOpen
  \bibfield  {author} {\bibinfo {author} {\bibfnamefont {John~L.}\ \bibnamefont
  {{Friedman}}},\ }\bibfield  {title} {\enquote {\bibinfo {title} {{Generic
  instability of rotating relativistic stars}},}\ }\href {\doibase
  10.1007/BF01202527} {\bibfield  {journal} {\bibinfo  {journal}
  {Communications in Mathematical Physics}\ }\textbf {\bibinfo {volume} {62}},\
  \bibinfo {pages} {247--278} (\bibinfo {year} {1978})}\BibitemShut {NoStop}%
\bibitem [{\citenamefont {{Chandrasekhar}}(1975)}]{1975RSPSA.343..289C}%
  \BibitemOpen
  \bibfield  {author} {\bibinfo {author} {\bibfnamefont {S.}~\bibnamefont
  {{Chandrasekhar}}},\ }\bibfield  {title} {\enquote {\bibinfo {title} {{On the
  Equations Governing the Perturbations of the Schwarzschild Black Hole}},}\
  }\href {\doibase 10.1098/rspa.1975.0066} {\bibfield  {journal} {\bibinfo
  {journal} {Proceedings of the Royal Society of London Series A}\ }\textbf
  {\bibinfo {volume} {343}},\ \bibinfo {pages} {289--298} (\bibinfo {year}
  {1975})}\BibitemShut {NoStop}%
\bibitem [{\citenamefont {Blanchet}(2014)}]{Blanchet:2013haa}%
  \BibitemOpen
  \bibfield  {author} {\bibinfo {author} {\bibfnamefont {Luc}\ \bibnamefont
  {Blanchet}},\ }\bibfield  {title} {\enquote {\bibinfo {title} {{Gravitational
  Radiation from Post-Newtonian Sources and Inspiralling Compact Binaries}},}\
  }\href {\doibase 10.12942/lrr-2014-2} {\bibfield  {journal} {\bibinfo
  {journal} {Living Rev. Rel.}\ }\textbf {\bibinfo {volume} {17}},\ \bibinfo
  {pages} {2} (\bibinfo {year} {2014})},\ \Eprint
  {http://arxiv.org/abs/1310.1528} {arXiv:1310.1528 [gr-qc]} \BibitemShut
  {NoStop}%
\bibitem [{\citenamefont {Most}(2024)}]{talk_with_Elias}%
  \BibitemOpen
  \bibfield  {author} {\bibinfo {author} {\bibfnamefont {Elias~R}\ \bibnamefont
  {Most}},\ }\href@noop {} {}\bibinfo {howpublished} {personal communication}
  (\bibinfo {year} {2024})\BibitemShut {NoStop}%
\bibitem [{\citenamefont {Mano}\ \emph {et~al.}(1996)\citenamefont {Mano},
  \citenamefont {Suzuki},\ and\ \citenamefont {Takasugi}}]{Mano_1996}%
  \BibitemOpen
  \bibfield  {author} {\bibinfo {author} {\bibfnamefont {S.}~\bibnamefont
  {Mano}}, \bibinfo {author} {\bibfnamefont {H.}~\bibnamefont {Suzuki}}, \ and\
  \bibinfo {author} {\bibfnamefont {E.}~\bibnamefont {Takasugi}},\ }\bibfield
  {title} {\enquote {\bibinfo {title} {Analytic solutions of the regge-wheeler
  equation and the post-minkowskian expansion},}\ }\href {\doibase
  10.1143/ptp.96.549} {\bibfield  {journal} {\bibinfo  {journal} {Progress of
  Theoretical Physics}\ }\textbf {\bibinfo {volume} {96}},\ \bibinfo {pages}
  {549–565} (\bibinfo {year} {1996})}\BibitemShut {NoStop}%
\end{thebibliography}%
